\documentclass[12pt]{article}
\usepackage{jheppub}
\usepackage{slashed}
\usepackage{multirow}
\usepackage{float}

\newcommand{\met}{$\not \!\! E_T$}

\title{Measuring Invisible Particle Masses Using a Single Short Decay Chain}
\author{Hsin-Chia Cheng and Jiayin Gu}

\affiliation{Department of Physics, University of California, Davis, CA 95616, U.S.A.}

\abstract{We consider the mass measurement at hadron colliders for a decay chain of two steps, which ends with a missing particle. Such a topology appears as a subprocess of signal events of many new physics models which contain a dark matter candidate. From the two visible particles coming from the decay chain, only one invariant mass combination can be formed and hence it is na\"ively expected that the masses of the three invisible particles in the decay chain cannot be determined from a single end point of the invariant mass distribution. We show that the event distribution in the $\log(E_{1T}/E_{2T})$ vs. invariant mass-squared plane, where $E_{1T}$, $E_{2T}$ are the transverse energies of the two visible particles, contains the information of all three invisible particle masses and allows them to be extracted individually. The experimental smearing and combinatorial issues pose challenges to the mass measurements. However, in many cases the three invisible particle masses in the decay chain can be determined with reasonable accuracies.}

\begin{document}
\maketitle

\section{Introduction}

Two major questions that particle physics are facing today are the origin of the electroweak symmetry breaking and the identity of the dark matter in the universe. The answers to both questions may lie on the physics at the TeV scale. For the first question the TeV scale is the scale of the electroweak symmetry breaking. It is widely expected that there will be new states at the TeV scale to stabilize the electroweak scale, solving the hierarchy problem of the Higgs sector. The second question is related to the TeV scale because the most promising dark matter candidate is a new stable weakly interacting massive particle (WIMP) at the weak scale. The thermal relic left from the Big Bang for such a particle provides the right amount of dark matter in the current universe. It is quite possible that these two questions are related and have their common origin from the TeV scale physics. The Large Hadron Collider (LHC) is currently running and searching for new physics at the TeV scale intensively. Discoveries may be made at any time to provide us important clues to these fundamental questions.

Even though we do not know the exact new theory at the TeV scale, a typical collider signature for new physics containing the dark matter is the missing transverse energy (\met) in an event. To ensure the stability of the dark matter particle, a new symmetry is often needed such that the dark matter particle is the lightest particle charged under this new symmetry while all standard model (SM) particles are neutral. There can be other new particles at the TeV scale which are charged under this new symmetry, as it is often the case in a more complete theory, including the most popular supersymmetric standard model (SSM) where the $R$-parity plays the role of the the new symmetry~\cite{Fayet:1977yc,Farrar:1978xj,Dimopoulos:1981dw,Dimopoulos:1981zb}. There are many other models which share the same feature, {\ e.g., universal extra dimensions (UEDs)~\cite{Appelquist:2000nn,Cheng:2002iz,Cheng:2002ab} and little Higgs theories with $T$-parity~\cite{Cheng:2003ju,Cheng:2004yc,Low:2004xc}. These new particles are necessarily pair-produced at the collider due to the new symmetry. After being produced, they decay down to the dark matter particles which escape the detector, yielding missing energies, together with SM particles coming from the decays. The standard signatures for this class of models are therefore jets and/or leptons with \met.

Missing transverse energy is an important channel to search for new physics at the colliders. Events with large missing transverse energies are easy to identify. However, once such signals beyond the SM backgrounds are discovered, it is a non-trivial task to identify what these new physics events are and to reconstruct their kinematics. With at least two missing particles in each event, in general one cannot reconstruct the full kinematics on an event-by-event basis without additional information. Some more sophisticated techniques need to be developed if one wants to determine the properties of the invisible particles including the dark matter particle which appear in the process. One of the first things that we would like to know about the new particles are their masses. Once the masses are determined, they can be used to reconstruct the kinematics of the events, and hence to help determining other properties of these new particles. In the past decade, there have been many techniques developed to determine the masses of the new particles in the decay chains which end up with missing particles. (See Ref.~\cite{Barr:2010zj} for a review.) The starting point is to isolate the signal events that  predominantly come from a particular topology\footnote{The event topology can be tested by looking at distributions of various kinematic variables such as the invariant masses~\cite{Bai:2010hd}.} and then find the appropriate kinematic variables or constraints which depend on the masses of the new particles. Some variables, e.g., the invariant mass end point~\cite{Paige:1996nx,Hinchliffe:1996iu} and the stransverse $M_{T2}$~\cite{Lester:1999tx,Barr:2003rg,Cheng:2008hk}) can be used in a wide variety of event topologies, but in general the best mass determination may require different techniques for different event topologies.

In recent years many studies have been focussed on symmetric decay chains~\cite{Cheng:2007xv,Cho:2007qv,Barr:2007hy,Cho:2007dh,Ross:2007rm,Barr:2008ba,Cheng:2008mg,Cheng:2009fw,Nojiri:2008hy,Nojiri:2008vq,Burns:2008va,Bai:2009it,Konar:2009wn,Nojiri:2010dk}. The advantage of considering symmetric decay chains is that there are additional kinematic constraints coming from mass equalities between the particles in the two decay chains and the total missing transverse momentum, which allowed accurate mass determination even for relatively short decay chains. On the other hand, in a model where many different decay patterns can happen, there are often many more events with asymmetric decay chains. There have been attempts to generalize the techniques developed for symmetric decay chains to asymmetric chains, but so far only to different mother or daughter particle masses between the two chains, and some prior knowledge about the asymmetry of the events is required~\cite{Barr:2009jv,Konar:2009qr}. To be complete general, it is worthwhile to find kinematic techniques or variables which can be applied to a single decay chain. Then they can be used even in a collection of different types of signal events as long as they contain one identical decay chain. In the case of limited statistics, they may give the first estimate of the masses before there are enough symmetric-chain events for analysis.

In fact, the early attempts of the mass determination in SUSY-like events were based on single decay chains~\cite{Allanach:2000kt,Kawagoe:2004rz,Gjelsten:2004ki,Gjelsten:2005aw,Miller:2005zp,Burns:2009zi,Matchev:2009iw}. However, to obtain enough kinematic constraints, long decay chains with at least 3 steps are required. There are many invariant mass combinations which can be formed with 3 or more visible particles, which provide enough mass relations among the invisible particles in the decay chain. So far there is no technique for  extracting the masses in a shorter decay chain with only two or less visible particles. Na\"ively one may imagine that this task is impossible. For example, consider a decay chain starting with a mother particle $Y$ which decays through an intermediate on-shell state $X$ and ends up with the missing particle $N$ as shown in Fig.~\ref{a0}. There is one visible particle from each step of the decays and the visible particles are labeled as 1 and 2 in the figure. 
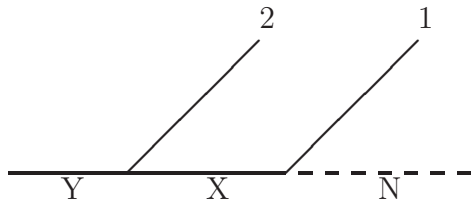
\begin{figure}
\centering
\begin{picture}(180,80)
  \thicklines
  \put(0,10){\line(1,0){100}}
  \multiput(100,10)(10,0){8}{\line(1,0){5}}
  \put(45,10){\line(1,1){50}}
  \put(105,10){\line(1,1){50}}
  \put(20,0){Y}
  \put(75,0){X}
  \put(140,0){N}
  \put(95,65){2}
  \put(155,65){1}
\end{picture}
\caption{Decay chain}
\label{a0}
\end{figure}
There are three invisible particles ($Y,\, X,\, N$) in the decay chain but there is only one invariant mass that can be formed from the two visible particles. One invariant end point certainly cannot determine three masses, but only provides one relation among them. It is also well known that the shape of the invariant mass distribution depends on  the spin of the intermediate particle $X$ and the chiralness of its couplings, but not the masses~\cite{Barr:2004ze,Smillie:2005ar,Datta:2005zs,Alves:2006df,Athanasiou:2006ef,Wang:2006hk,Smillie:2006cd,Kilic:2007zk,Csaki:2007xm,Burns:2008cp,Gedalia:2009ym,Ehrenfeld:2009rt,Cheng:2010yy}. One needs to find two more independent mass relations in order to solve for the three masses. In this work we make the first attempt in a model-independent way to measure the invisible particle masses of the short single decay chain of Fig.~\ref{a0}. We show that in certain cases, it is \emph{possible} to determine all three masses in the decay chain. We are not aiming for high accuracies. After all, this is a difficult case so even a 30\%--50\% measurement is infinitely better than not being able to determine them at all. Any information obtained or techniques developed here can also be used in more complicated event topologies which contain this decay chain as a subprocess to provide additional constraints. Furthermore, the variables used for mass determinations are often useful to separate signals from backgrounds and hence can be used in new physics search in the first place.

A crucial observation is that the masses are mostly determined by certain special events (and their nearby events). Those events often lie at the end point or the peak of some kinematic distribution. Mathematically, those special events make the kinematic constraint equations degenerate. In order for the degenerate equations to have solutions (since the correct masses should be compatible with those events), this implies certain relations among the coefficients of the constraint equations, which depend on the invisible particle masses. In this way we obtain mass relations among the invisible particles. It has been shown that the common variables such as the end points of the invariant mass and the transverse mass variables $M_T$, $M_{T2}$ can all be understood in this way~\cite{Kim:2009si}. There some constraint equations used are quadratic so the special events lie at the end point of a kinematic distribution. The regions near these points are often where signal events accumulate due to the phase space restriction, so sometimes they can also be used to select signals over backgrounds. 

For the decay chain shown in Fig.~\ref{a0}, the constraint equations are
\begin{eqnarray}
p_Y^2 &=& m_Y^2 ,\\
(p_Y - p_2)^2 &=& m_X^2 ,\\
(p_Y - p_2 - p_1)^2 &=& m_N^2,
\end{eqnarray}
where $p_2$ and $p_1$ are the 4-momenta of the visible particles from the decays of $Y$ and $X$ respectively and for simplicity we take them to be massless $p_2^2=p_1^2 =0$ which is a good approximation for most SM visible particles. Since we do not use the information from the other decay chain, there is no constraint from the missing transverse momentum. Taking the differences of these three equations, we can obtain two linear equations in the unknown momentum $p_Y$,
\begin{eqnarray}
2 p_2 \, p_Y = m_Y^2 - m_X^2 \equiv \Delta_2, \label{eq:lin2}\\
2 p_1 \, p_Y - 2 p_1 \, p_2 = m_X^2 - m_N^2 \equiv \Delta_1 \label{eq:lin1}, 
\end{eqnarray}
where $\Delta_{1,2}$ are defined as the corresponding mass-squared differences. Together with any one of the quadratic equation above, we have all the independent kinematic constraints. From the discussion in the previous paragraph we should look at the events which make these equations degenerate. (For the quadratic equation we just take the tangent on the curve.) The events with visible momenta $p_1,\, p_2$ which make the three equations (including the tangent of the quadratic equation) degenerate can be shown to satisfy
\begin{equation}
\label{eq:inv_mass}
(p_1 +p_2)^2 = \frac{(m_Y^2- m_X^2)(m_X^2 - m_N^2)}{m_X^2} = \frac{\Delta_1 \, \Delta_2}{m_X^2},
\end{equation}
which is nothing but the well-known end point of the invariant mass distribution of the visible particle system, $m^2_{12, \rm{max}}=(p_1+p_2)^2_{\rm max}$. By locating the end point of the invariant mass distribution, we obtain one relation among the three masses $m_Y,\, m_X,\, m_N$, eq.~(\ref{eq:inv_mass}). 

Because this decay chain can have any boost and orientation in the laboratory (lab) frame, one might think that the Lorentz-invariant mass is the only meaningful quantity in this process and eq.~(\ref{eq:inv_mass}) is the only mass-dependent kinematic variable that we can get. However, by examining the two linear equations, eqs.~(\ref{eq:lin2}),(\ref{eq:lin1}), one can see that there is another case where the two equations become degenerate, that is, when the two visible particles are parallel, $p_1 \propto p_2$. In this case the invariant mass vanishes, $(p_1+p_2)^2= 2 p_1 \, p_2 = 0$. If we take the ratio of the two linear equations in this case, we have
\begin{equation}
\label{eq:ratio}
\frac{E_1}{E_2} = \frac{\Delta_1}{\Delta_2},
\end{equation}
where $E_1$ and $E_2$ are the energies of the particle 1 and particle 2 respectively. Because the two 4-momenta are parallel, the ratio is invariant under any Lorentz transformation, and it gives the second relation among the three unknown masses.\footnote{This relation was also used in Ref.~\cite{Nojiri:2000wq}. We thank M. Nojiri for bringing it to our attention.} One may question whether such a co-linear degeneracy is useful in practice, because it rarely happens and depending on the nature of the visible particles, the two visible particles may not be resolvable in a real experiment. So, instead of using only the extremely rare special events when the two visible particles are parallel, we will look at the event distribution in the $E_1/E_2$ vs. $m^2_{12}$ space (more precisely $E_{1T}/E_{2T}$ where the subscript denotes the transverse energy as it is invariant under longitudinal boosts). We find that the two-dimensional distribution contains useful information and under certain circumstances, it can even provide one additional mass relation which allows us to solve for all three invisible particle masses. Even in the cases where three masses can not be independently and accurately determined, the two-dimensional event distribution provides information about the decay chain beyond what is contained in the one-dimensional invariant mass distribution.

This paper is organized as follows.
In section~\ref{ti} we analyze at the parton level the event distribution of the decay chain in the $\log(E_{1T}/E_{2T})$ vs. invariant mass-squared plane. We show that the masses of the three invisible particles can be obtained by fitting the distribution with a simple curve. Some more examples are presented in appendix~\ref{sec:other} to show that the method applies to a wide range of models and exceptions are included in appendix~\ref{sec:fail}. In section~\ref{real}, we consider the method in more realistic situations by including experimental smearing effects and combinatorial backgrounds. The case when the order of the two visible particles cannot be identified on an event-by-event basis is discussed in section~\ref{fold}. The combinatorial problems in the case when there are other visible particles identical to one of the visible particles from the decay chain in the same event are discussed in section~\ref{comb}. Conclusions are drawn in section~\ref{sec:conclusions}.

\section{The Kinematic Distribution for the Single Decay Chain}
\label{ti}

In the Introduction, we have argued that for the decay chain in Fig.~\ref{a0}, in addition to the invariant mass end point, a second mass relation can be obtained by looking at the energy ratio of the events where the two visible particles are parallel. However, such events are rare and may not be usable in real experiments. To avoid these problems, a natural thought is to look at the ``nearby'' events to see if one can extrapolate to the point that we are interested in. To do that we should examine the event distribution in the two-dimensional space of the energy ratio of the two visible particles and their invariant mass.

To study this with a concrete example, a sample of events from such a decay chain are generated with the SUSY LM2 point~\cite{LM2,Battaglia:2003ab} chosen as the underlying model. The masses of the particles in the decay chain are given in Table~\ref{td0}.\footnote{This point has been ruled out at the LHC~\cite{daCosta:2011qk,Khachatryan:2011tk,Chatrchyan:2011ek}. However, we just use this point for the purpose of illustration. The method that we develop is independent of the overall mass scale of the spectrum.}
\begin{table}
\begin{tabular}{|c|c|c|c|c|c|}
\hline
     & $Y$ & $X$ & $N$ & 2 & 1 \\ \hline \hline
   particle & left-handed down squark  & 2nd chargino & anti-sneutrino &  up quark & electron \\ \hline
   mass[GeV] & 777 & 465 & 292 & 0 & 0 \\ \hline
 \end{tabular}
 \caption{A summary of the decay chain studied in Section \ref{ti}.  The visible Standard Model particles are treated as being massless.}
 \label{td0}
 \end{table} 
$10^4$ events are generated using MadGraph 4.4.49~\cite{Alwall:2007st} at the parton level to reduce the statistical fluctuations. In this section we do not include experimental smearing and backgrounds, and assume no combinatorial problem. These issues will be discussed in the next section when we deal with realistic experimental situations. The invariant mass-squared distribution of the two visible particles is shown in Fig.~\ref{d1}. 
\begin{figure}
\centering
\includegraphics[width=8cm]{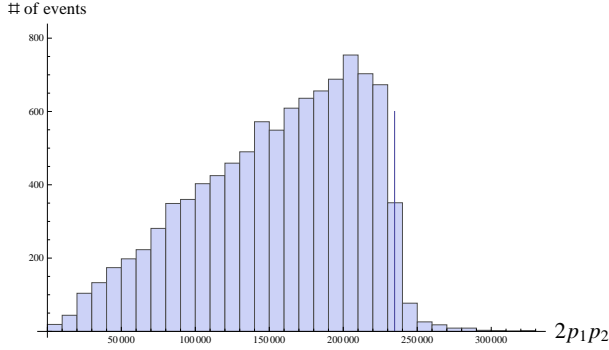}
\caption{The histogram of the invariant mass-squared of the two visible particles.  In the massless limit $(p_1+p_2)^2 = 2p_1p_2$.  The vertical line indicates the value of $\frac{\Delta_1\Delta_2}{m^2_X}$.}
\label{d1}
\end{figure}
The distribution has an end point which can be clearly identified. The triangular shape of the distribution is a characteristic of the spin-1/2 intermediate state (chargino) in the decay chain. Now we would like to look at the distribution of the energy ratios. 
We take the logarithm of the energy ratio to make it more symmetric between the two particles. The two-dimensional distribution in the space of $\log( E_{1}/E_{2})$ vs.\ invariant mass-squared $2 p_1\, p_2$ is shown in Fig.~\ref{fig:2d}.
\begin{figure}
\centering
\includegraphics[width=10cm]{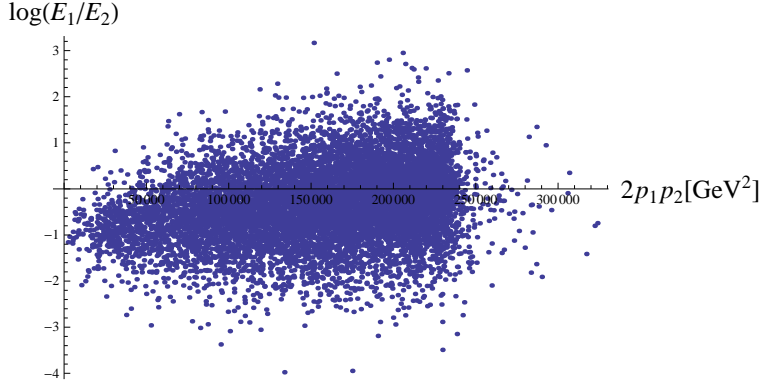}  
\caption{The scatter plot of $\log{(E_1/E_2)}$ vs. $2p_1p_2$ for all the events.}
\label{fig:2d}
\end{figure}
One can clearly see some interesting pattern of the distribution. As expected, the distribution of  $\log{(E_{1}/E_{2})}$ converges to a point at $\log (\Delta_1/\Delta_2)$ when the invariant mass goes to zero. Away from that point, the $\log{(E_{1}/E_{2})}$ distribution spread out and for a fixed invariant mass, it is more or less symmetric about some center point which moves up as the invariant mass increases. This qualitative feature already tells us some important information about the decay chain. It allows us to figure out which visible particle comes from the first step decay and which comes from the second in the case where the two visible particles are distinct, because if we switch $E_{1}$ and $E_{2}$ the distribution will move down instead. In our example it means that the quark jet being emitted before the lepton can be determined rather than assumed. 

To understand this distribution, let us rewrite eqs.~(\ref{eq:lin2}), (\ref{eq:lin1}) as
\begin{eqnarray}
2p_1p_Y &=& \Delta_1 + 2p_1p_2 \label{q1} ,\\
2p_2p_Y &=& \Delta_2 \label{q2} ,
\end{eqnarray}
and take the ratio between these two equations. We have
\begin{eqnarray}
\frac{\Delta_1 + 2p_1p_2}{\Delta_2} &=& \frac{p_1p_Y}{p_2p_Y} = \frac{E_1E_Y-\vec{p}_1\cdot\vec{p}_Y}{E_2E_Y-\vec{p}_2\cdot\vec{p}_Y} = \frac{E_1E_Y-|\vec{p}_1||\vec{p}_Y|\cos{\theta_{1Y}}}{E_2E_Y-|\vec{p}_2||\vec{p}_Y|\cos{\theta_{2Y}}} \bigg|_{\rm lab} \nonumber\\
&=& \frac{E_1E_Y(1-\beta_Y \cos{\theta_{1Y}})}{E_2E_Y(1-\beta_Y \cos{\theta_{2Y}})}\bigg|_{\rm lab}   \nonumber\\
&=& \frac{E_1(1-\beta_Y \cos{\theta_{1Y}})}{E_2(1-\beta_Y \cos{\theta_{2Y}})}\bigg|_{\rm lab}, \label{q3}
\end{eqnarray}
where the $ \cos{\theta_{1(2)Y}}$ is the angle between particle 1(2) and particle $Y$, the subscript ``lab" indicates that the angles are measured in the lab frame and $\beta_Y$ is the magnitude of the velocity (boost) of particle $Y$, defined as $\beta_Y = {|\vec{p}_Y|}/{E_Y}$.  It is easy to see that eq.~(\ref{q3}) reduces to the simple relation,  ${\Delta_1}/{\Delta_2}={E_1}/{E_2}$, when particle 1 \& 2 are parallel to each other. Now taking the logarithm of eq.~(\ref{q3}) we obtain
\begin{eqnarray}
\log{\frac{E_1}{E_2}} &=& \log{\frac{\Delta_1 + 2p_1p_2}{\Delta_2}} +\log{\frac{1-\beta_Y \cos{\theta_{2Y}}}{1-\beta_Y \cos{\theta_{1Y}}}}\bigg|_{\rm lab} \nonumber\\
&=&  \log{\frac{\Delta_1 + 2p_1p_2 }{\Delta_2}} +\log{\frac{1+\beta_Y \cos{\theta_{1Y}}}{1+\beta_Y \cos{\theta_{2Y}}}}\bigg|_{Y}, \label{q4}
\end{eqnarray}
where the subscript ``$Y$" denotes that the angles are measured in the rest frame of particle $Y$.  More specifically, $ \cos{\theta_{1(2)Y}}\big|_{Y}$ is the angle between particle 1(2)  (measured in the rest frame of particle $Y$) and particle $Y$ (measured in the lab frame).  From the first line to the second line of eq.~(\ref{q4}) we simply perform a Lorentz transformation and use (assuming particle 1 \& 2 are massless) 
\begin{equation}
 \cos{\theta_{1(2)Y}}\big|_{\rm lab} = \frac{ \cos{\theta_{1(2)Y}}+\beta_Y}{1+\beta_Y  \cos{\theta_{1(2)Y}}} \bigg|_Y .
 \end{equation}  
 The reason for writing the expression in terms of the angles in the rest frame of particle $Y$ is that $\cos{\theta_{2Y}}\big|_Y$ is directly related to the polarization of particle $Y$.

\begin{figure}
\centering
\includegraphics[width=11cm]{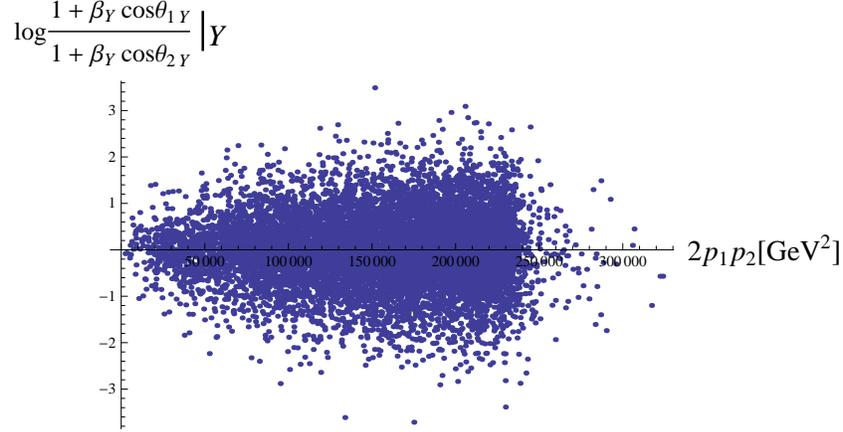}
\caption{The distribution of $\log{\frac{1+\beta_Y \cos{\theta_{1Y}}}{1+\beta_Y \cos{\theta_{2Y}}}}\big|_{Y}$ as a function of the invariant mass-squared.  The average of $\log{\frac{1+\beta_Y \cos{\theta_{1Y}}}{1+\beta_Y \cos{\theta_{2Y}}}}\big|_{Y}$ is $-0.011$.  It is roughly distributed evenly around zero and peaks at zero.}
\label{spread}
\end{figure}

The left-hand side of eq.~(\ref{q4}) can be measured experimentally. The first term on the right-hand side of the equation involves the unknown masses to be determined and the invariant mass of the visible particles which is also measurable. The second term on the right-hand side, on the other hand, is not measurable on an event-by-event basis. It involves the unknown momentum of the invisible particle $Y$. When we plot the events on the $\log (E_1/E_2)$ vs. $2 p_1 p_2$ plane, this term causes the spread in the vertical direction. Nevertheless, if the directions of particles 1 and 2 measured in the rest frame of $Y$ are not correlated with the direction of $Y$ itself, we expect that the second term will be evenly distributed around zero and peak at zero.
Fig.~\ref{spread} shows the distribution of $\log{\frac{1+\beta_Y \cos{\theta_{1Y}}}{1+\beta_Y \cos{\theta_{2Y}}}}\big|_{Y}$ as a function of the invariant mass-squared and one can see that this is roughly true for any invariant mass. In this case, we can fit the distribution with a two-parameter curve
\begin{equation}
\log \frac{E_1}{E_2} = \log \frac{\widetilde{\Delta}_1+2 p_1 p_2}{\widetilde{\Delta}_2},
\end{equation}
which has the least total $\chi^2$ measured in vertical distances. Consequently, the two fitted parameters $\widetilde{\Delta}_1$, $\widetilde{\Delta}_2$ give estimates of the two mass-squared differences $\Delta_1$, $\Delta_2$. As a result, $\Delta_1$ and $\Delta_2$ can be determined individually, not just their ratio. Combined with the end point of the invariant mass-squared, $\Delta_1 \Delta_2/ m_X^2$, they can be inverted to solve all 3 unknown masses. 

Actually, eq.~(\ref{q4}) can be applied in any longitudinal frame as we can give this system an arbitrary longitudinal boost. In particular, one can boost each event to a frame where $E_1/E_2 = E_{1T}/E_{2T}$. This in general removes a large longitudinal boost of the mother particle $Y$ which happens when the center of mass of the collision is highly boosted. Therefore, we can use the distribution in the $\log (E_{1T}/E_{2T})$ vs.\ $2 p_1 p_2$ space instead. Such a distribution is invariant under the longitudinal boosts and hence it is less sensitive to the center of mass of the production mechanism. It is also obvious for parallel massless visible particles, $E_{1T}/E_{2T} = E_1/E_2$. As shown at the end of this section, we find that the fit in the transverse energy ratio  space works somewhat better than in the total energy ratio space for most of the cases, so we will use the transverse energies in our variable $\log (E_{1T}/E_{2T})$ throughout the paper.

Fig.~\ref{d2} shows the fitted result of the scatter plot in the $\log (E_{1T}/E_{2T})$ vs.\ $2 p_1 p_2$ plane. The red curve is the function
\begin{equation}
\log \frac{E_{1T}}{E_{2T}} = \log \frac{\Delta_1 + 2 p_1 p_2}{\Delta_2}
\label{eq:master}
\end{equation}
for true $\Delta_1$ and $\Delta_2$.  The blue curve is the least square fit to the data of the same function (\ref{eq:master}) but 
treating $\Delta_1$, $\Delta_2$ as fitting parameters.
\begin{figure}
\centering
\includegraphics[width=11cm]{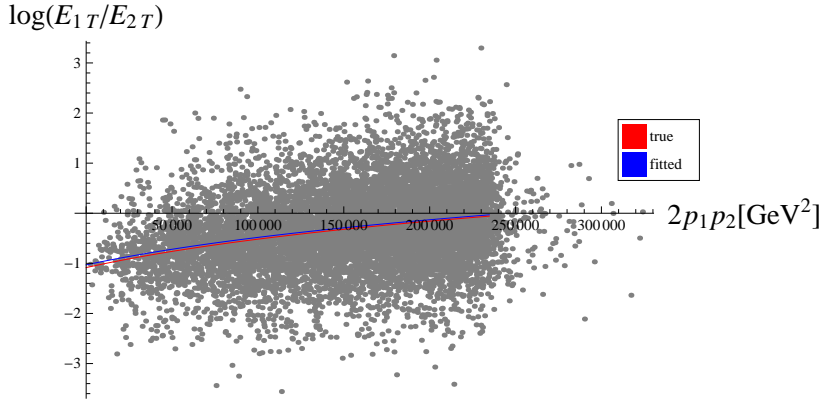}
\caption{The scatter plot of $\log{(E_{1T}/E_{2T})}$ vs. $2p_1p_2$ for all the events.  The red curve is $\log{(E_{1T}/E_{2T})}=\log{(\frac{\Delta_1+2p_1p_2}{\Delta_2})}$.  The blue curve is a least square fit to the data with the same function, treating $\Delta_1$ and $\Delta_2$ as unknown parameters.  The standard deviations for all the points are assumed to be the same.}
\label{d2}
\end{figure}
We can see that indeed the two curves are very close to each other. The term $\log{\frac{1+\beta_Y \cos{\theta_{1Y}}}{1+\beta_Y \cos{\theta_{2Y}}}}\big|_{Y}$ is indeed evenly distributed around zero in this frame for this case. Assuming that we can measure the invariant mass end point accurately, we can then reconstruct the 3 invisible particle masses $m_Y,\, m_X,\, m_N$ with the fitted values for $\Delta_1$ and $\Delta_2$. The result is shown in Table \ref{td1}. We see that, compared to the true values, the reconstructed masses are quite accurate.
\begin{table}
\begin{tabular}{|c|c|c|c|c|c|c|}
\hline
   & $\Delta_1[\mbox{GeV}^2]$ & $\Delta_2[\mbox{GeV}^2]$ & $\log{(\Delta_1/\Delta_2)}$ & $m_Y$[GeV] & $m_X$[GeV] & $m_N$[GeV] \\ \hline \hline
 true &  $1.310\times 10^5$  & $3.875\times 10^5$  & $-1.08$ & $777$ & $465$ & $292$ \\ \hline
 reconstructed &  $1.370\times 10^5$  & $3.838\times 10^5$  & $-1.03$ & $780$ & $473$ & $295$ \\ \hline
error &  $+4.6\%$  & $-0.96\%$  & $+5.5\%$ & $+0.34\%$ & $+1.8\%$ & $+1.0\%$ \\ \hline
 \end{tabular}
 \caption{True values, reconstructed values and the errors of the six quantities for the fit to all the events.  The errors are calculated using $\frac{\rm reconstructed-true}{\rm true}$ (except for $\log{(\Delta_1/\Delta_2)}$, which is ${\rm reconstructed-true}$) and do not represent the statistical fluctuation. In solving for the masses, we have used the true invariant mass end point value assuming that it can be accurately determined. The uncertainly in determining $\frac{\Delta_1\Delta_2}{m^2_X}$ will add additional error on the reconstructed values.}
 \label{td1}
 \end{table} 

We have checked this mass reconstruction method for models with different mass spectra and different particle spins in the decay chain, and we find that it works well for a wide range of different models and spectra. A few more examples can be found in Appendix~\ref{sec:other}.  However, there are two cases where our method falters:

1. The mother particle $Y$ is polarized and preferentially emits particle 2 in the forward or backward direction when it decays. In this case it is clear that $\cos \theta_{2Y} \big|_Y$ will have a nonzero average value. On the other hand, the direction of the particle 1 is less correlated with the polarization of the mother particle $Y$. The term $\log{\frac{1+\beta_Y \cos{\theta_{1Y}}}{1+\beta_Y \cos{\theta_{2Y}}}}\big|_{Y}$ will no longer be distributed evenly around zero, but has a bias depending on the $\cos \theta_{2Y} \big|_Y$ distribution. By Taylor expanding the expression
\begin{equation}
\log \frac{\Delta_1 + 2p_1 p_2}{\Delta_2} = \log \frac{\Delta_1}{\Delta_2} + \frac{2 p_1 p_2}{\Delta_1} + \cdots ,
\label{eq:expansion}
\end{equation}
we see that the curve has the intercept $\log (\Delta_1/\Delta_2)$ at $2 p_1 p_2 =0$ and the slope $1/\Delta_1$ to the first order. If $\langle \cos \theta_{2Y} \big|_Y \rangle >0$, we will obtain a fitted $\Delta_1$ (also $\Delta_2$ since their ratio is fixed by the intercept) greater than the true value, then the reconstructed masses will be too large, and vice versa.

2. The mass difference between two invisible particles (in particular between $Y$ and $X$) is small. The visible particle coming from the decay will be soft. There are two effects which can affect the mass determination. First, if the mass difference is not much larger than the width of one of the particles, the on-shell approximation does not work well. The off-shell effects can be asymmetric, which makes  the ``effective" value of $\Delta_2$ or $\Delta_1$ significantly different from the true value and introduces a bias to the term $\log{(\frac{\Delta_1+2p_1p_2}{\Delta_2})}$ in eq.~(\ref{q4}). Second,
if $\Delta_2$ is small and the typical energy of the visible particle 2 is not much larger than the $E_T$ cut (which is necessary in a real experiment), those emitted in the forward direction of particle $Y$ have a larger chance to pass the the trigger and the $E_T$ cut than those emitted in the backward direction. This introduces a ``fake" forward polarization to particle $Y$, which also causes a problem as we discussed in point 1. In practice, the mass determination is more of a problem for a small $\Delta_2$ than a small $\Delta_1$.

In these cases, our simple method does not give an accurate determination of the three invisible particle masses. However, the ratio $\Delta_1/\Delta_2$ is still usually well determined by the interception of the fitted curve with the axis of the zero invariant mass. If a third mass relation can be obtained in some other ways (e.g., double-chain events), this distribution can provide some other non-trivial information about this process such as the polarization of the particle $Y$. More detailed studies of these two cases will be given in Appendix~\ref{sec:fail}.

For the fit in Fig.~\ref{d2}, every event contributes with the same weight. In reality, backgrounds and noises are always present. There is no reason to expect that the background events will be distributed evenly around the $\log{(\frac{\Delta_1+2p_1p_2}{\Delta_2})}$ curve. The backgrounds can hence cause a bias in fitting. On the other hand, as long as background events are subdominant in the event sample, the central peak of the $\log (E_{1T}/E_{2T})$ distribution for a fixed $2 p_1 p_2$ may not be significantly affected. Therefore, a curve fitted to the peak locations of the $\log (E_{1T}/E_{2T})$ distributions for different invariant masses may be less sensitive to backgrounds than a direct $\chi^2$ fit to all events. We would like to check whether such a procedure gives the result as good as the result in Table~\ref{td1} in the absence of backgrounds first.

\begin{figure}
\centering
\includegraphics[width=10cm]{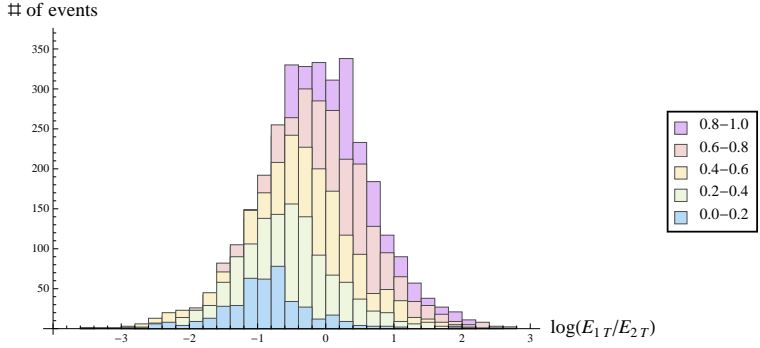}
\caption{Histograms of $\log{(E_{1T}/E_{2T})}$ for different ranges of invariant masses.  The chart on the righthand side indicates the range of invariant masses as a multiple of $\frac{\Delta_1\Delta_2}{m^2_X}$ for each histogram.  The histograms have approximate Gaussian shapes.}
\label{d3}
\end{figure}
In Fig.~\ref{d3} we divide the events into five sets with different ranges of invariant masses.  For each set of events, the histogram of $\log{(E_{1T}/E_{2T})}$ has an approximate Gaussian shape.  This indicates that we could divide the events according to their invariant masses and for each set we extract the peak value using a Gaussian fit.  After obtaining the peak point for each set, we then fit these peak points to the curve $\log{(E_{1T}/E_{2T})}=\log{(\frac{\Delta_1+2p_1p_2}{\Delta_2})}$.  The result of this fit is shown in Fig.~\ref{d4}, where the events are divided into 20 sets with equal width ($= 0.05\times {\Delta_1\Delta_2}/{m^2_X}$) in invariant mass-squareds.  The error bar of each point is estimated by the formula $\sigma[\bar{x}]={\sigma[x]}/{\sqrt{N}}$, where $\sigma[x]$ is obtained from the Gaussian fit and $N$ is the number of events in that set.  The reconstructed masses obtained from this fit is shown in Table~\ref{td2}. We see that the results are comparable to the ones in Table~\ref{td1}.
\begin{figure}
\centering
\includegraphics[width=11cm]{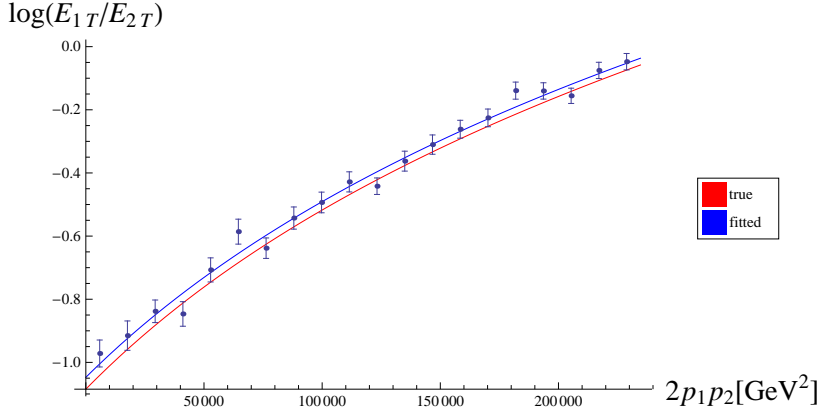}
\caption{The events are divide into 20 sets with equal width ($0.05\times \frac{\Delta_1\Delta_2}{m^2_X}$) of invariant mass.  For each set we extract the peak value by fitting with a Gaussian distribution.  The horizontal coordinate ($2p_1p_2$) of each point is the middle point of each division.  By doing this we have 20 points of peak values and we fit it with the curve $\log{(E_{1T}/E_{2T})}=\log{(\frac{\Delta_1+2p_1p_2}{\Delta_2})}$, treating $\Delta_1$ and $\Delta_2$ as unknown parameters.  The peak points and fitted curve are shown in blue.  The red curve is $\log{(E_{1T}/E_{2T})}=\log{(\frac{\Delta_1+2p_1p_2}{\Delta_2})}$ with true values of $\Delta_1$ and $\Delta_2$.  The error bar of each points is estimated by the formula $\sigma[\bar{x}]=\frac{\sigma[x]}{\sqrt{N}}$, where $\sigma[x]$ is obtained from the Gaussian fit and $N$ is the number of events in the set.}
\label{d4}
\end{figure}
\begin{table}
\begin{tabular}{|c|c|c|c|c|c|c|}
\hline
   & $\Delta_1[\mbox{GeV}^2]$ & $\Delta_2[\mbox{GeV}^2]$ & $\log{(\Delta_1/\Delta_2)}$ & $m_Y$[GeV] & $m_X$[GeV] & $m_N$[GeV] \\ \hline \hline
 true &  $1.310\times 10^5$  & $3.875\times 10^5$  & $-1.08$ & $777$ & $465$ & $292$ \\ \hline
 reconstructed &  $1.343\times 10^5$  & $3.827\times 10^5$  & $-1.05$ & $776$ & $468$ & $291$ \\ \hline
 error &  $+2.5\%$  & $-1.2\%$  & $+3.7\%$ & $-0.17\%$ & $+0.64\%$ & $-0.33\%$ \\ \hline
 \end{tabular}
 \caption{True values, reconstructed values and the errors of the six quantities for the fit to the peak points.  The errors are calculated using $\frac{\rm reconstructed-true}{\rm true}$ (except for $\log{(\Delta_1/\Delta_2)}$, which is ${\rm reconstructed-true}$) and do not represent the statistical fluctuation.}
 \label{td2}
 \end{table} 

 \begin{figure}
\centering
\includegraphics[width=7cm]{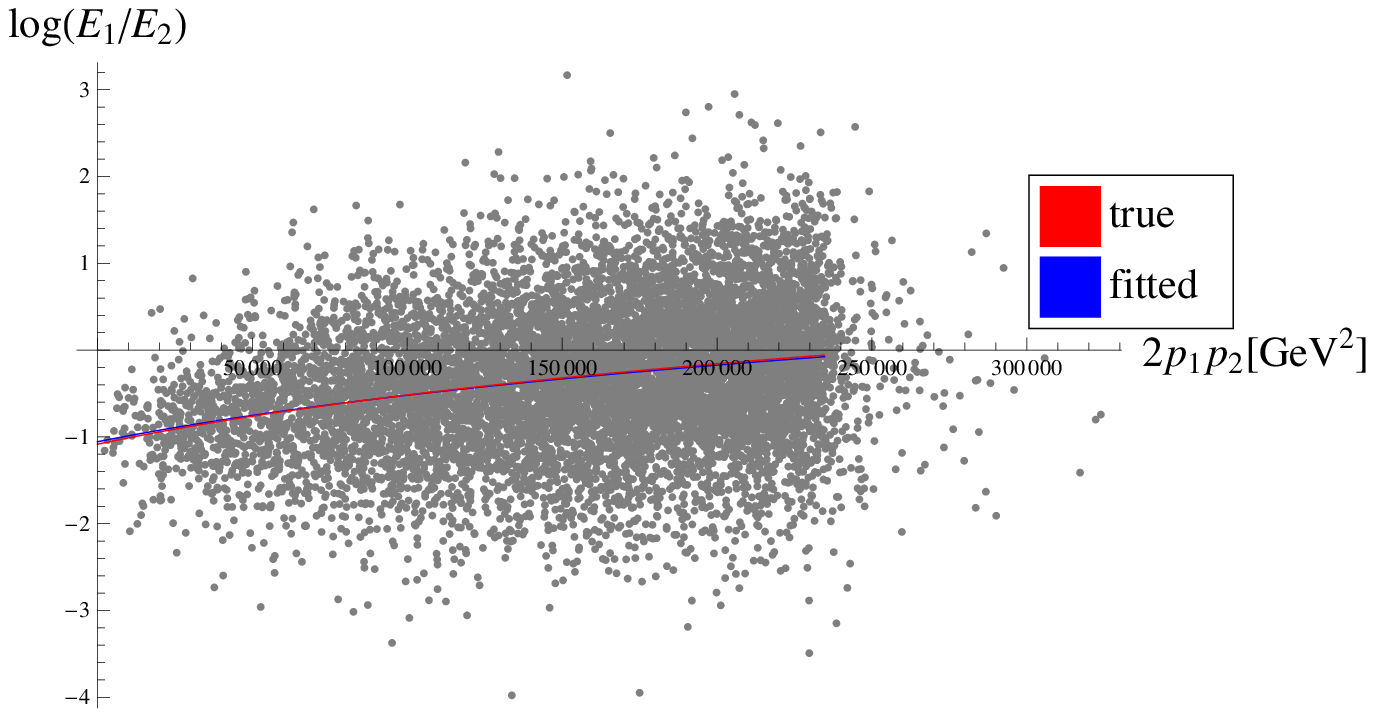}
\includegraphics[width=7cm]{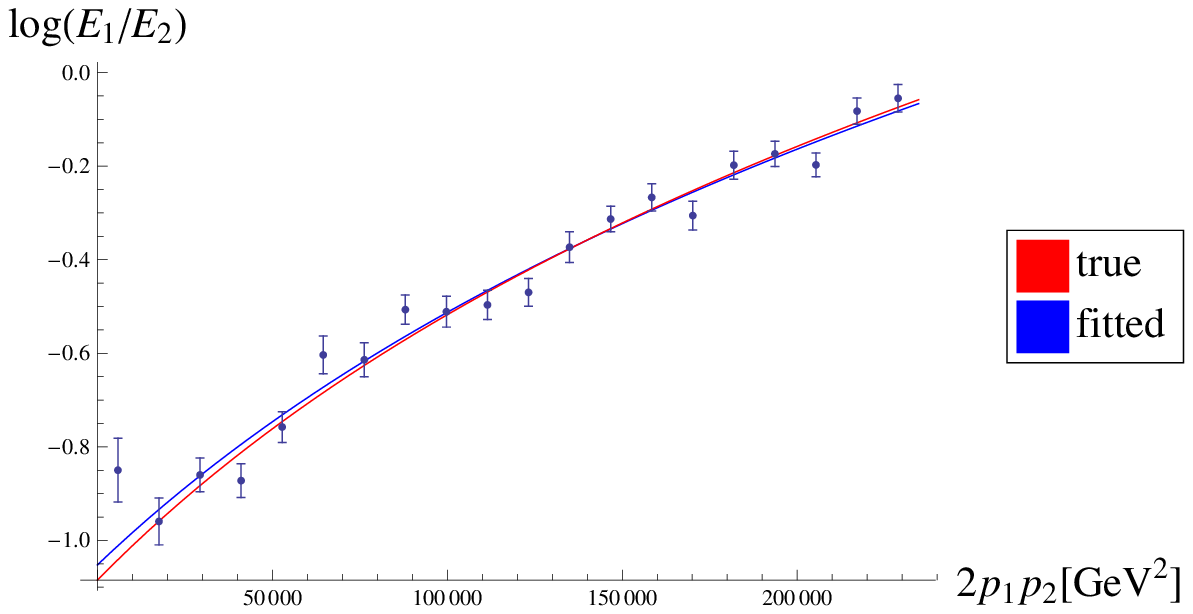}
\caption{The scatter plot (left) and the peak-point plot (right) of $\log{(E_{1}/E_{2})}$ vs. $2p_1p_2$ for the same set of events as in Fig.~\ref{d2} \& Fig.~\ref{d4}.  In both case the red curve is $\log{(E_{1}/E_{2})}=\log{(\frac{\Delta_1+2p_1p_2}{\Delta_2})}$ with true $\Delta_1$ and $\Delta_2$ and the blue curve is the fitted one.}
\label{d1s}
\end{figure}
 \begin{table}
 \centering
\begin{tabular}{|c|c|c|c|}
\hline
 &$m_Y$[GeV] & $m_X$[GeV] & $m_N$[GeV] \\ \hline 
 true & $777$ & $465$ & $292$ \\ \hline \hline
   \multicolumn{4}{|c|}{using transverse energies} \\ \hline
reconstructed & $764\pm 11$ & $456\pm 12$ & $280 \pm 11$ \\ \hline
error & $-1.6\%\pm 1.5\%$ & $-2.0\%\pm 2.6\%$ & $-4.2\%\pm 3.9\%$ \\ \hline \hline
  \multicolumn{4}{|c|}{using actual energies} \\ \hline
reconstructed & $821\pm 33$ & $511\pm 34$ & $336 \pm 32$ \\ \hline
error & $+5.7\%\pm 4.2\%$ & $+10.0\%\pm 7.4\%$ & $+15.1\%\pm 11.1\%$ \\ \hline
 \end{tabular}
 \caption{The reconstructed masses from 5 sets of events, each with $10^4$ events, in the form of mean $\pm$ standard deviation.  The reconstruction is obtained by fitting with the peak points, using both the transverse energies and the actual energies.  The standard deviations are estimated using the unbiased estimator (with a factor of $\sqrt{\frac{N}{N-1}}$) throughout this paper. The results suggest that using transverse energies results in smaller fluctuations in the reconstructed masses.}
 \label{td3}
 \end{table}  
 The errors in Table \ref{td1} and Table \ref{td2} could be a combination of both systematic and statistical errors.  Nevertheless, they are quite small and both reconstructions are good.  To further verify the goodness of our results, we repeated the event generation 5 times (with $10^4$ events for each set), and applied fits to the the peak points for the 5 sets of events. The mean and the standard deviation values of the reconstructed masses for the 5 sets are shown in Table~\ref{td3}.  We also compare with the results obtained from the same procedure on the same 5 sets of events, but using the ratio of the actual energies, $\log (E_1/E_2)$ in the lab frame, instead of the transverse energies. Fig.~\ref{d1s} shows the scatter plot and the fit to the peak points of $\log{(E_{1}/E_{2})}$ vs. $2p_1p_2$ for the same set of events as in Fig.~\ref{d2} \& Fig.~\ref{d4}.  The results from the 5 sets of events using the actual energies are also listed in Table~\ref{td3}.  We can see that using transverse energies gives better results and smaller fluctuations in the reconstructed masses.

\section{More Realistic Mass Measurements at Colliders}
\label{real}

In Sec.~\ref{ti} we saw that the event distribution in the transverse energy ratio vs. invariant mass-squared space can be used to determine all the invisible particle masses in the decay chain. It works very well under quite general conditions at the parton level. However, there are many complications in performing such a measurement in a real experiment. First, we must have a relatively clean sample of signal events to start with. Selecting signal events from backgrounds is non-trivial and depends on the type of signal events. This is beyond the scope of this paper and we assume that it can be achieved for the cases that we are interested in. Even if we have a pure sample of signal events, in general we have to face some combinatorial problems. If the two visible particles are not distinct or can appear in either order in the decay chain, { e.g.}, two leptons from a heavier neutralino decaying down to a lighter neutralino through a slepton in a SUSY theory, we do not know which order to take in the energy ratio $\log (E_{1T}/E_{2T})$. Another combinatorial issue is that there can be other experimentally identical particles appearing in the signal events and there is no absolute way to select the correct one. For example, if one of the visible particle is a jet, then we do expect other jets to be present in the same event, which can come from the other decay chain or initial state radiation (ISR).  We will discuss these two types of combinatorial problems in the following two subsections. In subsection~\ref{fold}, we consider an example of signal events with 2 leptons of the same flavor and opposite charges, which can appear in either order. We will see that mass determination still works for some cases but not for the others, depending on the mass parameters. In subsection~\ref{comb}, we consider a decay chain which produces 1 jet and 1 lepton with a definite order, but there are other jets coming from the other decay chain and/or from the initial state radiation in the same event. It is possible that both types of combinatorial problems are present simultaneously, as in the case when both visible particles in the decay chain are jets. Such cases will be difficult and we do not expect to achieve good mass measurements there.

Finally, experimental smearing of the visible particles from detector resolutions, fragmentation and hadronization in the case of a jet, will also deteriorate any mass measurement. The smearing effect is more important for jets than leptons. To take into account the experimental smearing effect, the parton level events are smeared according to the Gaussian errors listed in Table \ref{smt} in our studies in this section. They roughly correspond to the performance of the CMS detector~\cite{Ragusa:2007zz,cms1,cms2}.
\begin{table}
\centering
\begin{tabular}{|c|}
\hline
leptons: \\ \hline
$|\eta|<2.5$, $p_T>10$ GeV, \\
$\frac{\delta p_T}{p_T}=0.008 \oplus 0.00015 p_T$, \\
$\delta \theta =0.001$, $\delta \phi = 0.001$. \\ \hline \hline
jets: \\ \hline
$|\eta|<5.0$, $p_T>20$ GeV, \\
$\frac{\delta E_T}{E_T} =\bigg\{ \begin{matrix} \frac{5.6}{E_T} \oplus \frac{1.25}{\sqrt{E_T}} \oplus 0.033, \mbox{ for } |\eta|<1.4, \\
 \frac{4.8}{E_T} \oplus \frac{0.89}{\sqrt{E_T}} \oplus 0.043, \mbox{ for } |\eta|>1.4,  \end{matrix}$ \\
$\delta \eta = 0.03$, $\delta \phi =0.02$ for $ |\eta|<1.4$, \\
$\delta \eta = 0.02$, $\delta \phi =0.01$ for $ |\eta|>1.4$.\\ \hline
 \end{tabular}
 \caption{Parton level events are smeared according to the above Gaussian errors.  The cuts on $p_T$ and $\eta$ are consistent with the default cuts in MadGraph.  The observables of energy dimension are in GeV units and the angular and the rapidity variables are in radians.}
 \label{smt}
 \end{table}

\subsection{Combinatorial problems of the 2-lepton signal}
\label{fold}

In this subsection we consider the combinatorial problem between the two visible particles. As jets will have other problems, we consider that the two visible particles in the decay chain are leptons of the opposite charges and the same flavor.  This happens in SUSY if a heavier neutralino decays through a slepton to the lightest neutralino. The two leptons emitted can be in either charge order because the neutralinos are Majorana particles and the intermediate state can be either a slepton or an anti-slepton. To focus on this combinatorial problem, we assume that there is no other lepton of the same flavor in the event. Because we do not know which lepton is particle 1 and which is particle 2 on an event-by-event basis, we can not measure $\log(E_{1T}/E_{2T})$ but only its absolute value $|\log(E_{1T}/E_{2T})|$ ({i.e.}, the ratio between the larger $E_T$ and the smaller $E_T$).  In other words, the scatter plot in the $\log{(E_{1T}/E_{2T})}$ vs. $2p_1p_2$ plane is folded along the $\log{(E_{1T}/E_{2T})}=0$ axis.  This would certainly make the reconstruction more challenging.
 
Because the scatter plot is folded, the position of the distribution become crucial for reconstruction.  If the center of distribution is far away from the $\log{(E_{1T}/E_{2T})}=0$ line, the folding will not cause too much of a problem. One can still easily identify the peak of the $|\log(E_{1T}/E_{2T})|$ distribution of each invariant-mass interval, and the slope of the fitted curve connecting the peak points tells us the sign of $\log (\Delta_1/\Delta_2)$.  On the other hand, if the center of distribution is close to the $\log{(E_{1T}/E_{2T})}=0$ line, the folding makes it difficult to identify the pattern and the peak of the unfolded distribution, then reconstructing $\Delta_1$ and $\Delta_2$ individually by fitting a curve through the peak points may become impossible.

As the event distribution is folded, it is not good to fit a curve with minimal $\chi^2$ on the distribution directly. We will use the second technique discussed in the previous section by dividing the data into small intervals of invariant mass-squareds and then finding the peak point of the distribution for each set. To extract the peak point of the unfolded distribution, we fit each set with a folded Gassian disribution in $\log{(E_{1T}/E_{2T})}$, i.e., the Gaussian distribution also folded along the $\log{(E_{1T}/E_{2T})}=0$ point.  However, it is important to notice that the error for a Gaussian fit, $\sigma[\bar{x}]=\frac{\sigma[x]}{\sqrt{N}}$, is no longer a good estimate of the uncertainty if the distribution is folded.  If the center of the distribution is close to $\log{(E_{1T}/E_{2T})}=0$, the folded distribution may not be sensitive to the original peak position at all, and the uncertainty is much larger than that of the corresponding unfolded Gaussian distribution.  To estimate this uncertainty, we first use the maximum likelihood method to fit the distribution to a folded Gaussian distribution with two parameters, the mean $\mu$ and the standard deviation $\sigma$ of the unfolded Gaussian distribution.  We then find the contour $\log{L}=\log{L_{\rm max}}-{1}/{2}$ in the  $\mu$--$\sigma$ plane, and the tangent of this contour parallel to the $\sigma$-axis corresponds (approximately) to the boundary of the 68.3\% central confidence interval of $\mu$.  This method has some limitations, however, as the likelihood method requires knowledge of the full model up to a few free parameters. We do not know the exact shape of the original distribution, but simply treat it as approximately Gaussian. The estimations of the peak point and its uncertainty may be off if the original distribution is not close to a Gaussian.
 Once we obtain the estimations of the peak point and the uncertainty for each bin, we can fit the peak points with the folded curve $|\log{(E_{1T}/E_{2T})}|=|\log{(\frac{\Delta_1+2p_1p_2}{\Delta_2})}|$ to extract the values of $\Delta_1$ and $\Delta_2$.\footnote{To account for the asymmetric uncertainties but without too many complications, we assume a asymmetric Gaussian distribution for the peak point values and use a modified least square fit, i.e., the uncertainty used for the fit depends on whether the curve is above or below the point.}

\begin{table}
\centering
\begin{tabular}{|c|c|c|c|c|}
\hline
     & $m_Y$[GeV] & $m_X$[GeV] & $m_N$[GeV]  & $\log{\frac{\Delta_1}{\Delta_2}}$ \\ \hline
   case 1  & 468 & 187 & 140.5 & -2.5 \\ \hline
   case 2  & 468 & 237 & 140.5 & -1.5\\ \hline
   case 3 (LM2) & 468 & 304 & 140.5 & -0.56\\ \hline
   case 4  & 468 & 373 & 140.5 & 0.40\\ \hline
   case 5  & 468 & 407 & 140.5 & 1.0 \\ \hline
 \end{tabular}
 \caption{A summary of the mass spectra in subsection \ref{fold}.  The leptons (particles 1 \& 2) are treated as massless.}
 \label{fomass}
 \end{table} 
To examine how the mass determination depends on the value of $\log{{\Delta_1}/{\Delta_2}}$, we consider a SUSY process with a heavy neutralino (particle $Y$) decaying through an on-shell slepton (particle $X$) to the lightest neutralino (particle $N$), emitting a pair of leptons in the decays. The five examples of spectra that we choose to study are listed in Table~\ref{fomass}. We fix $M_Y$, $M_N$ and vary $M_X$ to obtain different values for $\log{{\Delta_1}/{\Delta_2}}$. The case 3 corresponds to the LM2 point with $Y$ being the 4th neutralino, but the actual model is not important. The events are generated through neutralino pair production and the polarization of the neutralino is close to zero. Since the neutralino is a Majorana particle, by the CP symmetry its decay is symmetric if all final states are included anyway. 

\begin{figure}
\centering
\includegraphics[width=7cm]{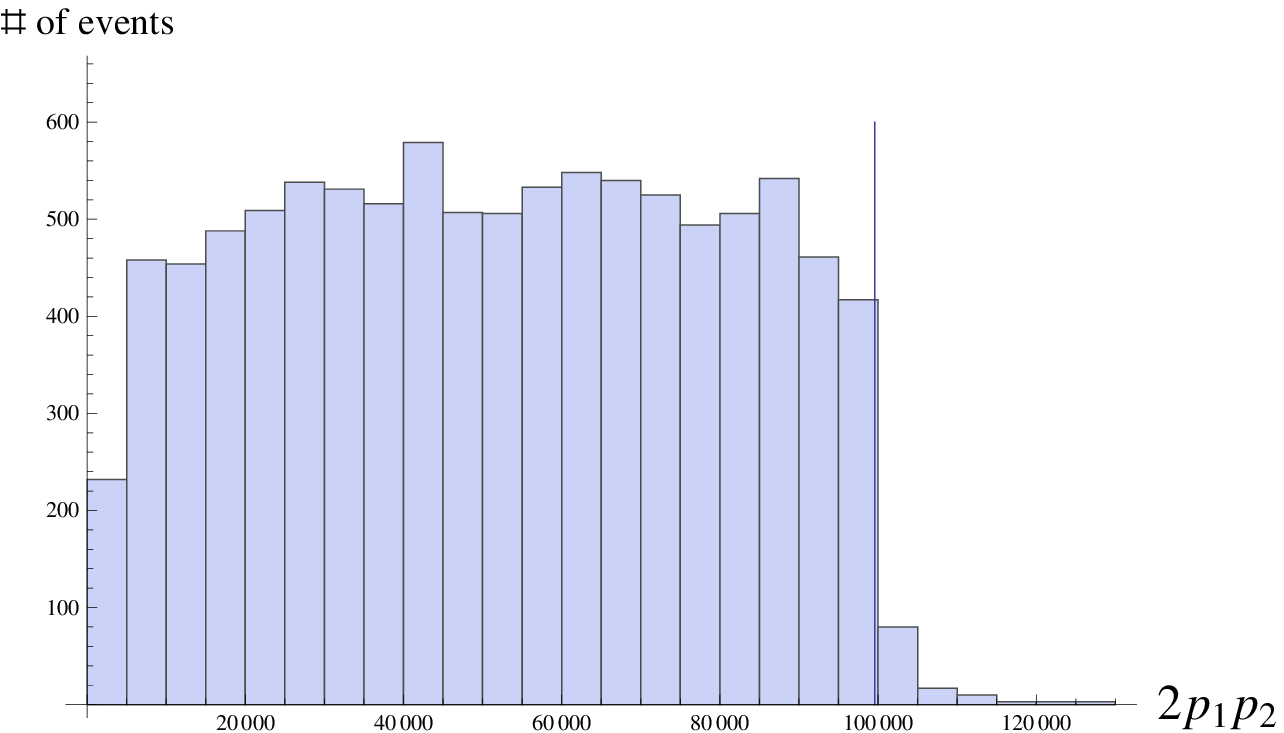}
\includegraphics[width=7cm]{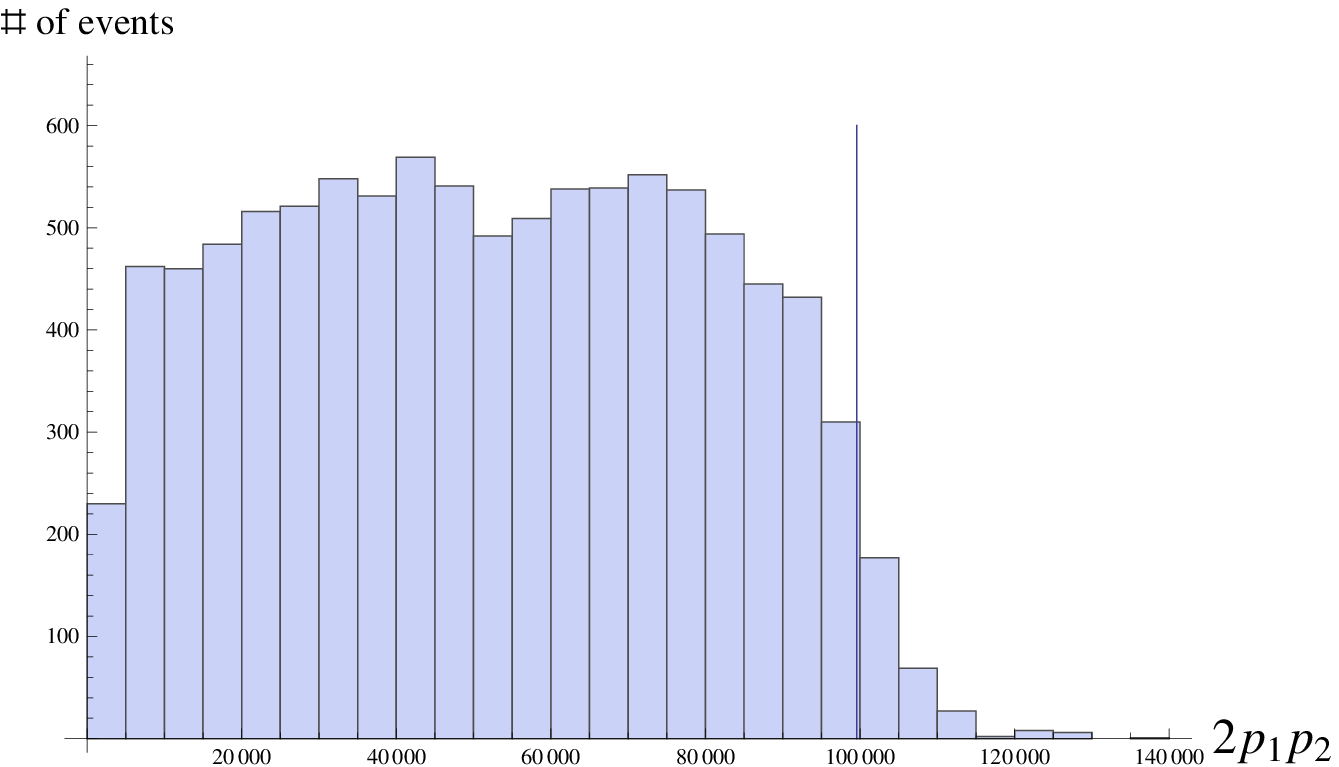}
\caption{The histogram of the invariant mass-squared of the two visible particles before (Left) and after smearing (Right) for case 3.  The vertical line indicates the value of $\frac{\Delta_1\Delta_2}{m^2_X}$.  Both visible particles are leptons.}
\label{oshist}
%
\centering
\includegraphics[width=7cm]{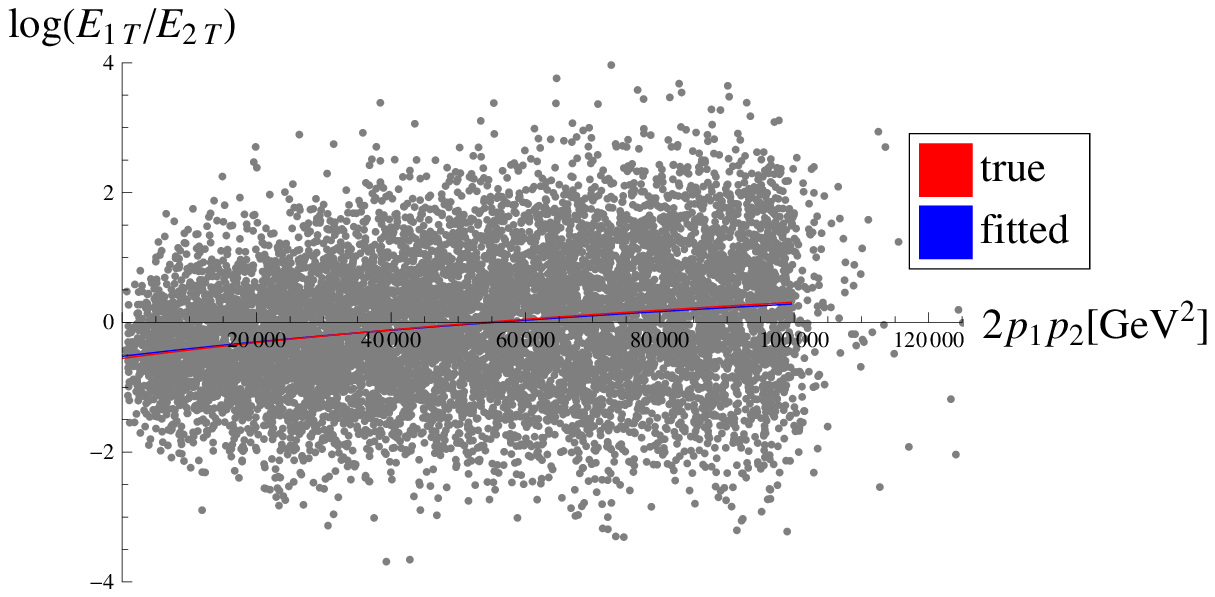}
\includegraphics[width=7cm]{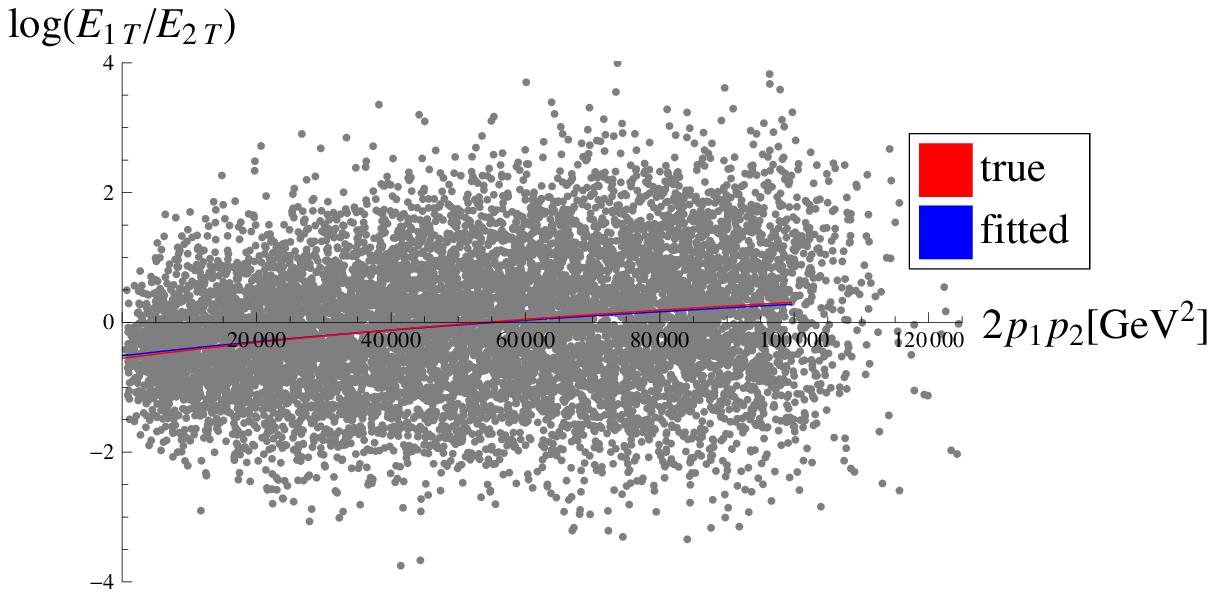}
\caption{The scatter plots in the $\log{(E_{1T}/E_{2T})}$ vs. $2p_1p_2$ plane before (Left) and after smearing (Right) for case 3.  In both case the red curve is $\log{(E_{1T}/E_{2T})}=\log{(\frac{\Delta_1+2p_1p_2}{\Delta_2})}$ with true $\Delta_1$ and $\Delta_2$ and the blue curve is the fitted one.}
\label{ossf}
%
\centering
\includegraphics[width=7cm]{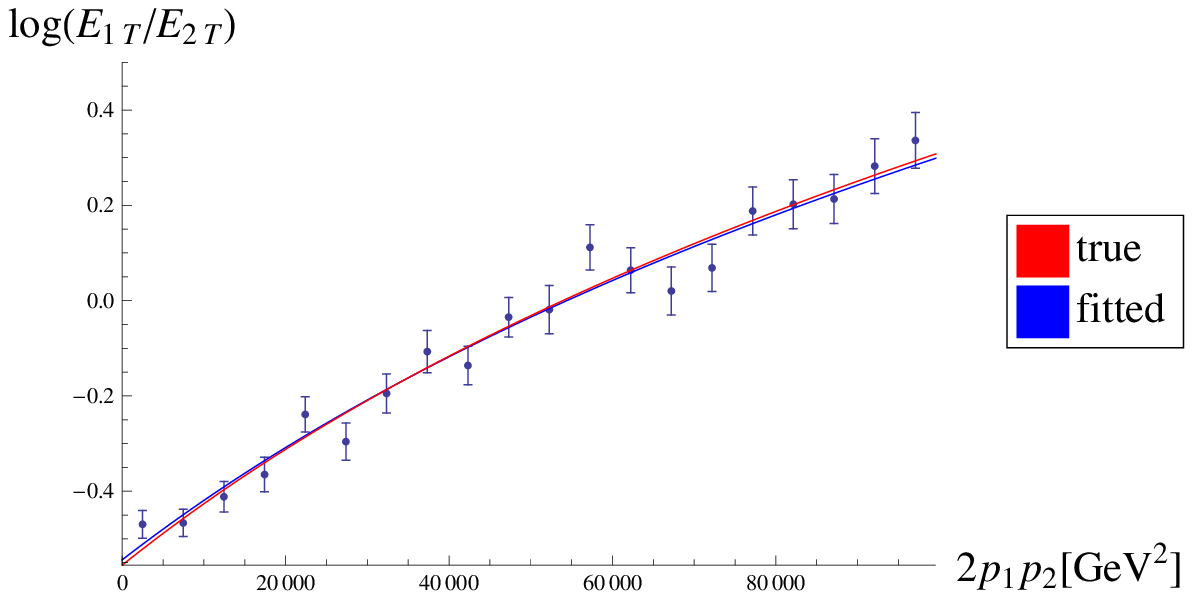}
\includegraphics[width=7cm]{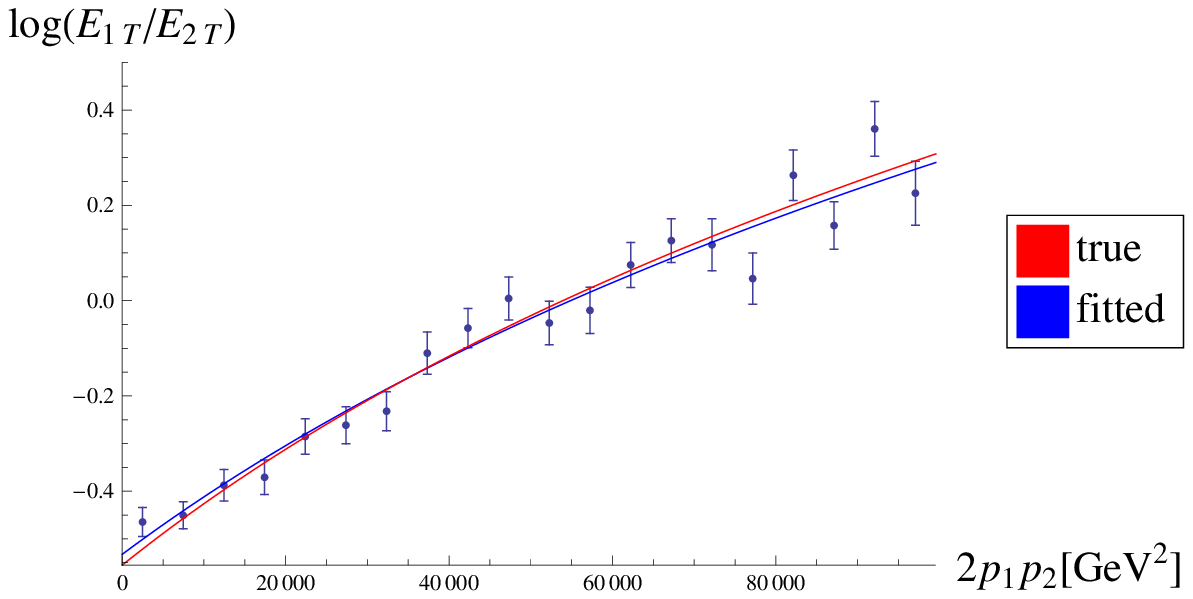}
\caption{The peak-point plots in the $\log{(E_{1T}/E_{2T})}$ vs. $2p_1p_2$ plane before (Left) and after smearing (Right) for case 3 (no folding).  In both case the red curve is $\log{(E_{1T}/E_{2T})}=\log{(\frac{\Delta_1+2p_1p_2}{\Delta_2})}$ with true $\Delta_1$ and $\Delta_2$ and the blue curve is the fitted one.}
\label{ospp}
\end{figure}
Before studying the combinatorial problem we first look at the effect of smearing. For case 3, Figs.~\ref{oshist}, \ref{ossf} and \ref{ospp} show the invariant mass distributions from the two visible particles, the scatter plots and the peak points plots in the $\log{(E_{1T}/E_{2T})}$ vs. $2p_1p_2$ plane before and after smearing.  We see that the invariant mass end point is less sharp after smearing but its existence is still eminent. The value of the end point may be obtained with a template fit and we assume that it can be accurately determined. Consequently we will use the true end point value in the mass reconstruction as we expect much larger uncertainties coming from other quantities. As the smearing effects are small for leptons, the distribution in the $\log{(E_{1T}/E_{2T})}$ vs. $2p_1p_2$ plane does not change much. The fitted curve also agrees with the true curve quite well after smearing, if the ordering of the 2 visible particles were known.

We now study the folded distributions for the five cases listed in Table \ref{fomass}.  We first look at the distribution of $10^4$ events for each case to reduce the statistical fluctuations.  The experimental smearings are included but they do not have a significant effect as we see in the above discussion. 
\begin{figure}
\centering
\includegraphics[width=7cm]{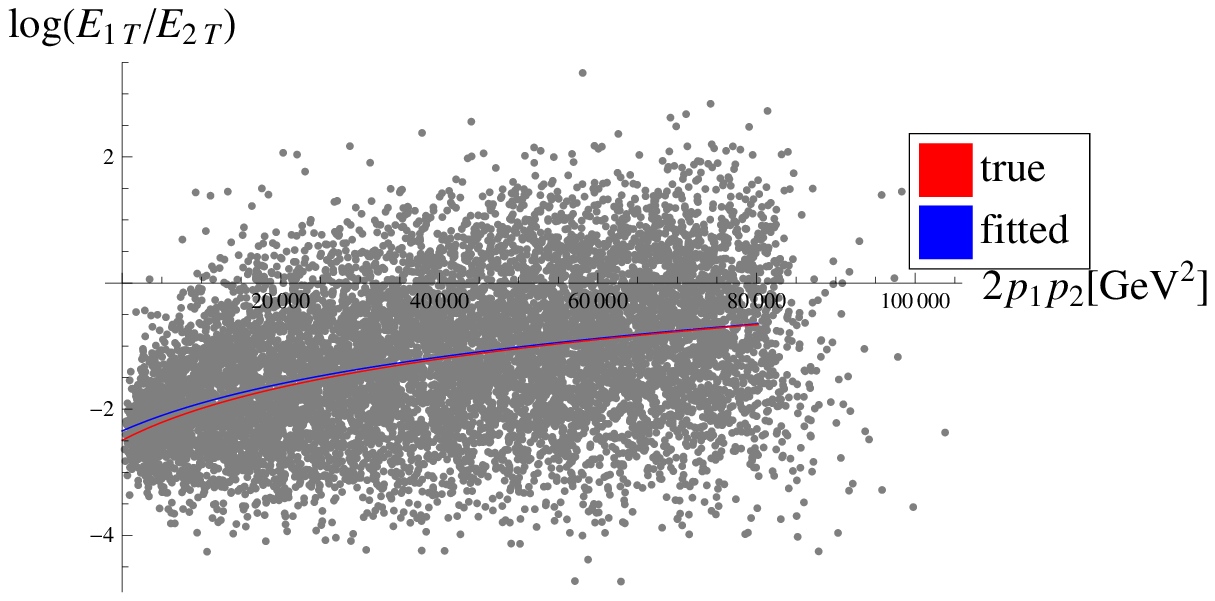}
\includegraphics[width=7cm]{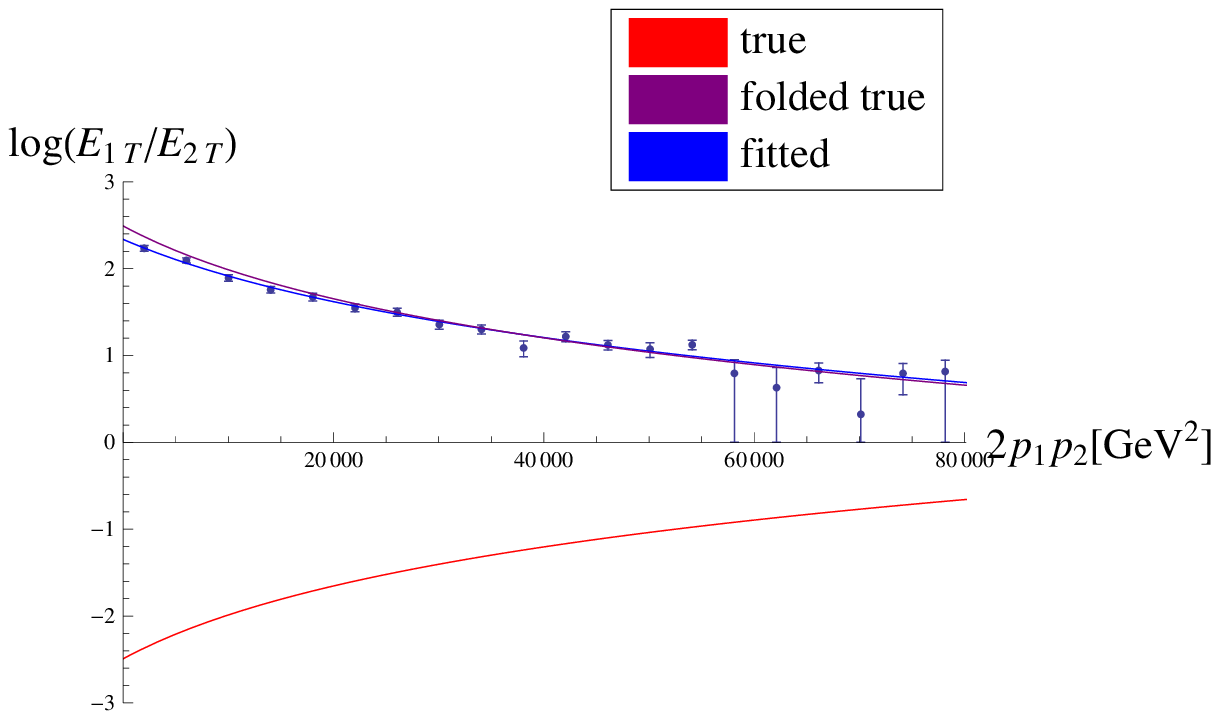}
\caption{The scatter plot (Left) and the folded peak-point plot (Right) in the $\log{(E_{1T}/E_{2T})}$ vs. $2p_1p_2$ plane for case 1  ($\log{\frac{\Delta_1}{\Delta_2}}=-2.5$).  In both case the red curve is $\log{(E_{1T}/E_{2T})}=\log{(\frac{\Delta_1+2p_1p_2}{\Delta_2})}$ with true $\Delta_1$ and $\Delta_2$ and the blue curve is the fitted one.  In the folded plot the purple curve is the folded true curve.}
\label{b8p}
\end{figure}
\begin{figure}
\centering
\includegraphics[width=7cm]{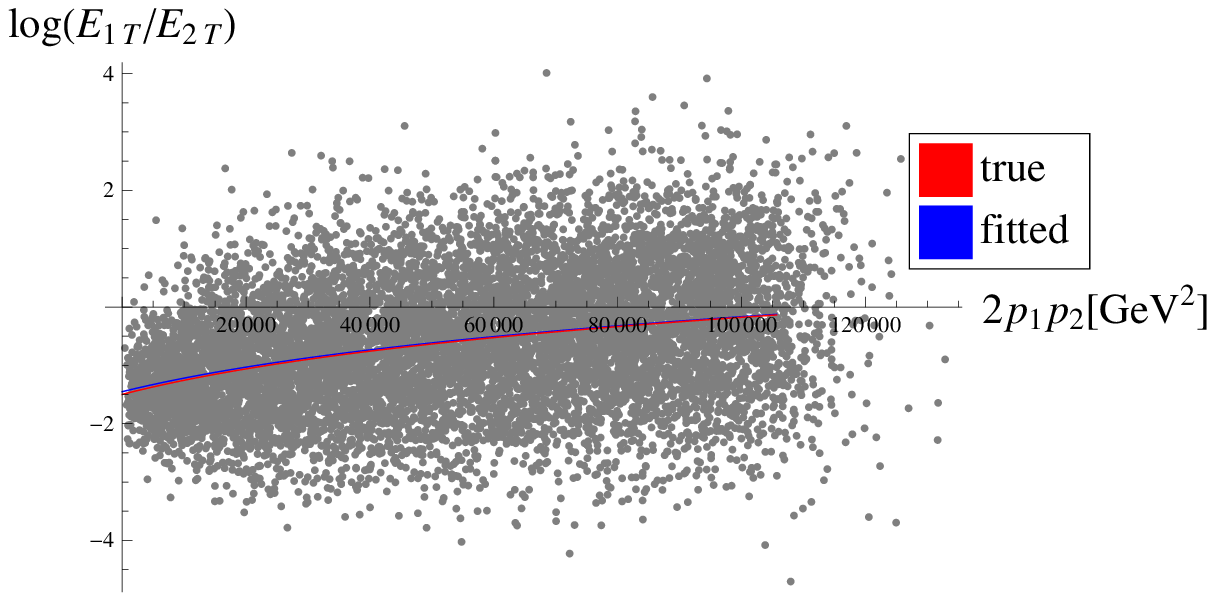}
\includegraphics[width=7cm]{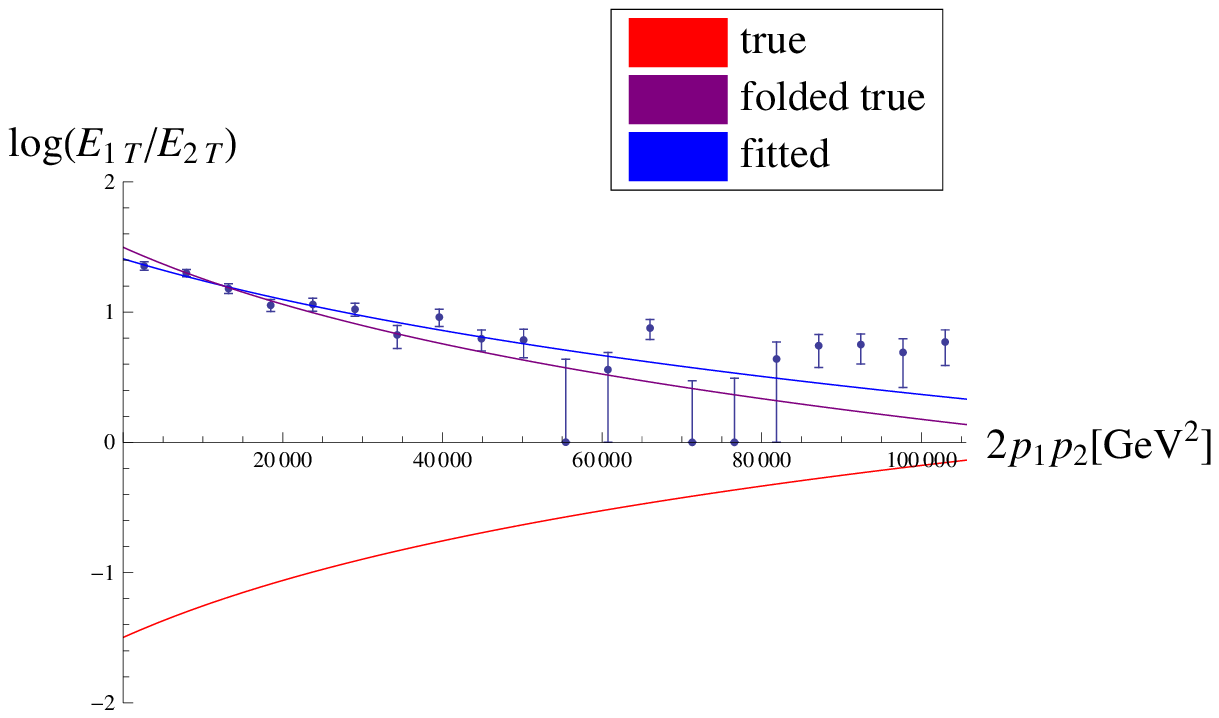}
\caption{The scatter plot (Left) and the folded peak-point plot (Right) in the $\log{(E_{1T}/E_{2T})}$ vs. $2p_1p_2$ plane for case 2  ($\log{\frac{\Delta_1}{\Delta_2}}=-1.5$).  In both case the red curve is $\log{(E_{1T}/E_{2T})}=\log{(\frac{\Delta_1+2p_1p_2}{\Delta_2})}$ with true $\Delta_1$ and $\Delta_2$ and the blue curve is the fitted one.  In the folded plot the purple curve is the folded true curve.}
\label{a2p}
\end{figure}
\begin{figure}
\centering
\includegraphics[width=7cm]{s1sf.eps}
\includegraphics[width=7cm]{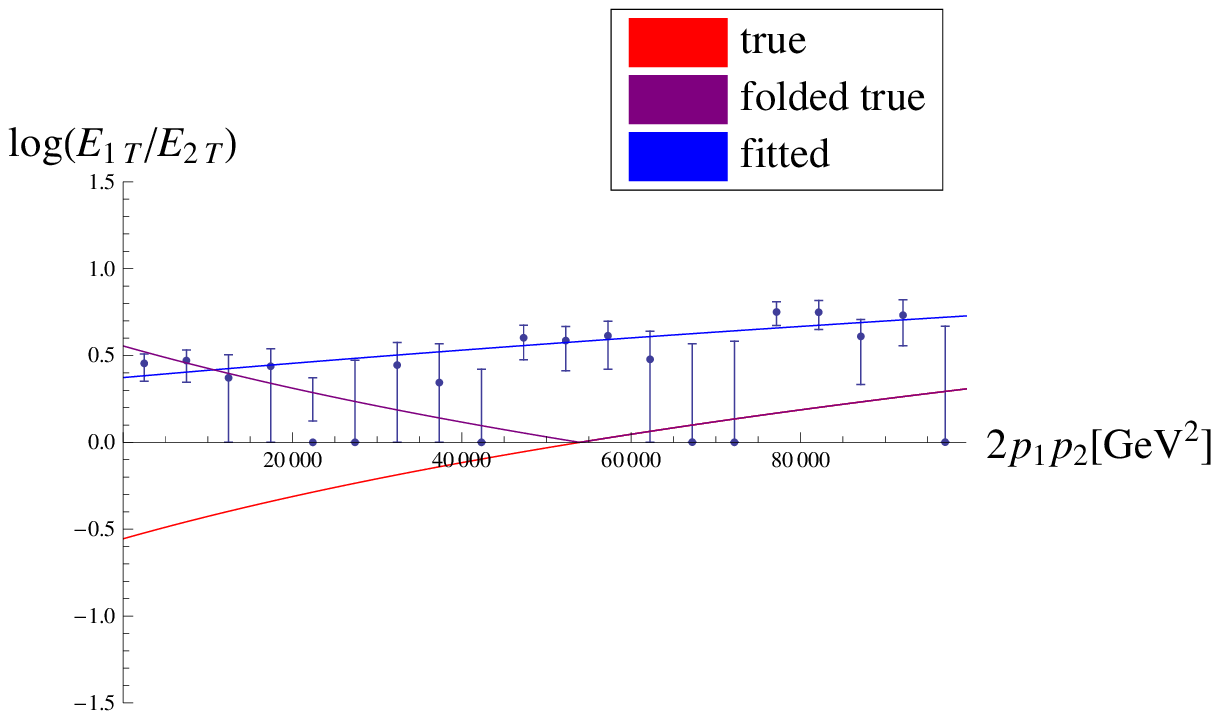}
\caption{The scatter plot (Left) and the folded peak-point plot (Right) in the $\log{(E_{1T}/E_{2T})}$ vs. $2p_1p_2$ plane for case 3  ($\log{\frac{\Delta_1}{\Delta_2}}=-0.56$).  In both case the red curve is $\log{(E_{1T}/E_{2T})}=\log{(\frac{\Delta_1+2p_1p_2}{\Delta_2})}$ with true $\Delta_1$ and $\Delta_2$ and the blue curve is the fitted one.  In the folded plot the purple curve is the folded true curve.}
\label{s1p}
\end{figure}
\begin{figure}
\centering
\includegraphics[width=7cm]{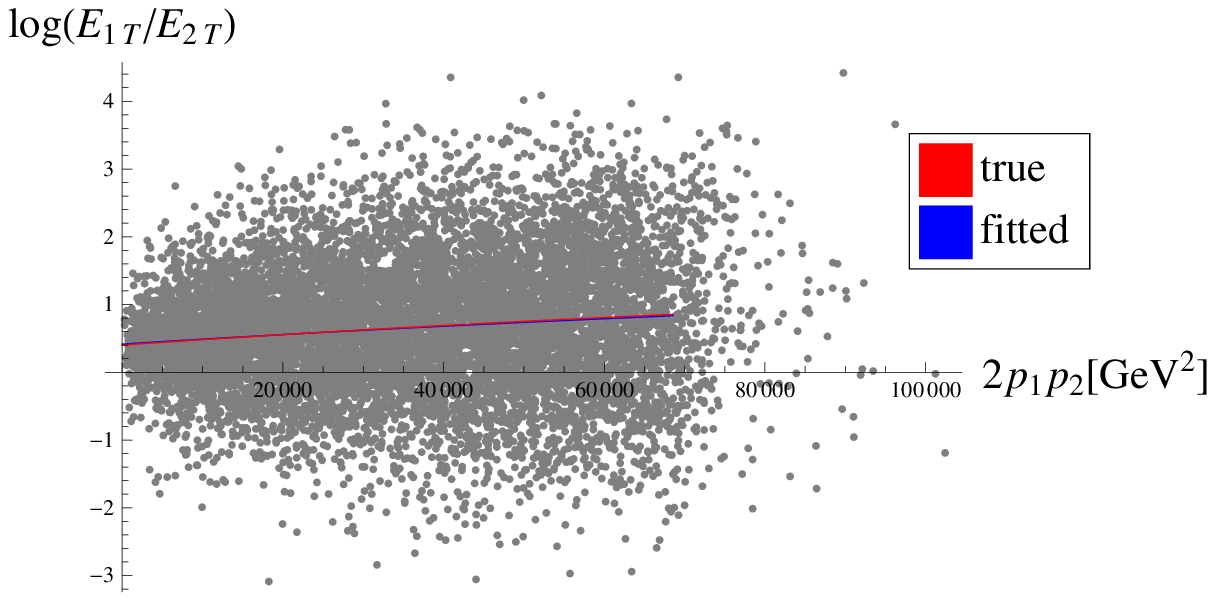}
\includegraphics[width=7cm]{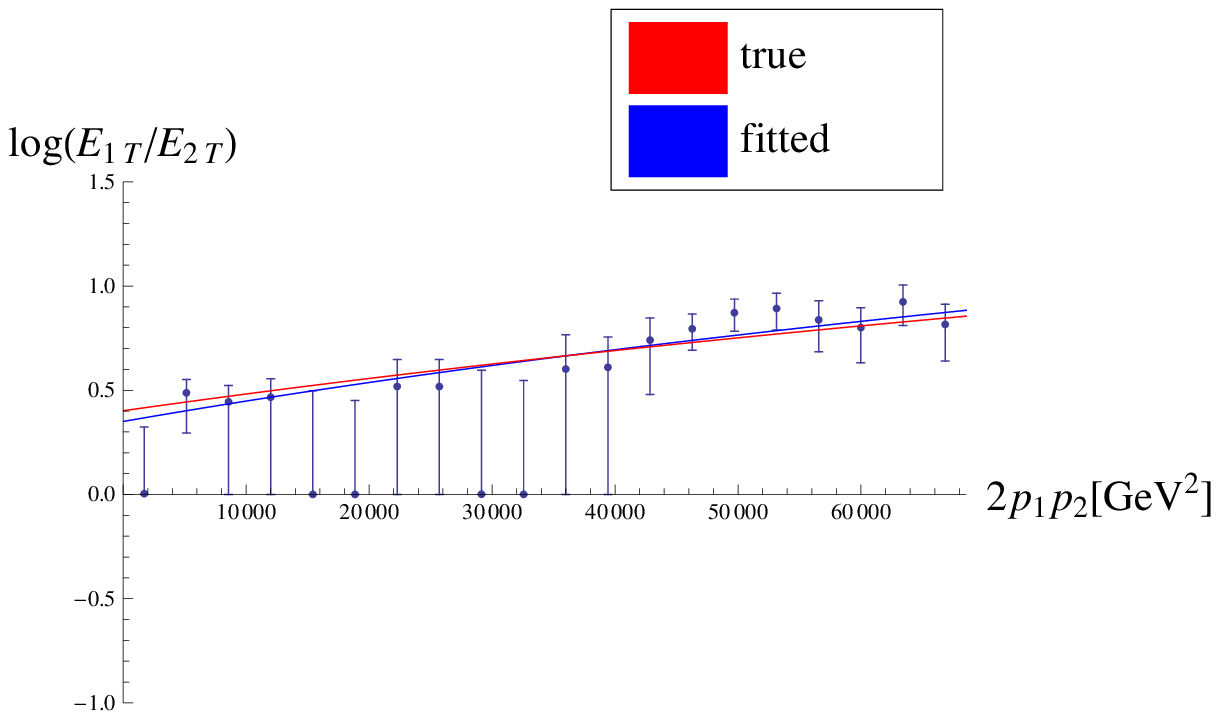}
\caption{The scatter plot (Left) and the folded peak-point plot (Right) in the $\log{(E_{1T}/E_{2T})}$ vs. $2p_1p_2$ plane for case 4  ($\log{\frac{\Delta_1}{\Delta_2}}=0.40$).  In both case the red curve is $\log{(E_{1T}/E_{2T})}=\log{(\frac{\Delta_1+2p_1p_2}{\Delta_2})}$ with true $\Delta_1$ and $\Delta_2$ and the blue curve is the fitted one.  In the folded plot the purple curve is the folded true curve.}
\label{d8p}
\end{figure}
\begin{figure}
\centering
\includegraphics[width=7cm]{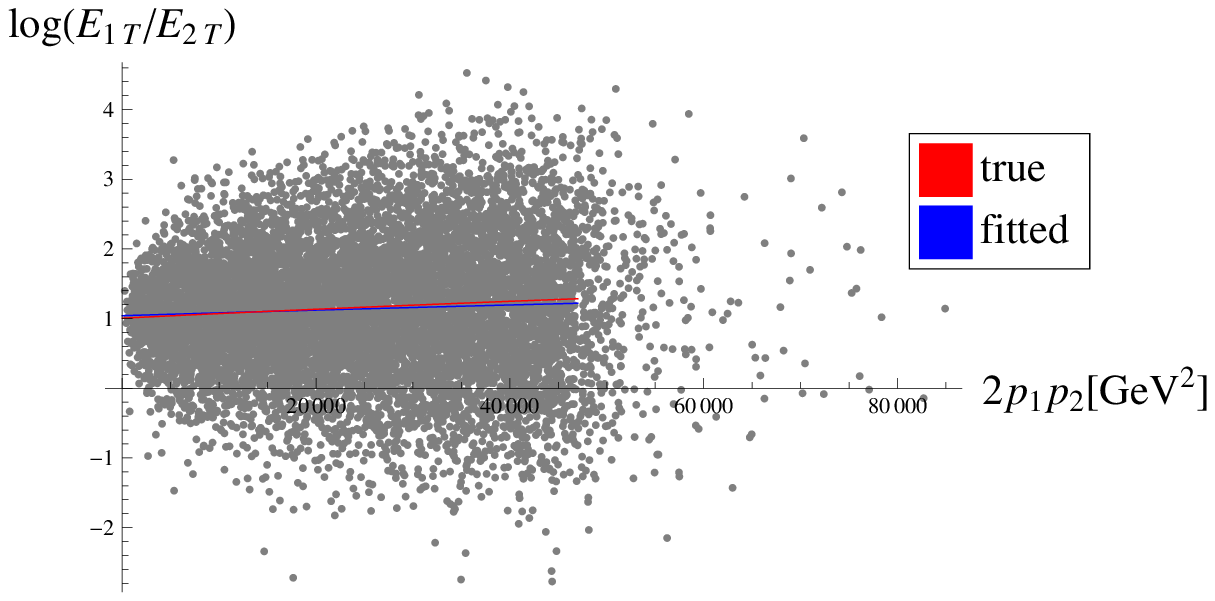}
\includegraphics[width=7cm]{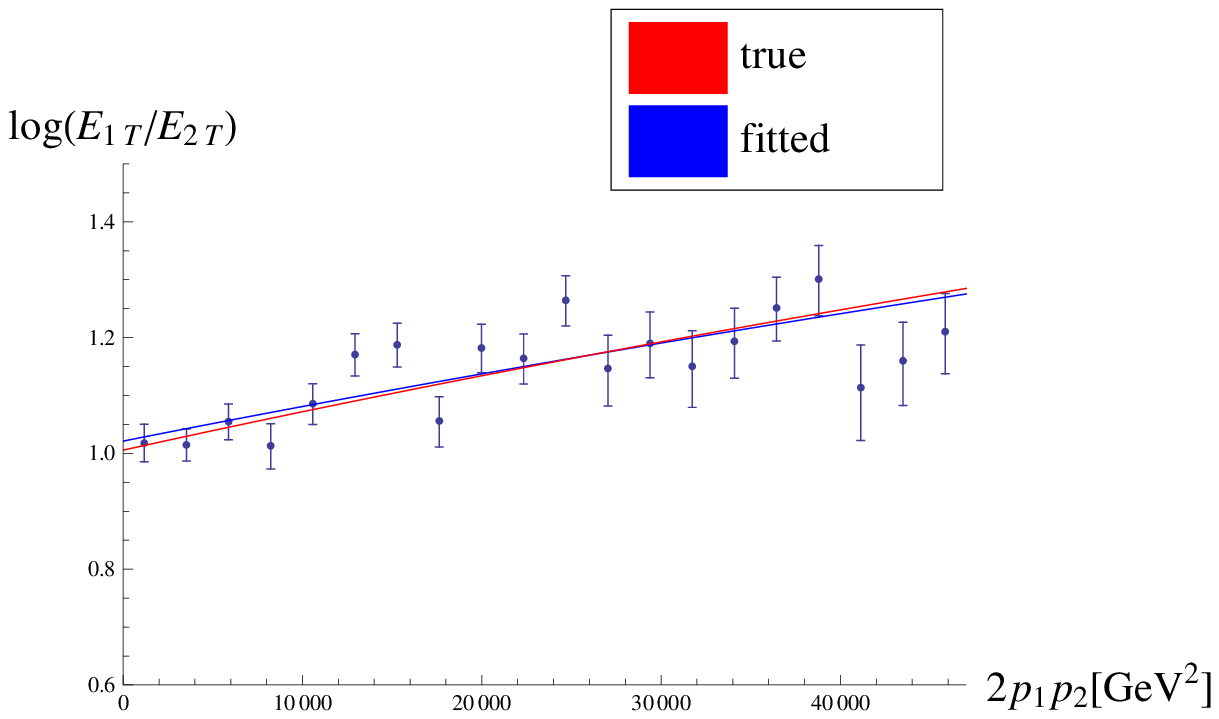}
\caption{The scatter plot (Left) and the folded peak-point plot (Right) in the $\log{(E_{1T}/E_{2T})}$ vs. $2p_1p_2$ plane for case 5  ($\log{\frac{\Delta_1}{\Delta_2}}=1.0$).  In both case the red curve is $\log{(E_{1T}/E_{2T})}=\log{(\frac{\Delta_1+2p_1p_2}{\Delta_2})}$ with true $\Delta_1$ and $\Delta_2$ and the blue curve is the fitted one.  In the folded plot the purple curve is the folded true curve. (Note: The origin is not at 0 in the peak plot.)}
\label{c8p}
\end{figure}

For case 1 ($\log{{\Delta_1}/{\Delta_2}}=-2.5$) the scatter plot and the folded peak-point plot are shown in Fig.~\ref{b8p}.  In the scatter plot one can see that a small portion of the events is above the $\log{(E_{1T}/E_{2T})}=0$ axis, which will be folded. Nevertheless, as the center of distribution is far away from the folding line, the folded Gaussian fit works well for all bins.  The error bars are larger and some touch the folding axis for the last few points because the peak points are closer to the folding axis. In this case a fit to all the peak points is quite close to the true curve and we can get a good determination of $\Delta_1$ and $\Delta_2$.

Fig.~\ref{a2p} shows the plots for case 2 ($\log{{\Delta_1}/{\Delta_2}}=-1.5$).  Compared to case 1 the center of the distribution moves closer to the folding axis. As a result, the error bars are larger in the folded Gaussian fits especially for large invariant masses, as shown in the right panel of Fig.~\ref{a2p}. It is interesting to notice that there are a few points at large invariant mass that are far away from the true curve and have relatively small error bars, while the true peak points should be close to zero.  We check their distributions and it turns out that their unfolded distributions are somewhat asymmetric and hence are not Gaussian-like. Therefore, the likelihood method assuming a Gaussian distribution did not work very well for these points.  The fit will deteriorate if these points are included.  However, as the nearby points have large errors and touch the folding axis, the center of the distribution for this range of invariant mass should be quite close to the folding axis. As the points with large error bars are not very useful in the fitting, we choose to fit the curve only in the ``good region," which contains the first ten points from the left, and a reasonable fit can be obtained. 

Fig.~\ref{s1p} shows the plots for case 3 ($\log{{\Delta_1}/{\Delta_2}}=-0.56$).  The center of the distribution is quite close to the folding line for most range of the invariant mass and the fitted points behaves quite badly, as one can see in the plot that most points have very large error bars.  A na\"ive fit to all points even gives the wrong sign of $\log{{\Delta_1}/{\Delta_2}}$, which may be figured out by a more careful examination of the folded distribution. Anyway, this is a bad case for our mass determination method. 

Fig.~\ref{d8p} shows the plots for case 4 ($\log{{\Delta_1}/{\Delta_2}}=0.40$).  Now the center of the distribution is above the folding axis. The peak points are closer to the folding axis for small invariant masses but the corresponding distributions are also less spread.  Although many points have large error bars as shown in the figure, a curve fit to all points seems to give quite good results.

Fig.~\ref{c8p} shows the plots for case 5 ($\log{{\Delta_1}/{\Delta_2}}=1.0$).  The center of distribution is quite far away from the folding axis and the effects of folding are unimportant.  However, in this case $\Delta_2$ is small and hence particle 2 is quite soft. As discussed in Appendix~\ref{o4} our reconstruction method starts to produce some bias even without the folding. 

The results of mass determination for the five cases are shown in Table~\ref{fore}.  For cases 1, 2, 4 \& 5 we also wish to know how well the method works with a smaller sample size.  It turns out that in terms of statistical fluctuations, for cases 1 and 2 the method works reasonably well with $10^3$ events.  The results are presented in Table \ref{fore2}.
\begin{table}
\begin{tabular}{|c|c|c|c|c|c|c|}
\hline
   & $\Delta_1[\mbox{GeV}^2]$ & $\Delta_2[\mbox{GeV}^2]$ & $\log{(\Delta_1/\Delta_2)}$ & $m_Y$[GeV] & $m_X$[GeV] & $m_N$[GeV] \\ \hline \hline
   \multicolumn{7}{|c|}{case 1} \\ \hline
 true &  $1.523\times 10^4$  & $1.841\times 10^5$  & $-2.49$ & $468$ & $187$ & $140.5$ \\ \hline
 reconstructed &  $1.907\times 10^4$  & $1.974\times 10^5$  & $-2.34$ & $494$ & $217$ & $167$ \\ \hline
error &  $+25\%$  & $+7.3\%$  & $+15\%$ & $+5.6\%$ & $+16\%$ & $+19\%$ \\ \hline \hline

   \multicolumn{7}{|c|}{case 2} \\ \hline
 true &  $3.643\times 10^4$  & $1.629\times 10^5$  & $-1.50$ & $468$ & $237$ & $140.5$ \\ \hline
 reconstructed &  $5.456\times 10^4$  & $2.233\times 10^5$  & $-1.41$ & $582$ & $340$ & $247$ \\ \hline
error &  $+50\%$  & $+37\%$  & $+8.8\%$ & $+24\%$ & $+43\%$ & $+75\%$ \\ \hline \hline

   \multicolumn{7}{|c|}{case 3} \\ \hline
 true &  $7.268\times 10^4$  & $1.266\times 10^5$  & $-0.555$ & $468$ & $304$ & $140.5$ \\ \hline
 reconstructed &  $2.336\times 10^5$  & $1.609\times 10^5$  & $0.37$ & $734$ & $614$ & $379$ \\ \hline
error &  $+221\%$  & $+27\%$  & $+93\%$ & $+57\%$ & $+102\%$ & $+170\%$ \\ \hline \hline

   \multicolumn{7}{|c|}{case 4} \\ \hline
 true &  $1.194\times 10^5$  & $7.990\times 10^4$  & $0.402$ & $468$ & $373$ & $140.5$ \\ \hline
 reconstructed &  $9.732\times 10^4$  & $6.856\times 10^4$  & $0.35$ & $407$ & $312$ & $0.0$ \\ \hline
error &  $-18\%$  & $-14\%$  & $-5.1\%$ & $-13\%$ & $-16\%$ & $-100\%$ \\ \hline \hline

   \multicolumn{7}{|c|}{case 5} \\ \hline
 true &  $1.459\times 10^5$  & $5.338\times 10^4$  & $1.01$ & $468$ & $407$ & $140.5$ \\ \hline
 reconstructed &  $1.625\times 10^5$  & $5.853\times 10^4$  & $1.02$ & $511$ & $450$ & $200$ \\ \hline
error &  $+11\%$  & $+9.7\%$  & $+1.6\%$ & $+9.1\%$ & $+11\%$ & $+42\%$ \\ \hline

 \end{tabular}
 \caption{The results of mass reconstruction from fitting the folded peak points for the 5 cases in Section \ref{fold} with $10^4$ events.  The errors are calculated using $\frac{\rm reconstructed-true}{\rm true}$ (except for $\log{(\Delta_1/\Delta_2)}$, which is ${\rm reconstructed-true}$) and do not represent the statistical fluctuation.}
 \label{fore}
 \end{table} 

For cases 4 \& 5, some care needs to be taken.  As seen in eq.~(\ref{eq:expansion}), the slope corresponds to the value of ${1}/{\Delta_1}$ at the first order.  When $\Delta_1 \gg \Delta_2$, the slope is small and a small fluctuation in the slope could cause a large fluctuation in the value of $\Delta_1$  (also $\Delta_2$ since their ratio can be well determined) and the mapping is nonlinear.  As a result, a larger statistics  for cases 4 \& 5 is needed to get a good measurement.  Furthermore, a Gaussian-like uncertainty of the reconstructed slope will not result in a Gaussian-like uncertainty in the reconstructed $\Delta_1$ when the slope is close to zero due to the nonlinear mapping. In particular, the value of $\Delta_1$ goes to infinity when the slope goes to zero.  This behavior shows up when we look at case 4 with 50 sets of $10^3$ events.  Among the 50 sets of reconstructed masses, 44 of them are relatively close to each other but 6 sets has very large reconstructed masses ($m_Y >$ 1 TeV).  Case 5 has a similar behavior, and we found that a larger data sample ($2\times 10^3$ events) is needed for a reasonable reconstruction.  Among 25 sets of $2\times 10^3$ events, 5 sets has $m_Y >$ 1 TeV and the other 20 sets are relatively close to each other.  As the uncertainty is non-Gaussian, it would be better to quote the central 68.3\% confidence region instead of mean $\pm$ standard deviation.  However, it is hard to obtain the form of the probability distribution function without generating a large number of sets of events.  Here we simply present the means and standard deviations of the reconstructed masses excluding the bad sets in Table~\ref{fore2}.
 \begin{table}
 \centering
\begin{tabular}{|c|c|c|c|}
\hline
 &$m_Y$[GeV] & $m_X$[GeV] & $m_N$[GeV] \\ \hline  \hline
 \multicolumn{4}{|c|}{case 1 ($10^3$ events)} \\ \hline
 true &  $468$ & $187$ & $140.5$ \\ \hline
reconstructed & 527$\pm38 $ & 245$\pm34 $ & $195 \pm33 $ \\ \hline
error & $+13\%\pm 8\%$ & $+31\%\pm 18\%$ & $+39\%\pm 23\%$ \\ \hline \hline

 \multicolumn{4}{|c|}{case 2 ($10^3$ events)} \\ \hline
 true &  $468$ & $237$ & $140.5$ \\ \hline
reconstructed & $616\pm 169$ & $373\pm 160$ & $ 280\pm 161$ \\ \hline
error & $+32\%\pm 36\%$ & $+57\%\pm 67\%$ & $+99\%\pm 115\%$ \\ \hline \hline

 \multicolumn{4}{|c|}{case 4 ($10^3$ events, 44 out of 50 sets have $m_Y<1$ TeV)} \\ \hline
 true &  $468$ & $373$ & $140.5$ \\ \hline
reconstructed & $438\pm62 $ & $347\pm62$ & $ 26\pm90$ \\ \hline
error & $-6.4\%\pm 13\%$ & $-6.9\%\pm 17\%$ & $-81\%\pm 64\%$ \\ \hline \hline

 \multicolumn{4}{|c|}{case 5 ($2\times10^3$ events, 20 out of 25 sets  have $m_Y<1$ TeV)} \\ \hline
 true &  $468$ & $407$ & $140.5$ \\ \hline
reconstructed & $498\pm119 $ & $437\pm119 $ & $ 151\pm160 $ \\ \hline
error & $+6.5\%\pm 25\%$ & $+7.5\%\pm 29\%$ & $+7.8\%\pm 114\%$ \\ \hline \hline

 \end{tabular}
 \caption{The results of reconstruction of case 1, 2, 4 \& 5.  The errors are in the form ``bias$\pm$uncertainty".  The means and uncertainties are estimated from a total number of $5\times10^4$ events. (i.e. For case 1 \& 2, we divided the sample into 50 sets, each with $10^3$ events, reconstructed the masses for each set and obtained the means and standard deviations of the reconstructed masses.  For case 4 we did the same but we only used the 44 ``good" sets out of 50 sets.  For case 5 we divided the sample into 25 sets, each with $2\times10^3$ events, and only used the 20 ``good" sets among the 25 sets.)  For case 2, only the first ten points (counting from the left, see Fig.~\ref{a2p}) are used for fitting.}
 \label{fore2}
 \end{table}

\subsection{Combinatorial problems of the 1 jet + 1 lepton signal}
\label{comb}

In this subsection we study another type of the combinatorial problem when there are other particles which can not be distinguished experimentally from one of the particles from the decay chain in the event. A typical example is that the two visible particles from the decay chain are one jet and one lepton. In general we expect there will be other jets present in the same event. They can come from the other decay chain(s) in the event or simply the initial and final state radiations. In this case we face the problem of not knowing which jet is the correct one to be paired with the lepton in the same decay chain. It is possible that there are also other leptons in the same event to cause further confusions. To understand the basic issues of this type of the combinatorial problem we will keep it simple by restricting ourselves to the case where only additional jets are present. The analysis should be readily generalizable to more complicated situations.
 
The jets coming from decays of some heavy particles and the jets coming from initial state radiation have very different characteristics. To consider both possibilities, we generated the following process using MadGraph. A left-handed down squark and its anti-particle are pair produced with the left-handed down squark decaying through the same chain studied in Sec.~\ref{ti} (with the spectrum shown in Table~\ref{td0}), while the anti left-handed down squark decays to an anti-down quark and a lightest neutralino. In addition, an extra gluon is produced, either through ISR or FSR or elsewhere.  The events are smeared according to the Gaussian errors listed in Table~\ref{smt}.  The signal of this process is 3 jets + 1 lepton + $\not \!\! E_T$ and we do not know which jet is particle 2.  This case is, of course, an over simplified version compared with the real processes at the LHC, but it should help us understand how to apply our method to the realistic cases.

Fig.~\ref{zhist} shows the invariant mass-squared distributions for different combinations before and after smearing.  For convenience we label the anti-down quark from the other chain particle 3 and the extra gluon particle 4.  The invariant mass-squared distribution of the three different combinations are shown in the same plot.  In reality we will not be able to distinguish them, but only obtain a distribution in which all three are stacked together.\footnote{The ISR jet may be identified with some probabilities through some sophisticated methods~\cite{Krohn:2011zp,Gallicchio:2011xq}, but will not be attempted here.} 
\begin{figure}
\centering
\includegraphics[width=7.5cm]{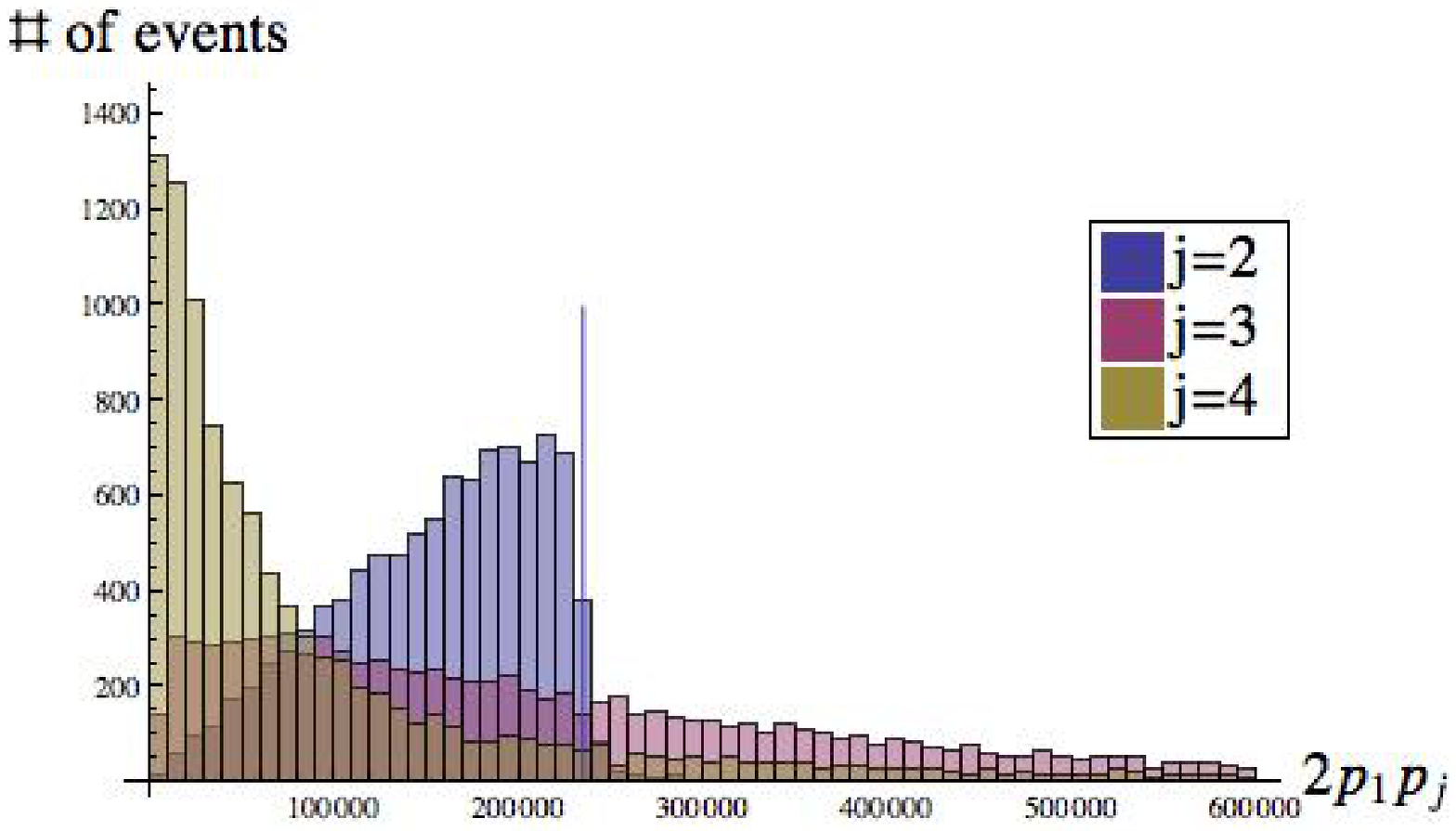}
\includegraphics[width=7.5cm]{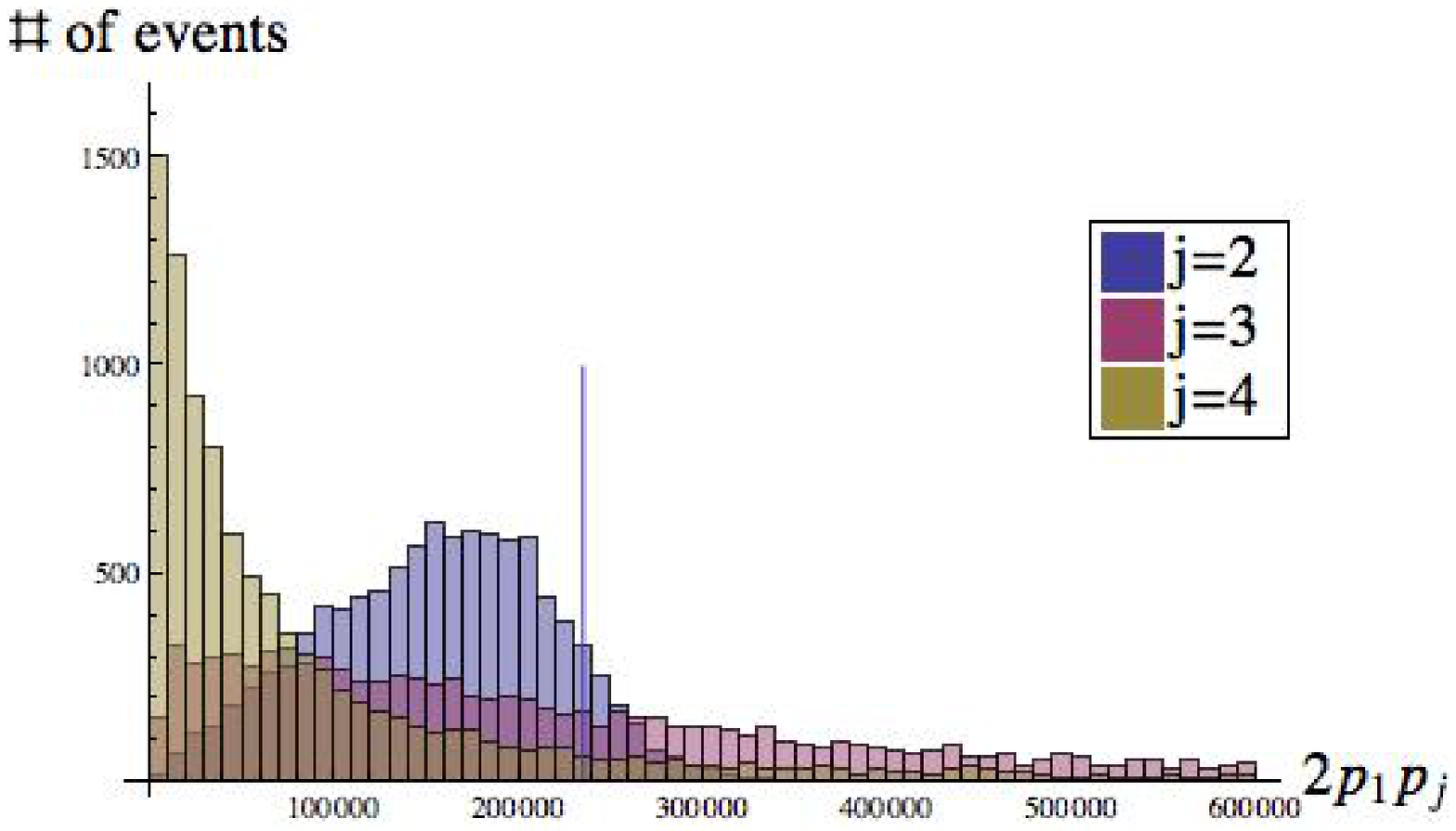}
\caption{The histogram of the invariant mass-squared of the three possible combinations before smearing (left) and after smearing (right).  The vertical line indicates the value of $\frac{\Delta_1\Delta_2}{m^2_X}$.}
\label{zhist}
\end{figure}
Fig.~\ref{zs} shows the scatter plot in the $\log{(E_{1T}/E_{Tj})}$ vs. $2p_1p_j$ plane where $j=2,3,4$ denotes the jet is particle 2, 3 (from the other decay chain) or 4 (extra gluon).  For better visualization only 1000 events are plotted.  For $j=4$, the events are mostly in the upper left region due to the fact that most gluons are soft.  For $j=3$ the events are more spread on the $2p_1p_j$ axis and have less correlation between $\log{(E_{1T}/E_{jT})}$ and $2p_1p_j$.  
\begin{figure}
\centering
\includegraphics[width=7.5cm]{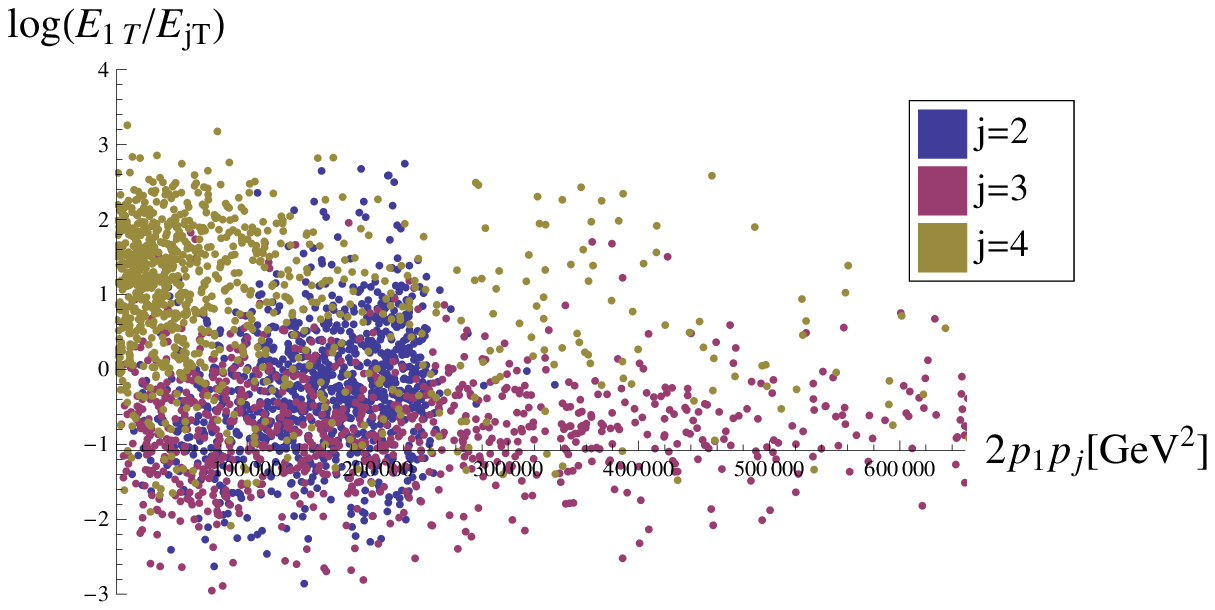}
\includegraphics[width=7.5cm]{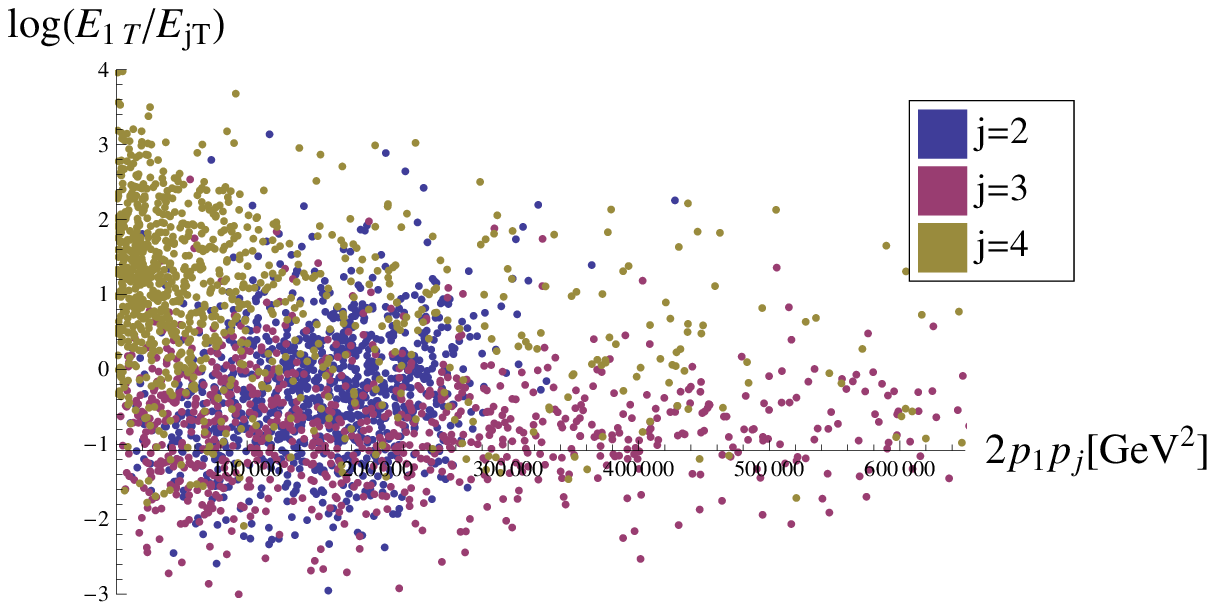}
\caption{The scatter plots in the  $\log{(E_{1T}/E_{jT})}$ vs. $2p_1p_2$ plane before smearing (left) and after smearing (right) where $j=2,3,4$ denotes the jet is particle 2, 3 (from the other decay chain) or 4 (extra gluon).  1000 events are plotted.  No cut (apart from the default cut in MadGraph) is applied.}
\label{zs}
\end{figure}

Before smearing, the invariant mass-squared distributions have two distinct features.  First, for the correct combinations ($p_1$,  $p_2$), the distribution has a sharp edge at $\frac{\Delta_1\Delta_2}{m^2_X}$, while the wrong combinations do not have such an edge.  Second, for invariant mass below the edge, the correct combination is more likely to have large invariant mass than that of the wrong combinations. The shape of the invariant mass-squared is model-dependent. In particular, it depends on the spin of the particle $X$ and the chiralness of its couplings.  If particle $X$ is a scalar, the invariant mass-squared will have a flat distribution below the edge, but compared with the distributions of the wrong combinations, the two features still hold. On the other hand, if the left-handed down squark in the above process is replaced by an up anti-squark, then the quark and the lepton in the decay chains will have opposite helicities (in contrast to the same helicity in our example). Consequently they tend to go in the same direction and have a small invariant mass. In this case there will be no sharp edge at the end point. However, given that we need a good measurement of $\frac{\Delta_1\Delta_2}{m^2_X}$ from the end point, we will only consider cases where a sharp edge is present, then generally the second feature also holds. Based on these two features we will use a simple and na\"ive way to select the jet-lepton pair by choosing the combination with the largest invariant mass below the edge. 

Because jets suffer more experimental smearing than leptons, the edge is more washed out compared to the pure leptonic case in the previous subsection. Nevertheless, the existence of an edge in the invariant mass-squared distribution should still be identifiable from the right panel of Fig.~\ref{zhist}. We again assume that the location of $\frac{\Delta_1\Delta_2}{m^2_X}$ can be obtained from a template fit. Any uncertainty for this quantity will add to the total uncertainties of the reconstructed masses but we do not expect it to be the main source of the final uncertainties. We will not include it in our following discussion.

To select the correct jet, we first make the following cuts:
\begin{itemize}
\item A $p_T$ cut on the jets is imposed. Such a $p_T$ cut is inevitable experimentally to reduce the QCD backgrounds. It mostly help to remove the ISR jets. However, if the $p_T$ cut is too high, it may create a fake polarization for the mother particle $Y$ as it will favor the events with $Y$ decaying to particle 2 in the forward direction, and hence causes a bias in the mass determination. In our example we require jets to have $p_T > 50$~GeV. 

\item We require the jet when combined with the lepton to have an invariant mass-squared smaller than
the position of the edge, $\frac{\Delta_1\Delta_2}{m^2_X}$. This dominantly reduces the wrong combinations coming from the jet of the other decay chain. Because of smearing a small fraction of the correct combinations will be removed too.
\end{itemize}
If no jet survives the above cuts, the event is dropped. If there are more than one jets passing both cuts, we select the jet which forms the largest invariant mass together with the lepton. 

After selecting the jet in each event, we can proceed as usual by dividing the data points according to their invariant mass-squareds and finding the peak location in the $\log (E_{1T}/E_{2T})$ distribution for each invariant mass-squared bin.
The peak-point plot after smearing is shown in Fig.~\ref{z2com}.  The blue points are for all correct combinations and the green points are for selected combinations using the method we described above.  From the blue points we can see the effects of smearing: most points at lower invariant massed are shifted slightly upwards while the points with invariant mass close to $\frac{\Delta_1\Delta_2}{m^2_X}$ are shifted downwards. The wrong combinations mostly come from the extra gluon at low invariant masses and the jet from the other decay chain at high invariant masses. We see that the green points at small invariant masses are shifted up significantly due to the softness of the extra gluon. At large invariant masses close to the edge, the green points are further shifted down compared with the blue points because in this example the jet from the other chain on average has a larger energy than the correct jet does in the large invariant mass region.
\begin{figure}
\centering
\includegraphics[width=12cm]{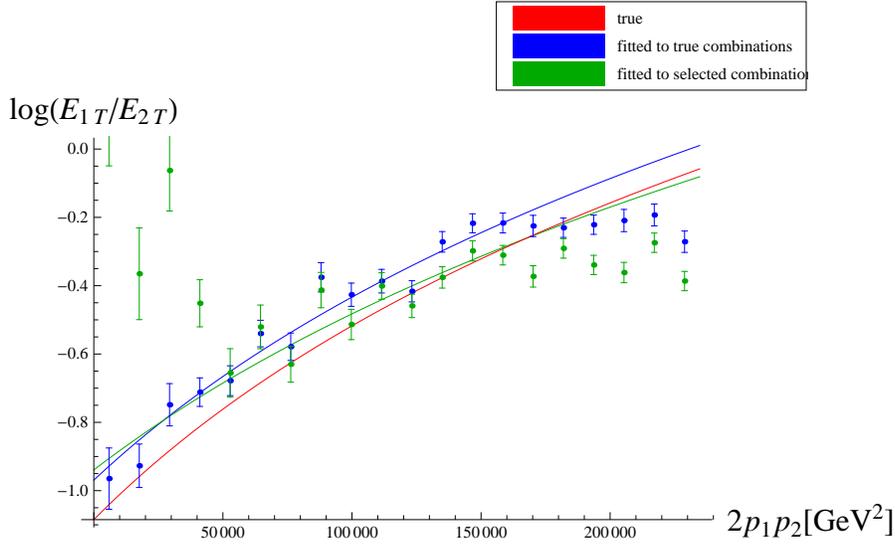}
\caption{The peak-point plot of $\log{(E_{1T}/E_{2T})}$ vs. $2p_1p_2$ after smearing. The events are divide into 20 sets with equal width ($0.05\times \frac{\Delta_1\Delta_2}{m^2_X}$) of invariant mass and there are 20 points in the plot. The red curve is $\log{(E_{1T}/E_{2T})}=\log{(\frac{\Delta_1+2p_1p_2}{\Delta_2})}$ with true $\Delta_1$ and $\Delta_2$.  The blue points are for all correct combinations and the blue curve is the fitted curve with last 5 points dropped.  The green points and curve are for the selected combinations with the method described in Section~\ref{comb}.  For the green curve, point 7 to 13 are used in the fit.}
\label{z2com}
\end{figure}

The wrong combinations make the mass determination more challenging.  We see that the peak points at both ends of the invariant mass range deviate from the true values significantly. Because the distribution of the ISR jets mostly lies at the upper left corner and the distribution of the jets from the other decay chain is flat in the $\log (E_{1T}/E_{2T})$ vs. $2 p_1 p_2$ plane, the effects of the wrong combinations from both sources will make the extracted slope smaller than the true value, which will result in too big reconstructed masses. To reduce the bias caused by wrong combinations, one may want to use only points in the middle part of the invariant mass-squared interval to perform the fit. However, the good  range for the fit seems to be model-dependent when we check models with different spins and spectra. A fixed range in the invariant mass-squared interval ({e.g.}, middle 1/3  of points) which works for some models does not work for the others. The best strategy that we come up with so far is to have a fixed size of the invariant mass-squared range, but keep the upper and lower ends floating to maximize the fitted slope. This is motivated by the fact that wrong combinations tend to reduce the slope. The range used for fitting should not be too small to avoid a too large slope coming from the statistical fluctuations, especially when the sample size is small. In Fig.~\ref{z2com} we fix the length of the fitting range to be 7 points and   find that point 7 to 13 (counting from the left) gives the largest slope after scanning through all possible choices. The result obtained from fitting these points is shown in Table~\ref{tz2}.  
\begin{table}
\begin{tabular}{|c|c|c|c|c|c|c|}
\hline
   & $\Delta_1[\mbox{GeV}^2]$ & $\Delta_2[\mbox{GeV}^2]$ & $\log{(\Delta_1/\Delta_2)}$ & $m_Y$[GeV] & $m_X$[GeV] & $m_N$[GeV] \\ \hline 
    true &  $1.310\times 10^5$  & $3.875\times 10^5$  & $-1.08$ & $777$ & $465$ & $292$ \\ \hline\hline
  \multicolumn{7}{|c|}{true combinations} \\ \hline
 reconstructed &  $1.411\times 10^5$  & $3.719\times 10^5$  & $-0.97$ & $772$ & $473$ & $287$ \\ \hline
 error &  $+7.8\%$  & $-4.0\%$  & $+12\%$ & $-0.68\%$ & $+1.7\%$ & $-1.6\%$ \\ \hline \hline
  \multicolumn{7}{|c|}{selected combinations} \\ \hline
 reconstructed &  $1.727\times 10^5$  & $4.418\times 10^5$  & $-0.939$ & $876$ & $570$ & $390$ \\ \hline
 error &  $+32\%$  & $+14\%$  & $+15\%$ & $+13\%$ & $+23\%$ & $+34\%$ \\ \hline
 \end{tabular}
 \caption{Results of reconstruction from the fit to the peak points after smearing with true and selected combinations.  The errors are calculated using $\frac{\rm reconstructed-true}{\rm true}$ (except for $\log{(\Delta_1/\Delta_2)}$, which is ${\rm reconstructed-true}$) and do not represent the statistical fluctuation.}
 \label{tz2}
 \end{table} 

The fitting method is somewhat heuristic so one would like to know whether the errors obtained in Table~\ref{tz2} is typical and how they depend on the size of the event sample. To study that we generate $10^5$ events
which are smeared according to the Gaussian errors listed in Table~\ref{smt}.  We first divide the events into 10 sets with $10^4$ events in each set, then reconstruct the masses for each set using the above method.  The results are shown in Table~\ref{tz3}. In practice, $10^4$ is a very large number in terms of number of events which can obtained in real experiments, so we also check how well the method works with less events.  To do this, we divide the events into 50 sets, each containing $2000$ events. The results of the mass reconstruction in this case are shown in Table \ref{tz4} using the same method, except that the length of the fitting range is increased to 8 points to compensate for the effect of the larger fluctuations due to the smaller sample size.
 \begin{table}
 \centering
\begin{tabular}{|c|c|c|c|}
\hline
 &$m_Y$[GeV] & $m_X$[GeV] & $m_N$[GeV] \\ \hline  \hline
 true & $777$ & $465$ & $292$ \\ \hline
reconstructed & $931\pm 100$ & $624\pm 103$ & $445 \pm 99$ \\ \hline
error & $+20\%\pm 13\%$ & $+34\%\pm 22\%$ & $+52\%\pm 34\%$ \\ \hline
 \end{tabular}
  \caption{The reconstructed masses from 10 sets of events, each with $10^4$ events, in the form of mean $\pm$ standard deviation.  The reconstruction is obtained by fitting with 7 consecutive peak points which maximize the fitted slope.}
 \label{tz3}
 \end{table}  
 
 \begin{table}
 \centering
\begin{tabular}{|c|c|c|c|}
\hline
 &$m_Y$[GeV] & $m_X$[GeV] & $m_N$[GeV] \\ \hline  \hline
 true & $777$ & $465$ & $292$ \\ \hline
reconstructed & $754\pm 157$ & $442\pm 162$ & $265 \pm 162$ \\ \hline
error & $-2.9\%\pm 20\%$ & $-4.9\%\pm 35\%$ & $-9.4\%\pm 55\%$ \\ \hline
 \end{tabular}
 \caption{The reconstructed masses from 50 sets of events, each with $2000$ events, in the form of mean $\pm$ standard deviation.  The reconstruction is obtained by fitting with 8 consecutive peak points which maximize the fitted slope.}
 \label{tz4}
 \end{table}  
 
The uncertainties of the results in Table~\ref{tz4} are quite substantial, though they already represent significant progress as there was no existing method which can determine the invisible particle masses for this topology.  We have used a rather simple and somewhat heuristic method in treating the combinatorial problem and the results may be improved if more sophisticated techniques are employed. Some possible improvements include:
\begin{enumerate}
\item We use a na\"ive method to select the jet to pair with the lepton. We expect that the correct rate can be improved with a better algorithm. The ISR jet may be better identified with more sophisticated methods~\cite{Krohn:2011zp,Gallicchio:2011xq}. One can also employ a full likelihood method to distinguish the correct jet from the wrong ones using more variables. With better selections we may extend the fitting range to reduce statistical uncertainties and model dependence. 
\item Our heuristic method in choosing the fitting range is concerted in such a way that it can be  applied to a wide range of models. If more detailed information is available about the process and the impostor particles from other part of the event which contains the decay chain, one can design a fitting procedure best suit for the specific set of signal events using all available information and hence obtain better results. 
\end{enumerate}

\section{Conclusions}
\label{sec:conclusions}

In searching for new physics at colliders there are different approaches. Given a specific model, one can design cuts and strategies to obtain the maximal reach for that particular model, and perform the most likelihood fit to extract model parameters. However, they may not be suitable for a different model. As there is a vast number of possibilities for new physics at the TeV scale, one can only perform specific searches for a limited number of models and there is no reason to expect that they are favored over the other models. On the other hand, 
model-independent global searches compare data and standard model predictions in a global sample of high-$p_T$ collision events and look for excess in the high-$p_T$ tails as signals for new physics~\cite{Aaltonen:2007dg}. They do not require specification of any particular model. The drawback is that new physics signals may be subtle and hide in the correlations of observables, then the global searches are not effective to uncover them.

An intermediate approach is to focus on the possible event topologies of new physics without completely specifying the underlying model.  
An example in this direction is the simplified model approach for new physics searches~\cite{Alves:2011wf}. For any given event topology, one can look for kinematic variables and relations which exhibit particular features for that topology, so that one can use them to distinguish new physics from the standard model backgrounds and also extract relevant parameters of new physics. Because many different models can produce signal events of the same topologies, the study of any particular event topology can apply to a wide range of possible new physics models. 

In this paper we consider the event topology of a 2-step decay chain ending with a missing particle. This topology appears in many models which contain a stable or long-lived neutral particle that escapes detection. Out of the two visible particles coming from the decay chain, the obvious kinematic variable to look at is the invariant mass combination of the 4-momenta of the two visible particles, and one might thought that it is the only relevant kinematic variable for this topology. We show that the logarithm of the transverse energy ratio of the two visible particles is also a useful variable. The event distribution in the  $\log(E_{1T}/E_{2T})$ vs. invariant mass-squared space carry useful information beyond what is contained in the one-dimensional invariant mass distribution. In many cases it allows the extraction of masses of all three invisible particles in the decay chain. The shape of the distribution may also be used to distinguish signals from backgrounds.

The method of searching for new useful kinematic variables may be generalized to other event topologies. Digging new physics signals out 
of the tremendous backgrounds at hadron colliders is always a challenging task. Finding the most effective kinematic variables for new physics signals will improve the search reaches and help us to figure out the properties of the underlying theory once it is discovered.

\section*{Acknowledgments}
We would like to thank Spencer Chang, Max Chertok, Zhenyu Han, Ian-Woo Kim, Markus Luty and Jesse Thaler for discussion. H.-C. C. would like to thank the Kavli Institute for Theoretical Physics and CERN TH-LPCC Summer Institute where part of this work was done. This work was supported by the US Department of Energy under contract DE-FG02-91ER406746.

\appendix

\section{More Parton-level Examples}
\label{sec:other}

In this Appendix we present a few more parton-level examples for our mass determination method. It works well for a variety of models with different mass spectra and spins. In all examples in this Appendix, the mother particle $Y$ decays isotropically and the order of the visible particles are assumed to be known. Anisotropic decays of particle $Y$ will cause a bias in mass determination which will be discussed in Appendix~\ref{sec:fail}. The combinatorial problems of indistinguishable particles or orders are discussed in sec.~\ref{real}.
 
 \subsection*{Example 1}
 
In this example we consider a SUSY process with a heavy neutralino (particle $Y$) decaying through an on-shell slepton (particle $X$) to the lightest neutralino (particle $N$).  The mass spectrum and spin configuration are summarized in Table~\ref{o1sm}.  The invariant mass distribution is shown in Fig.~\ref{o1inv}.  The scatter plot and the peak-point plot are shown in Fig.~\ref{op1}.  The result of mass reconstruction from fitting the peak points is shown in Table.~\ref{o1reg}.

 \begin{table}[H]
 \centering
 \begin{tabular}{|c|c|c|c|}
 \hline
& $Y$ & $X$ & $N$ \\ \hline
spin & 1/2 & 0 & 1/2  \\ \hline
 mass[GeV]  & 468 & 304 & 140.5 \\ \hline
  \end{tabular}
 \caption{The summary of the mass spectrums and spin configurations of Example 1 in Appendix~\ref{sec:other}.}
 \label{o1sm}
 \end{table}
 \begin{figure}[H]
\centering
\includegraphics[width=7cm]{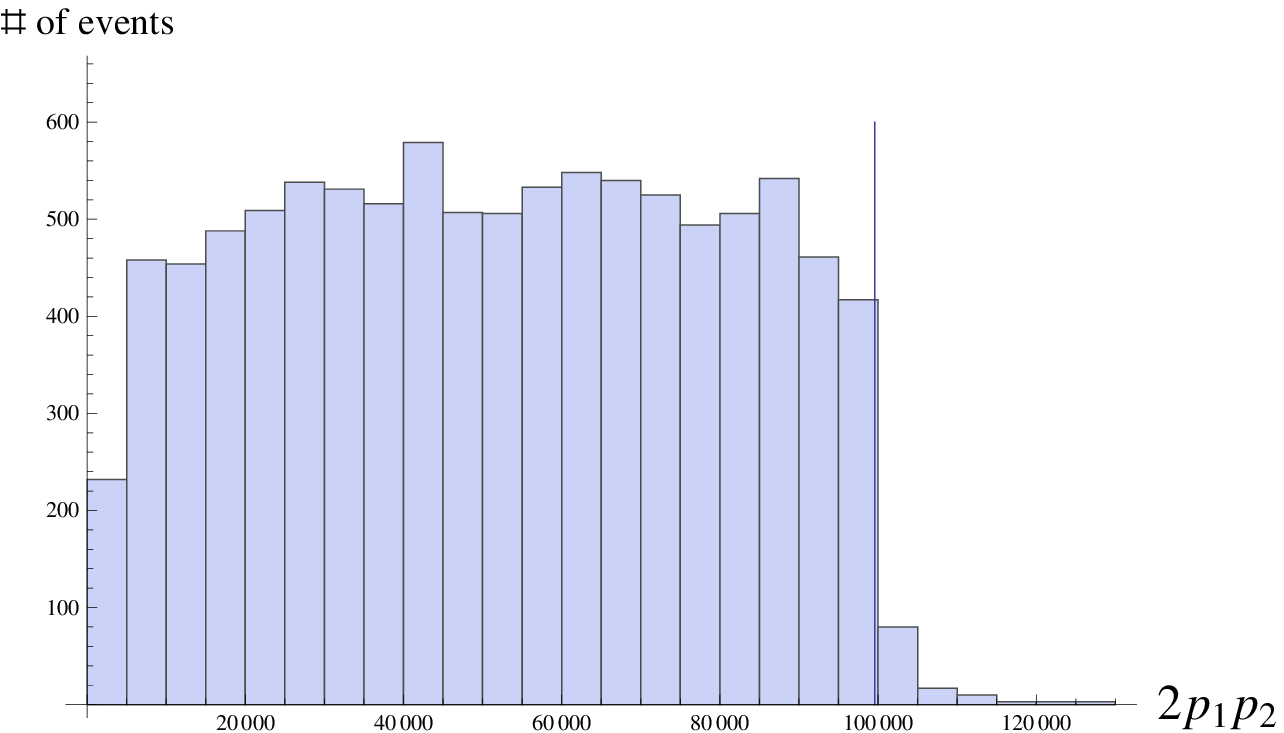}
\caption{The histogram of the invariant mass-squared of the two visible particles for Example 1 in Appendix~\ref{sec:other}.  In the massless limit $(p_1+p_2)^2 = 2p_1p_2$.  The vertical line indicates the value of $\frac{\Delta_1\Delta_2}{m^2_X}$.}
\label{o1inv}
\end{figure}
\begin{figure}[H]
\centering
\includegraphics[width=7cm]{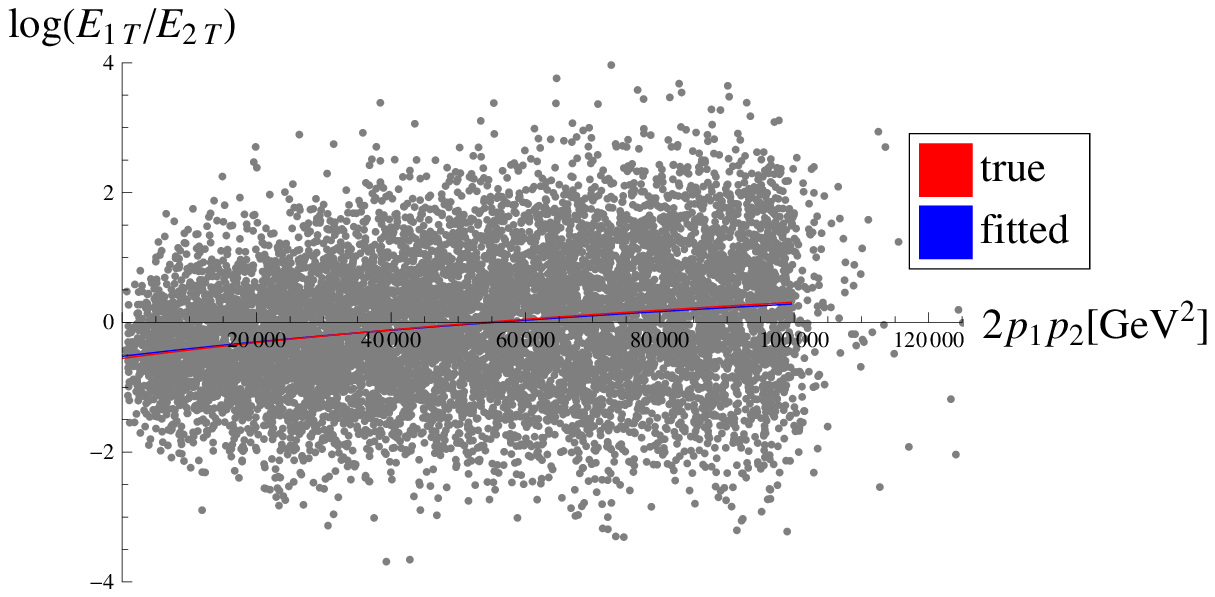}
\includegraphics[width=7cm]{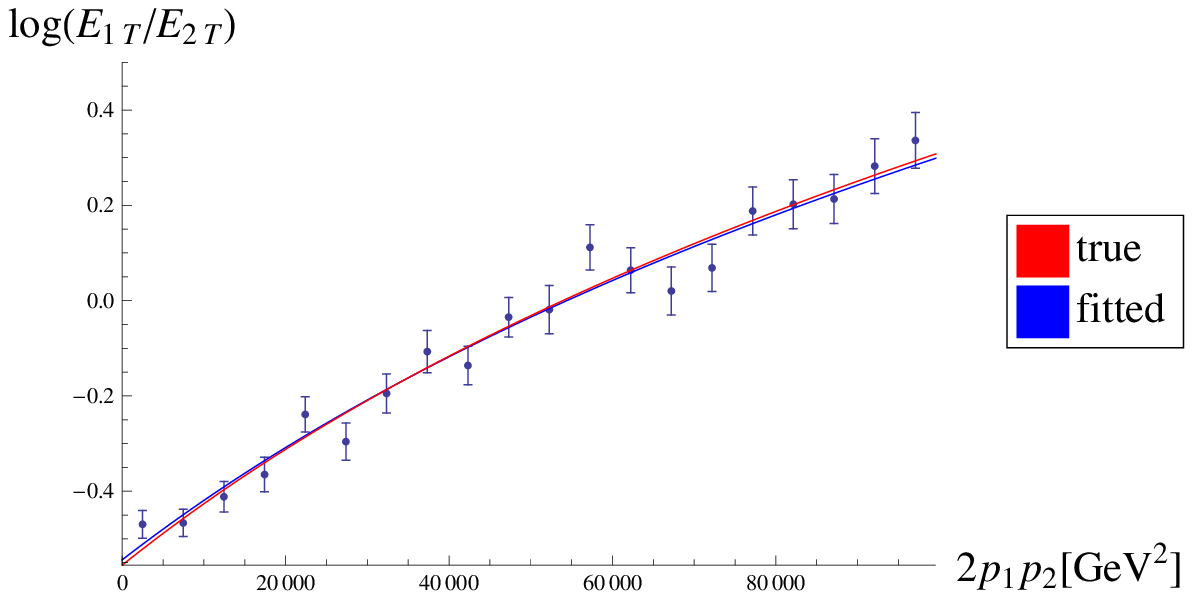}
\caption{The scatter plot (left) and the peak-point plot (right) of $\log{(E_{1T}/E_{2T})}$ vs. $2p_1p_2$ for Example 1 in Appendix~\ref{sec:other}.  In both case the red curve is $\log{(E_{1T}/E_{2T})}=\log{(\frac{\Delta_1+2p_1p_2}{\Delta_2})}$ with true $\Delta_1$ and $\Delta_2$ and the blue curve is the fitted one.}
\label{op1}
\end{figure}
\begin{table}[H]
\begin{tabular}{|c|c|c|c|c|c|c|}
\hline
   & $\Delta_1[\mbox{GeV}^2]$ & $\Delta_2[\mbox{GeV}^2]$ & $\log{(\Delta_1/\Delta_2)}$ & $m_Y$[GeV] & $m_X$[GeV] & $m_N$[GeV] \\ \hline \hline
 true &  $7.268\times 10^4$  & $1.266\times 10^5$  & $-0.555$ & $468$ & $304$ & $140.5$ \\ \hline
 reconstructed &  $7.517\times 10^4$  & $1.295\times 10^5$  & $-0.544$ & $477$ & $313$ & $150.4$ \\ \hline
error &  $+3.4\%$  & $+2.3\%$  & $+1.1\%$ & $+1.9\%$ & $+2.9\%$ & $+7.1\%$ \\ \hline
 \end{tabular}
 \caption{The result of mass reconstruction from fitting the peak points for Example 1 in Appendix~\ref{sec:other}.  The errors are calculated using $\frac{\rm reconstructed-true}{\rm true}$ (except for $\log{(\Delta_1/\Delta_2)}$, which is ${\rm reconstructed-true}$) and do not represent the statistical fluctuation.  The statistical fluctuation are expected to be at the same order as the example in Section \ref{ti}.}
 \label{o1reg}
 \end{table}

 \subsection*{Example 2}

This is a simplified model. The interactions are vector-like so that there is no preferred direction for the particle $Y$ decay. The mass spectrum and spin configuration of the particles in the decay chain are summarized in Table~\ref{u1sm}.  The invariant mass distribution is shown in Fig.~\ref{u1inv}.  The scatter plot and the peak-point plot are shown in Fig.~\ref{up1}.  The result of mass reconstruction from fitting the peak points is shown in Table.~\ref{u1reg}.

 \begin{table}[H]
 \centering
 \begin{tabular}{|c|c|c|c|}
 \hline
 & $Y$ & $X$ & $N$ \\ \hline
spin &  1 & 1/2 & 1   \\ \hline
 mass[GeV]  &  1000 & 800 & 500 \\ \hline
  \end{tabular}
 \caption{The summary of the mass spectrums and spin configurations of Example 2 in Appendix~\ref{sec:other}.}
 \label{u1sm}
 \end{table}
 \begin{figure}[H]
\centering
\includegraphics[width=7cm]{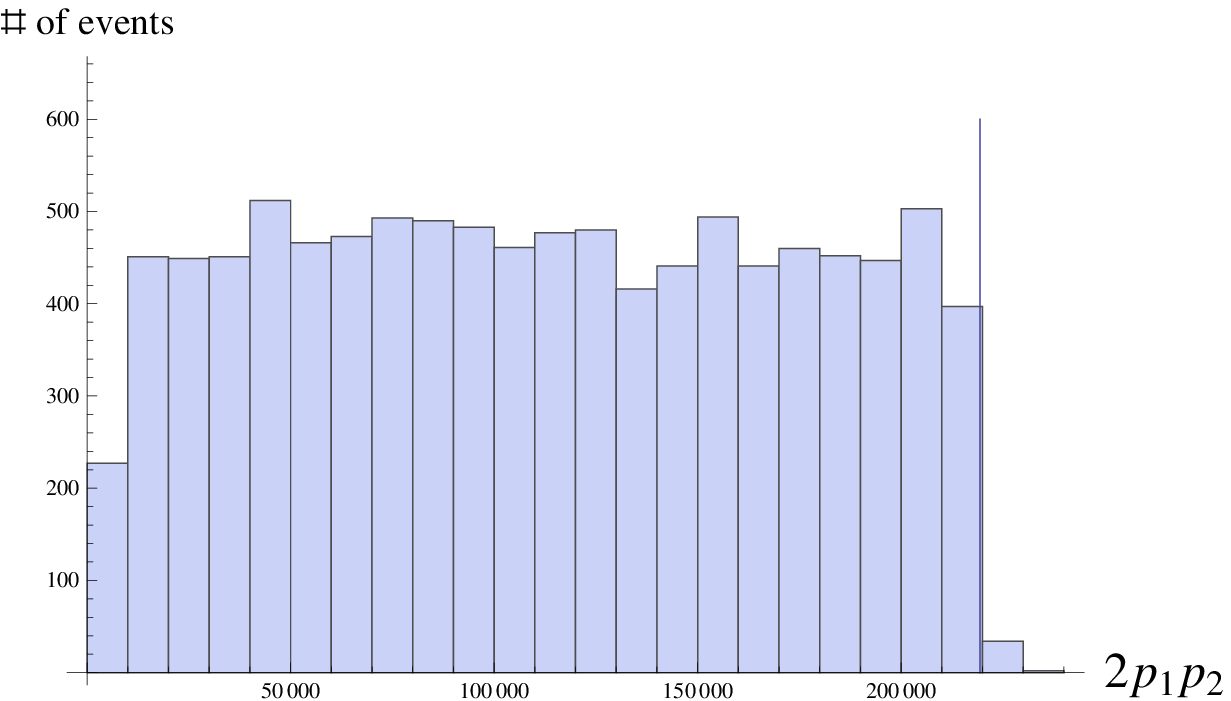}
\caption{The histogram of the invariant mass-squared of the two visible particles for Example 2 in Appendix~\ref{sec:other}.  In the massless limit $(p_1+p_2)^2 = 2p_1p_2$.  The vertical line indicates the value of $\frac{\Delta_1\Delta_2}{m^2_X}$.}
\label{u1inv}
\end{figure}
\begin{figure}
\centering
\includegraphics[width=7cm]{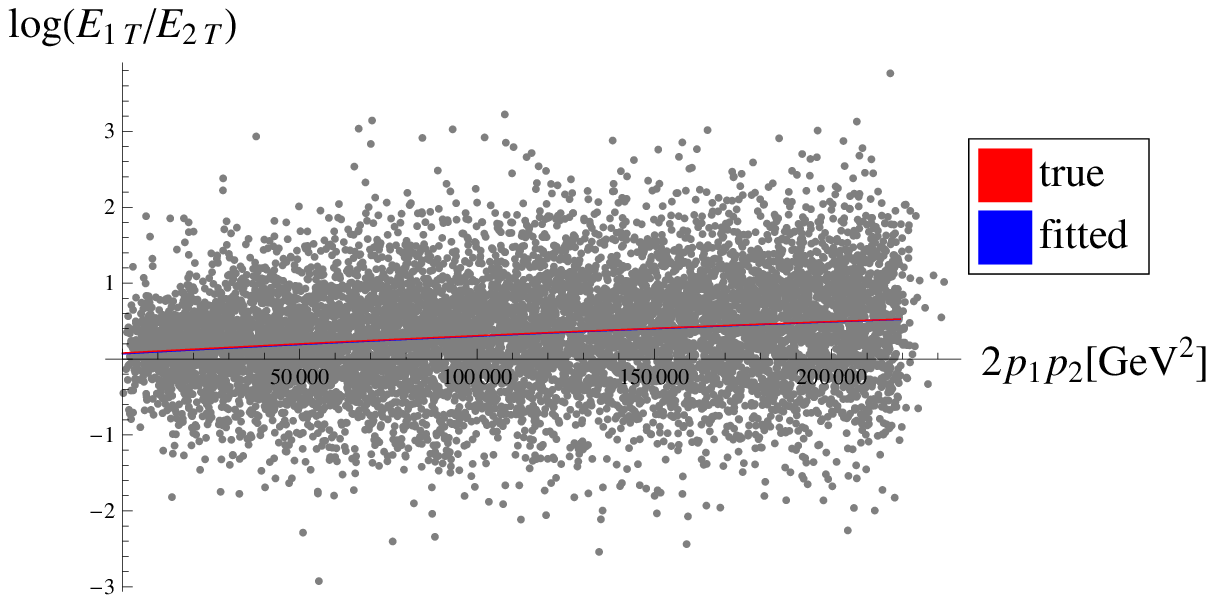}
\includegraphics[width=7cm]{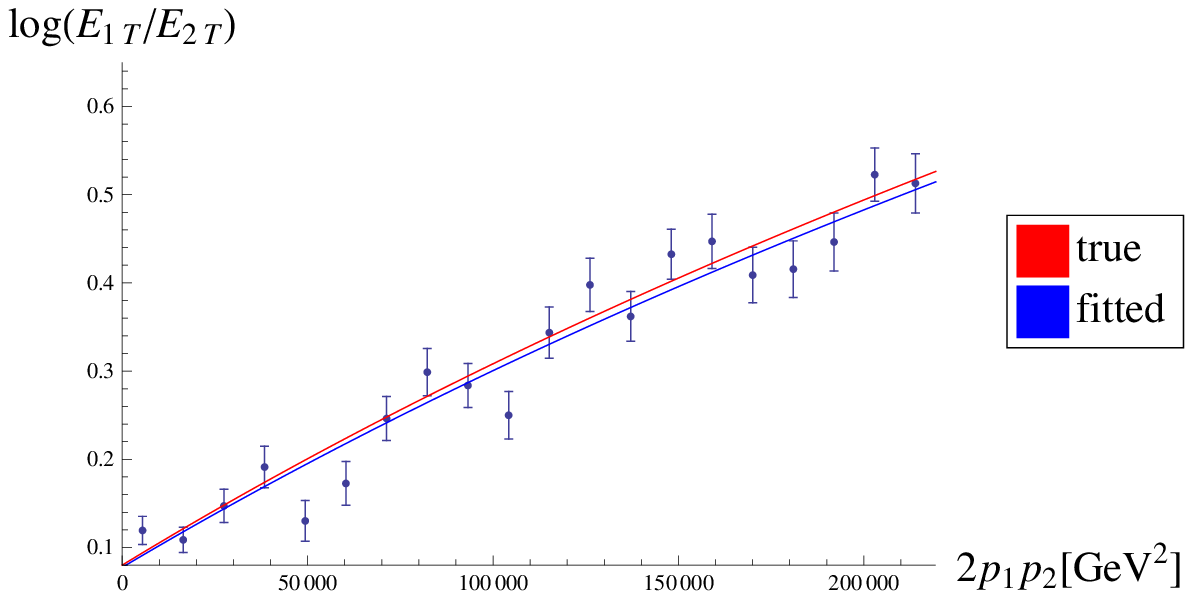}
\caption{The scatter plot (left) and the peak-point plot (right) of $\log{(E_{1T}/E_{2T})}$ vs. $2p_1p_2$ for Example 2 in Appendix~\ref{sec:other}.  In both case the red curve is $\log{(E_{1T}/E_{2T})}=\log{(\frac{\Delta_1+2p_1p_2}{\Delta_2})}$ with true $\Delta_1$ and $\Delta_2$ and the blue curve is the fitted one.}
\label{up1}
\end{figure}
\begin{table}[H]
\begin{tabular}{|c|c|c|c|c|c|c|}
\hline
   & $\Delta_1[\mbox{GeV}^2]$ & $\Delta_2[\mbox{GeV}^2]$ & $\log{(\Delta_1/\Delta_2)}$ & $m_Y$[GeV] & $m_X$[GeV] & $m_N$[GeV] \\ \hline \hline
 true &  $3.900\times 10^5$  & $3.600\times 10^5$  & $0.080$ & $1000$ & $800$ & $500$ \\ \hline
 reconstructed &  $4.003\times 10^5$  & $3.704\times 10^5$  & $0.078$ & $1023$ & $822$ & $525$ \\ \hline
error &  $+2.6\%$  & $+2.9\%$  & $-0.25\%$ & $+2.3\%$ & $+2.8\%$ & $+5.0\%$ \\ \hline
 \end{tabular}
 \caption{The result of mass reconstruction from fitting the peak points for Example 2 in Appendix~\ref{sec:other}.  The errors are calculated using $\frac{\rm reconstructed-true}{\rm true}$ (except for $\log{(\Delta_1/\Delta_2)}$, which is ${\rm reconstructed-true}$) and do not represent the statistical fluctuation.  The statistical fluctuation are expected to be at the same order as the example in Section \ref{ti}.}
 \label{u1reg}
 \end{table} 

 \subsection*{Example 3}

This is also a simplified model with vector-like couplings.  The mass spectrum and spin configuration are summarized in Table~\ref{v2sm}.  The invariant mass distribution is shown in Fig.~\ref{v2inv}.  The scatter plot and the peak-point plot are shown in Fig.~\ref{vp2}.  The result of mass reconstruction from fitting the peak points is shown in Table.~\ref{v2reg}.

 \begin{table}[H]
 \centering
 \begin{tabular}{|c|c|c|c|}
 \hline
 & $Y$ & $X$ & $N$ \\ \hline
spin &  1/2 & 1 & 1/2   \\ \hline
 mass[GeV]  &  800 & 300 & 100 \\ \hline
  \end{tabular}
 \caption{The summary of the mass spectrums and spin configurations of Example 3 in Appendix~\ref{sec:other}.}
 \label{v2sm}
 \end{table}
 \begin{figure}[H]
\centering
\includegraphics[width=7cm]{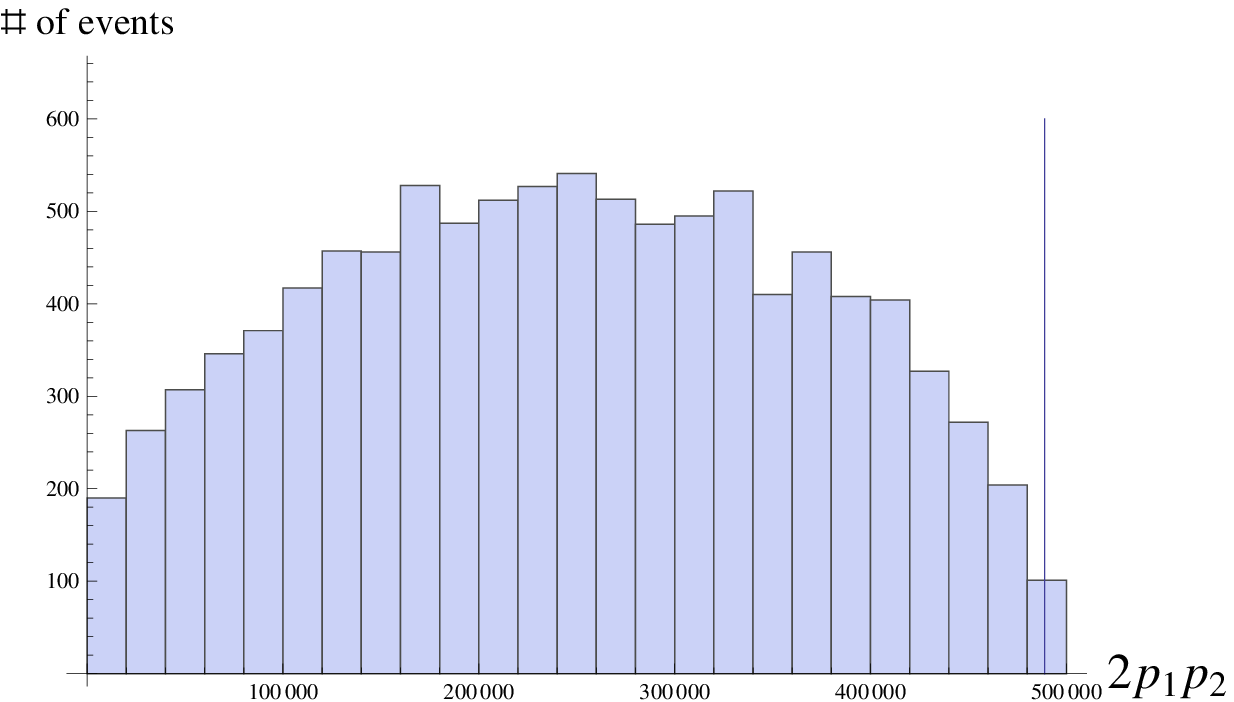}
\caption{The histogram of the invariant mass-squared of the two visible particles for Example 3 in Appendix~\ref{sec:other}.  In the massless limit $(p_1+p_2)^2 = 2p_1p_2$.  The vertical line indicates the value of $\frac{\Delta_1\Delta_2}{m^2_X}$.}
\label{v2inv}
\end{figure}
\begin{figure}[H]
\centering
\includegraphics[width=7cm]{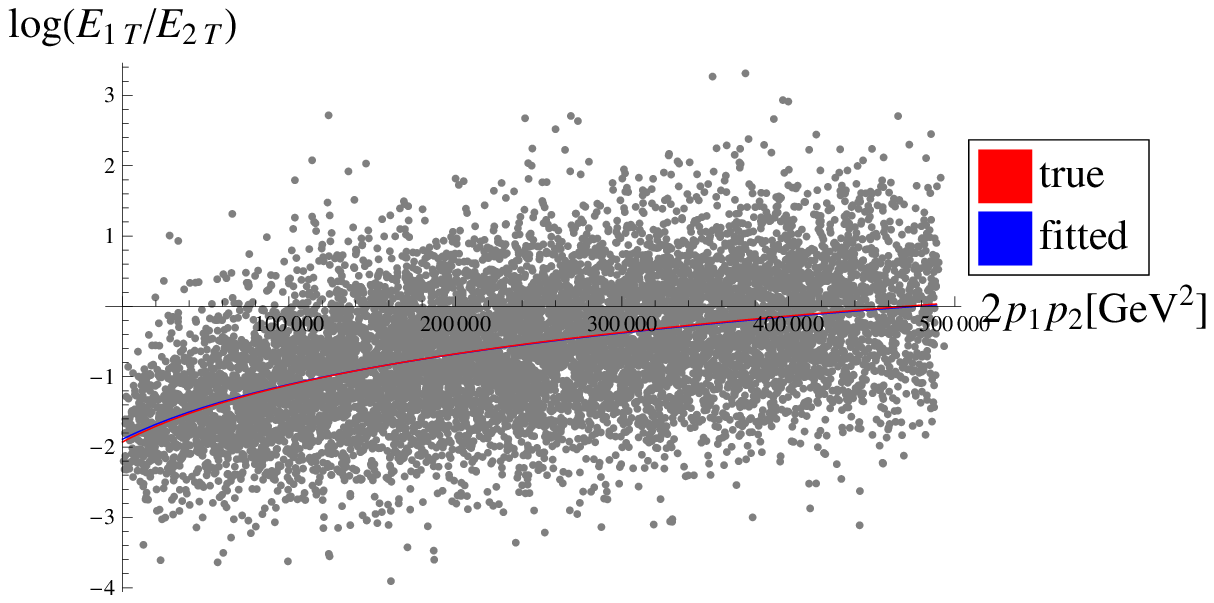}
\includegraphics[width=7cm]{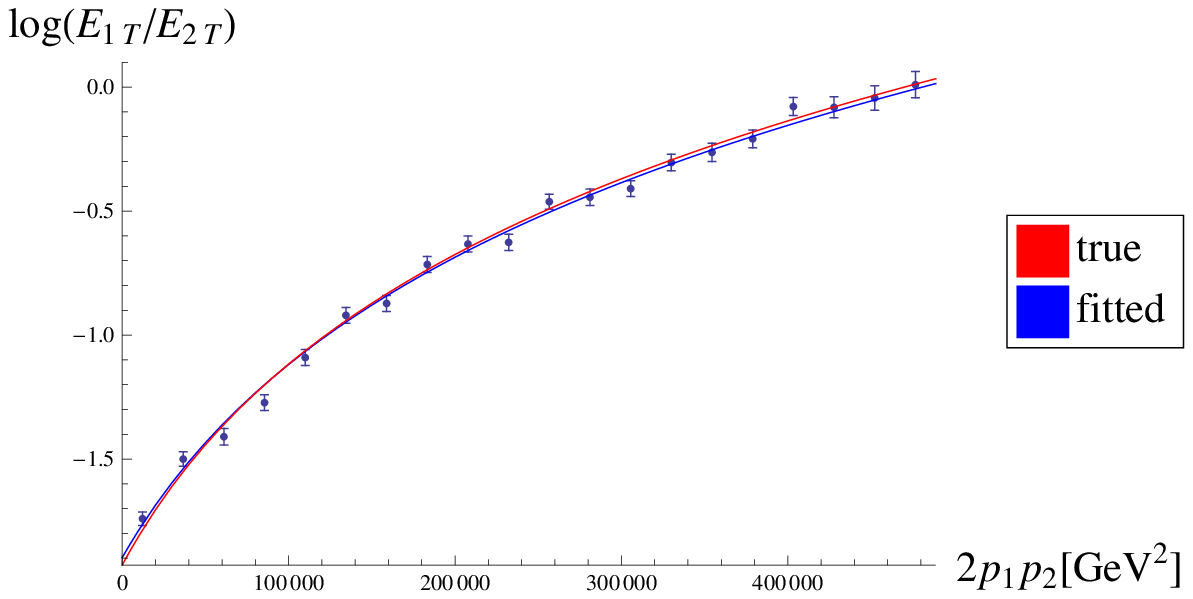}
\caption{The scatter plot (left) and the peak-point plot (right) of $\log{(E_{1T}/E_{2T})}$ vs. $2p_1p_2$ for Example 3 in Appendix~\ref{sec:other}.  In both case the red curve is $\log{(E_{1T}/E_{2T})}=\log{(\frac{\Delta_1+2p_1p_2}{\Delta_2})}$ with true $\Delta_1$ and $\Delta_2$ and the blue curve is the fitted one.}
\label{vp2}
\end{figure}
\begin{table}[H]
\begin{tabular}{|c|c|c|c|c|c|c|}
\hline
   & $\Delta_1[\mbox{GeV}^2]$ & $\Delta_2[\mbox{GeV}^2]$ & $\log{(\Delta_1/\Delta_2)}$ & $m_Y$[GeV] & $m_X$[GeV] & $m_N$[GeV] \\ \hline \hline
 true &  $8.000\times 10^4$  & $5.500\times 10^5$  & $-1.93$ & $800$ & $300$ & $100$ \\ \hline
 reconstructed &  $8.480\times 10^4$  & $5.656\times 10^5$  & $-1.90$ & $815$ & $313$ & $115$ \\ \hline
error &  $+6.0\%$  & $+2.8\%$  & $+3.0\%$ & $+1.8\%$ & $+4.4\%$ & $+15\%$ \\ \hline
 \end{tabular}
 \caption{The result of mass reconstruction from fitting the peak points for Example 3 in Appendix~\ref{sec:other}.  The errors are calculated using $\frac{\rm reconstructed-true}{\rm true}$ (except for $\log{(\Delta_1/\Delta_2)}$, which is ${\rm reconstructed-true}$) and do not represent the statistical fluctuation.  The statistical fluctuation are expected to be at the same order as the example in Section \ref{ti}.}
 \label{v2reg}
 \end{table} 

\section{Problematic Cases for Mass Determination}
\label{sec:fail}

In this Appendix we discuss cases where the mass determination method does not work even at the parton level. The two cases mentioned in sec.~\ref{ti} are discussed in more details below.

\subsection{The mother particle $Y$ is polarized and decays anisotropically}
\label{polar}

Our mass determination method relies on that the distribution of $\log{\frac{1+\beta_Y \cos{\theta_{1Y}}}{1+\beta_Y \cos{\theta_{2Y}}}}\big|_{Y}$ is symmetric around zero. This is obviously not the case if the decay of particle $Y$ is not isotropic and preferentially emits particle 2 in the forward (or backward) direction. It can happen if the particle $Y$ has nonzero spin and is polarized. Even if $Y$ is polarized, sometimes a P or CP symmetry may still ensure the decay to be symmetric when all final states are included, as in the case of a Majorana neutralino, so some chiral structure of the process needs to be present. In the minimal supersymmetric standard model, the only particles that can have asymmetric decays are the charginos, so the study case that we consider is a decay chain initiated by a chargino. The particles in the decay chain are summarized in Table~\ref{tpc2}.  The underlying model is SUSY LM2, with the mass of the heavy chargino adjusted by hand so that the decay chain can occur. To illustrate the polarization effect we restrict the heavy chargino to be produced only from the $t$-channel processes to obtain a larger polarization.
\begin{table}
\centering
\begin{tabular}{|c|c|c|c|}
\hline
     & $Y$ & $X$ & $N$ \\ \hline
   particle & 2nd chargino  & left handed anti up squark & 1st neutralino  \\ \hline
   mass[GeV] & 1000 & 770 & 140.5 \\ \hline
 \end{tabular}
 \caption{A summary of the decay chain in Appendix~\ref{polar}.}
 \label{tpc2}
 \end{table} 

The average value of the cosine of the transverse angle between particle 2 in the rest frame of $Y$ and particle $Y$ in the lab frame $\langle\cos \theta_{2Y} \big|_Y\rangle$ is $0.124$ (while $\langle\cos \theta_{1Y} \big|_Y\rangle = -0.033$).  
The distribution of $\log{\frac{1+\beta_Y \cos{\theta_{1Y}}}{1+\beta_Y \cos{\theta_{2Y}}}}\big|_{Y}$ after we boost each event to the frame where ${E_1}/{E_2}={E_{1T}}/{E_{2T}}$ is shown in Fig.~\ref{c2log}.  The distribution has a bias and the average value is $-0.136$.  The scatter plot and the peak-point plot (right) are shown in Fig.~\ref{pc2}.  The ``peak points" are clearly shifted downwards.  The slope of the fitted curve is smaller than the true slope, but the intercept is still close to the true value.    The result of mass reconstruction shown in Table~\ref{tc2} exhibits large deviations.
 \begin{figure}
\centering
\includegraphics[width=7cm]{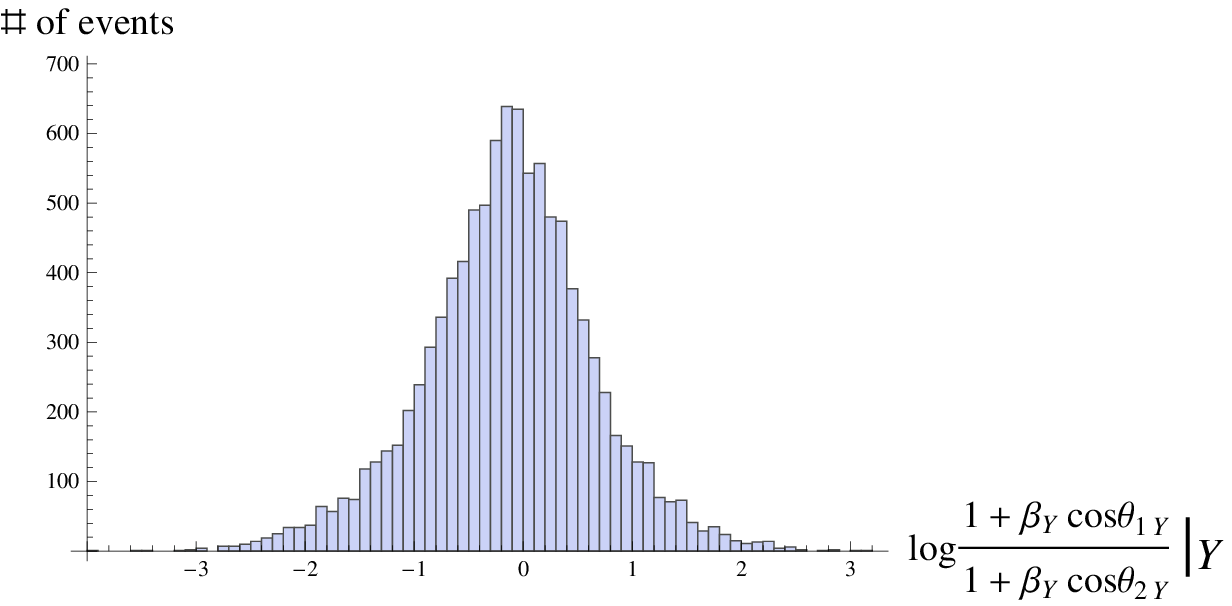}
\includegraphics[width=7cm]{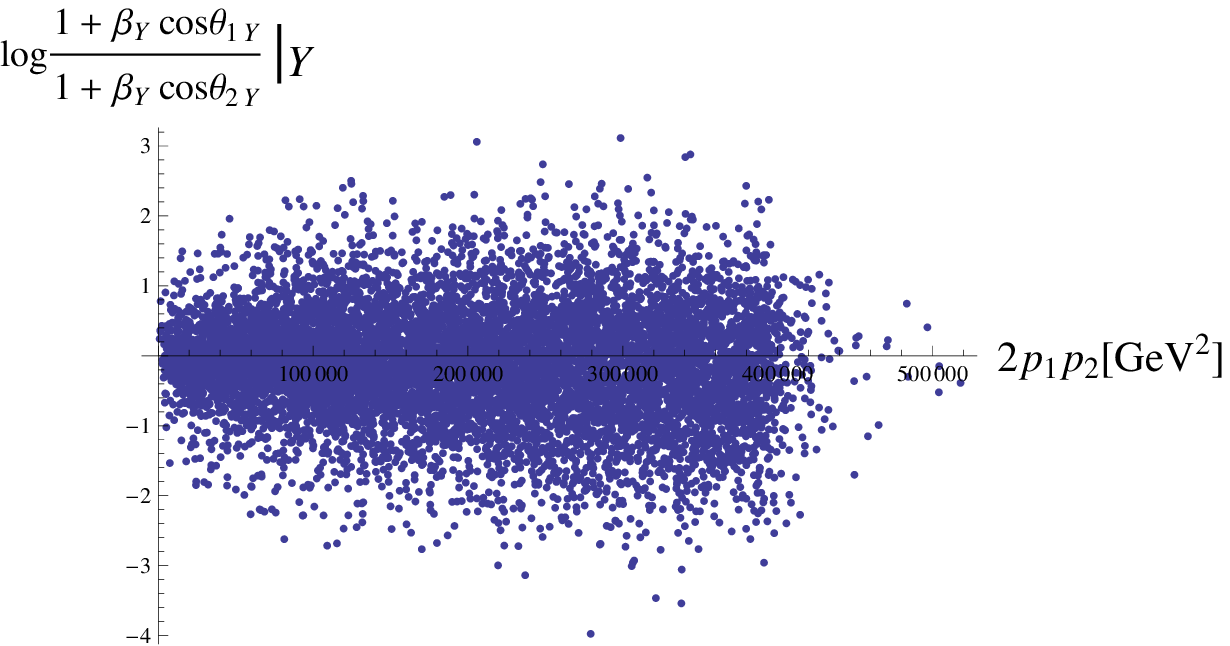}
\caption{The distribution of $\log{\frac{1+\beta_Y \cos{\theta_{1Y}}}{1+\beta_Y \cos{\theta_{2Y}}}}\big|_{Y}$ (Left) and $\log{\frac{1+\beta_Y \cos{\theta_{1Y}}}{1+\beta_Y \cos{\theta_{2Y}}}}\big|_{Y}$ vs. invariant mass-squared (Right) of Appendix~\ref{polar}.  Each event is boosted longitudinally to the frame in which $\frac{E_1}{E_2}=\frac{E_{T1}}{E_{T2}}$.  The mean value of $\log{\frac{1+\beta_Y \cos{\theta_{1Y}}}{1+\beta_Y \cos{\theta_{2Y}}}}\big|_{Y}$ is $-0.136$.}
\label{c2log}
\end{figure}
\begin{figure}
\centering
\includegraphics[width=7cm]{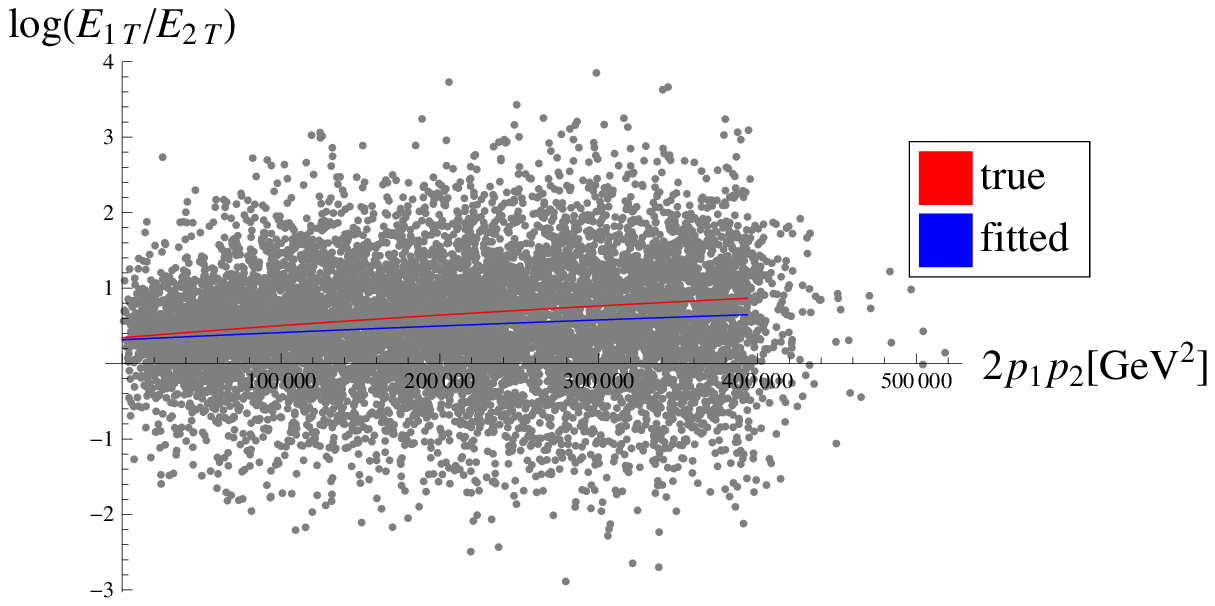}
\includegraphics[width=7cm]{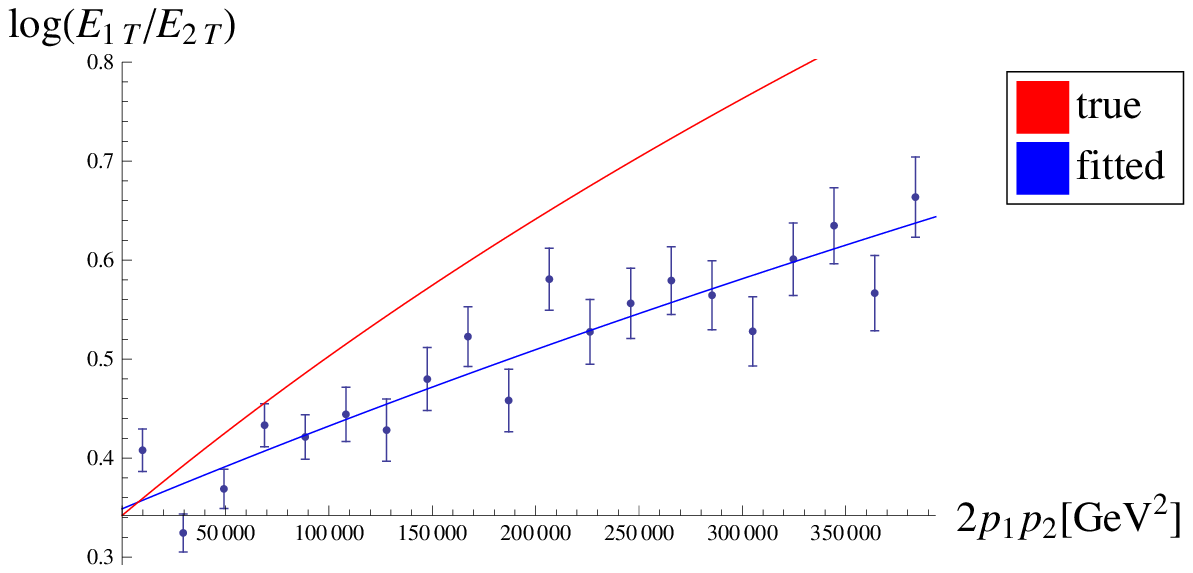}
\caption{The scatter plot (left) and the peak-point plot (right) of $\log{(E_{1T}/E_{2T})}$ vs. $2p_1p_2$ for Appendix~\ref{polar}.  In both case the red curve is $\log{(E_{1T}/E_{2T})}=\log{(\frac{\Delta_1+2p_1p_2}{\Delta_2})}$ with true $\Delta_1$ and $\Delta_2$ and the blue curve is the fitted one.}
\label{pc2}
\end{figure}
\begin{table}
\begin{tabular}{|c|c|c|c|c|c|c|}
\hline
   & $\Delta_1[\mbox{GeV}^2]$ & $\Delta_2[\mbox{GeV}^2]$ & $\log{(\Delta_1/\Delta_2)}$ & $m_Y$[GeV] & $m_X$[GeV] & $m_N$[GeV] \\ \hline \hline
 true &  $5.732\times 10^5$  & $4.071\times 10^5$  & $0.342$ & $1000$ & $770$ & $140.5$ \\ \hline
 reconstructed &  $1.147\times 10^6$  & $8.092\times 10^5$  & $0.349$ & $1780$ & $1536$ & $1101$ \\ \hline
 error &  $+100\%$  & $+99\%$  & $+0.67\%$ & $+78\%$ & $+99\%$ & $+683\%$ \\ \hline
 \end{tabular}
 \caption{The result of mass reconstruction from fitting the peak points for Appendix~\ref{polar}.  The errors are calculated using $\frac{\rm reconstructed-true}{\rm true}$ (except for $\log{(\Delta_1/\Delta_2)}$, which is ${\rm reconstructed-true}$) and do not represent the statistical fluctuation.  The statistical fluctuation are expected to be at the same order as the example in Section \ref{ti}.}
 \label{tc2}
 \end{table} 

\subsection{Soft visible particles and off-shell effects}
\label{o4}

As mentioned in sec.~\ref{ti}, if the mass difference of two neighboring invisible particles in the decay chain is small, the off-shell effects and the ``fake polarization'' created by the $E_T$ cut may cause our mass determination method to fail. Here we examine these effects in more details.  We first consider a decay chain with a small $\Delta_2$. The spectrum is summarized in Table~\ref{tpo4}.  The on-shell approximation does not seem to work well for particle $Y$, which can be seen in Fig.~\ref{o4width}.  The distribution of $\delta\log{(\frac{\Delta_1+2p_1p_2}{\Delta_2})}$ (defined as the effective value minus the true value, where the effective value for each event is calculated using $p^2_Y$ and $p^2_X$ while the true value is calculated using $m^2_Y$ and $m^2_X$) vs. invariant mass-squared is shown in Fig.~\ref{o4dld}.   One could see that it is asymmetric especially near $2p_1p_2 = \frac{\Delta_1\Delta_2}{m^2_X}$.  This is due to a strong correlation between the maximum invariant mass and the effective values of $\Delta_1$ and $\Delta_2$.
\begin{table}
\centering
\begin{tabular}{|c|c|c|c|}
\hline
     & $Y$ & $X$ & $N$ \\ \hline
   particle & heavy neutralino & slepton & lightest neutralino  \\ \hline
   mass[GeV] & 468 & 441.9 & 140.5 \\ \hline
 \end{tabular}
 \caption{A summary of the decay chain in Appendix~\ref{o4}.}
 \label{tpo4}
 \end{table} 
\begin{figure}
\centering
\includegraphics[width=7cm]{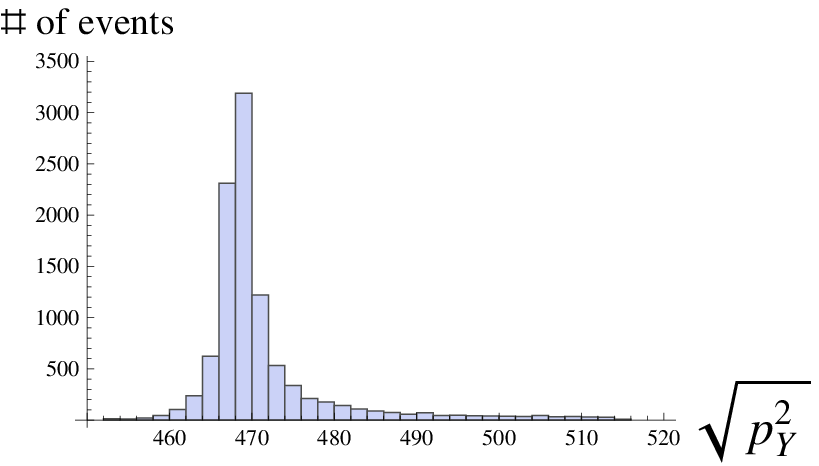}
\includegraphics[width=7cm]{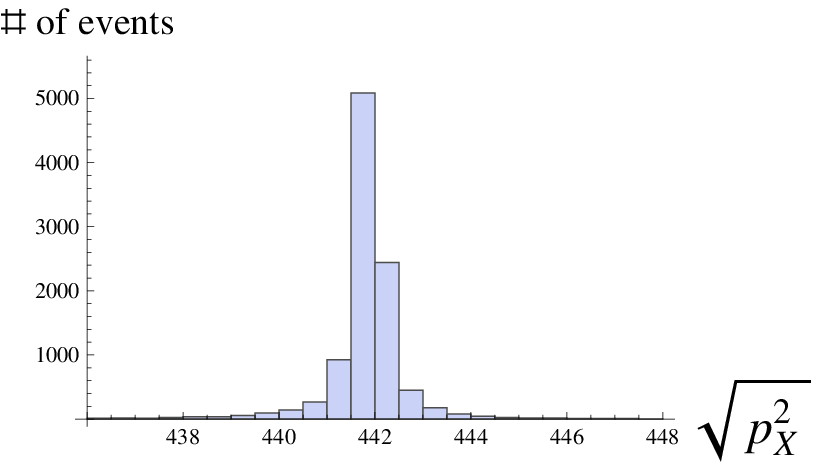}
\caption{The histograms of $\sqrt{p^2_Y}$ and $\sqrt{p^2_X}$ of Appendix~\ref{o4}.  The mass and width of $Y$ are 468 GeV and 3.115 GeV.  The mass and width of $X$ are 441.9 GeV and 0.4076 GeV.}
\label{o4width}
\end{figure}
\begin{figure}
\centering
\includegraphics[width=10cm]{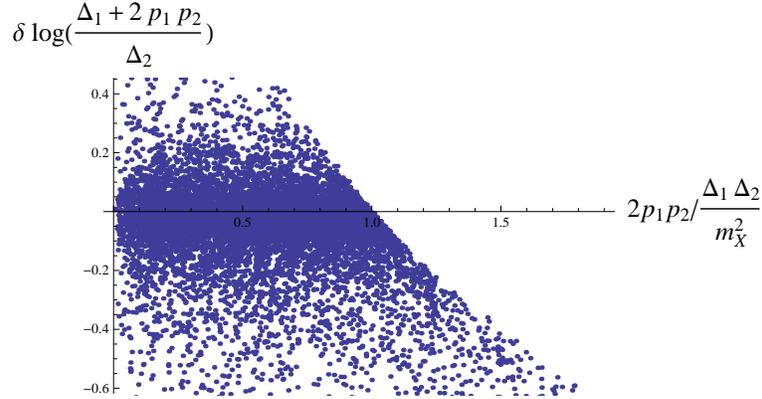}
\caption{The scatter plot of $\delta\log{(\frac{\Delta_1+2p_1p_2}{\Delta_2})}$ vs. $2p_1p_2/\frac{\Delta_1\Delta_2}{m^2_X}$.  $\delta\log{(\frac{\Delta_1+2p_1p_2}{\Delta_2})}$ is defined as the effective value minus the true value, where the effective value for each event is calculated using $p^2_Y$ and $p^2_X$ while the true value is calculated using $m^2_Y$ and $m^2_X$.  The distribution is asymmetric, especially near $2p_1p_2/\frac{\Delta_1\Delta_2}{m^2_X}=1$.  This is due to a strong correlation between the maximum invariant mass and the effective values of $\Delta_1$ and $\Delta_2$.}
\label{o4dld}
\end{figure}

In addition, particle 2 tends to be very soft in this example and the cut on $E_T$ introduces a fake polarization for the particle $Y$ decay.  In this case, a 10~GeV  $E_T$ cut is applied to the visible particles and the resulting mean transverse angle between particle 2 in the rest frame of $Y$ and particle $Y$ in the lab frame ($\cos \theta_{2Y} \big|_Y$) is $0.0406$.  To demonstrate this fake polarization induced by the $E_T$ cut, we generated a few more examples with the same spectrum but different $E_T$ cuts.  The results are shown in Table~\ref{etcut}.  It is clear that when the visible particles are soft, a large $E_T$ cut will introduce a large fake polarization which could result in poor mass determination.

\begin{table}
\centering
\begin{tabular}{|c|c|c|}
\hline
$E_T$ cut (GeV) & $\langle\cos \theta_{2Y} |_Y\rangle$ (transverse) & $\langle\cos \theta_{2Y} |_Y\rangle$ (3-D) \\ \hline\hline
0 & -0.025 & -0.039 \\ \hline
10 & 0.041 & 0.011 \\ \hline
20 & 0.269 & 0.176 \\ \hline
30 & 0.439 & 0.315 \\ \hline
\end{tabular}
 \caption{$\langle\cos \theta_{2Y} |_Y\rangle$ (transverse and 3-D) for the spectrum in Appendix~\ref{o4} with different  $E_T$ cuts.}
 \label{etcut}
\end{table} 

The scatter plot and the peak-point plot are shown in Fig.~\ref{op4}.  It is clear that the distribution is shifted downwards.  Furthermore, because the slope is very small, the reconstructed values are very sensitive to fluctuations which brings additional difficulties to the reconstruction.  The result of mass reconstruction is shown in Table~\ref{to4}.

As a comparison, we look at the same case, but with the decay widths changed to very small values to force on-shell decays, and also with no $E_T$ cut.  The result is shown in Fig.~\ref{op0}.  It is clear that the bias presented in the previous case is removed.  This shows that indeed the $E_T$ cut and the off-shell effects are the sources of the bias.

\begin{figure}
\centering
\includegraphics[width=7cm]{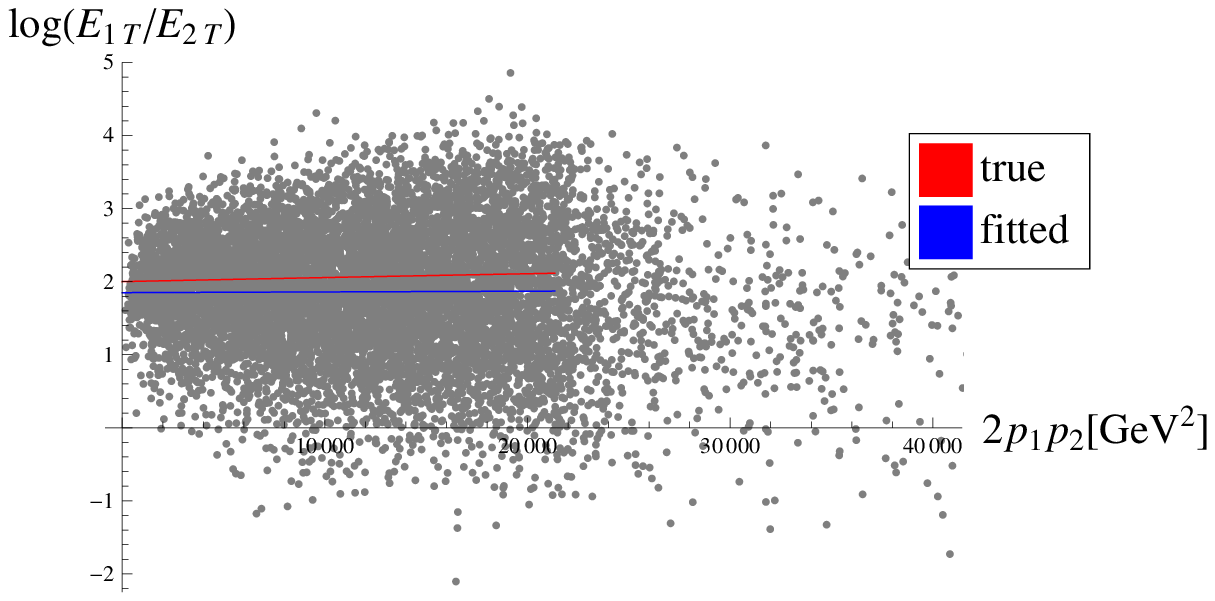}
\includegraphics[width=7cm]{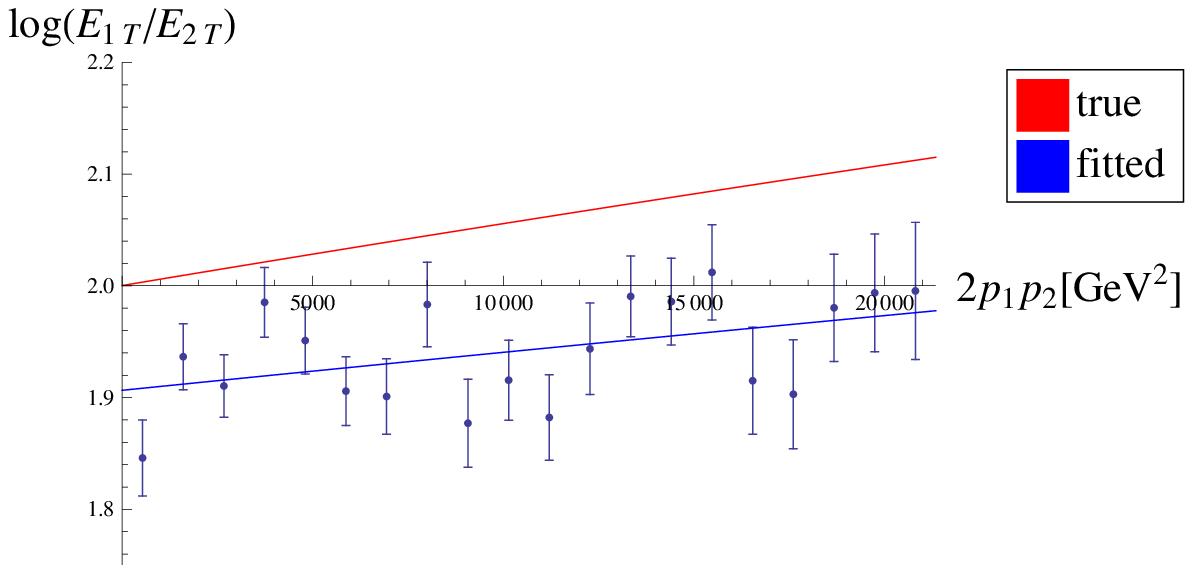}
\caption{The scatter plot (left) and the peak-point plot (right) of $\log{(E_{1T}/E_{2T})}$ vs. $2p_1p_2$ for Appendix~\ref{o4}.  In both case the red curve is $\log{(E_{1T}/E_{2T})}=\log{(\frac{\Delta_1+2p_1p_2}{\Delta_2})}$ with true $\Delta_1$ and $\Delta_2$ and the blue curve is the fitted one.}
\label{op4}
\end{figure}
\begin{figure}
\centering
\includegraphics[width=7cm]{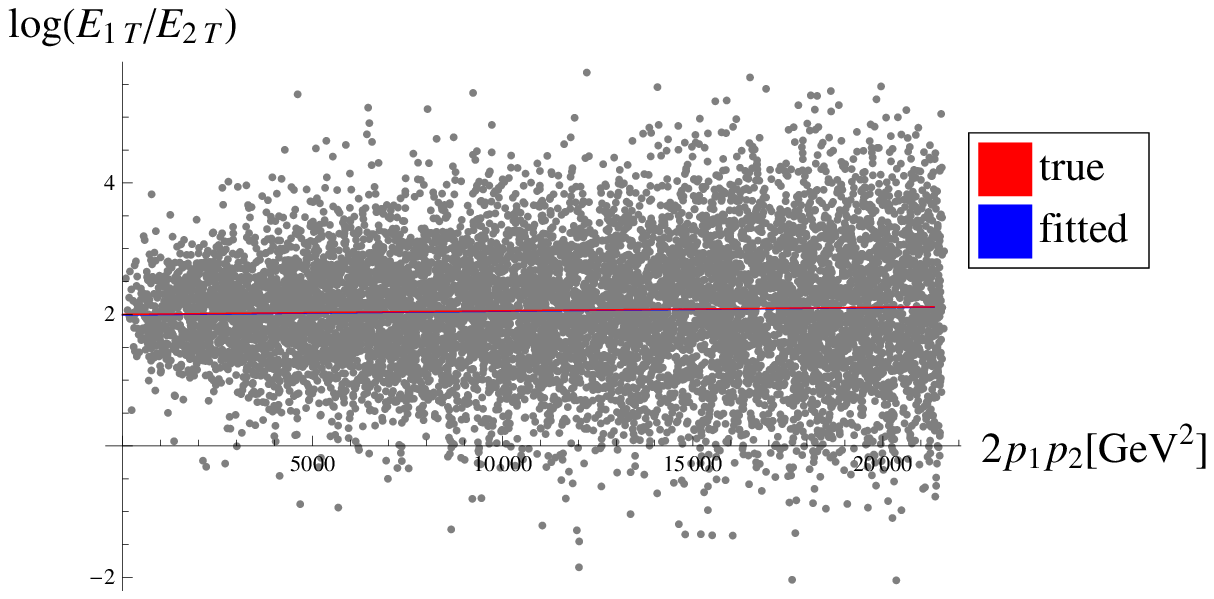}
\includegraphics[width=7cm]{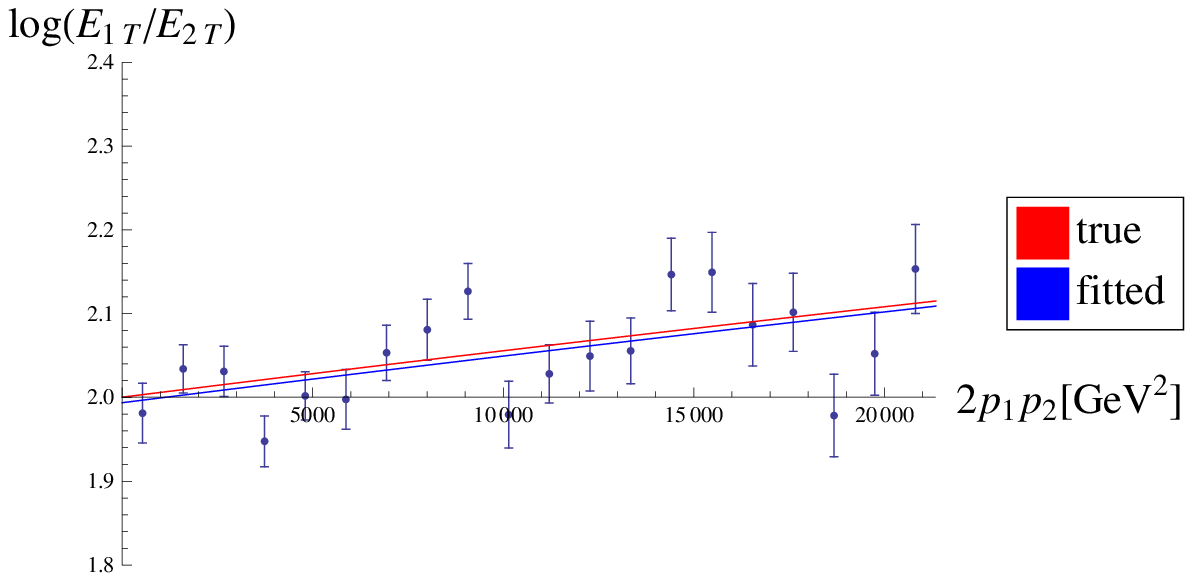}
\caption{The scatter plot (left) and the peak-point plot (right) of $\log{(E_{1T}/E_{2T})}$ vs. $2p_1p_2$ of the case with very narrow width and no $E_T$ cut in Appendix~\ref{o4}.  In both case the red curve is $\log{(E_{1T}/E_{2T})}=\log{(\frac{\Delta_1+2p_1p_2}{\Delta_2})}$ with true $\Delta_1$ and $\Delta_2$ and the blue curve is the fitted one.  The fit to the scatter plot works well.  For the peak-point plot, the fluctuation is large due to a small slope.  However, it is clear that the bias in Fig.~\ref{op4} is not present in these two examples.}
\label{op0}
\end{figure}
\begin{table}
\begin{tabular}{|c|c|c|c|c|c|c|}
\hline
   & $\Delta_1[\mbox{GeV}^2]$ & $\Delta_2[\mbox{GeV}^2]$ & $\log{(\Delta_1/\Delta_2)}$ & $m_Y$[GeV] & $m_X$[GeV] & $m_N$[GeV] \\ \hline \hline
   
 true &  $1.755\times 10^5$  & $2.375\times 10^4$  & $2.00$ & $468$ & $441.9$ & $140.5$ \\ \hline
 reconstructed &  $3.051\times 10^5$  & $4.478\times 10^4$  & $1.92$ & $827$ & $800$ & $579$ \\ \hline
error &  $+74\%$  & $+89\%$  & $-8.2\%$ & $+77\%$ & $+81\%$ & $+312\%$ \\ \hline

 \end{tabular}
 \caption{The result of mass reconstruction from fitting the peak points for Appendix~\ref{o4}.  The errors are calculated using $\frac{\rm reconstructed-true}{\rm true}$ (except for $\log{(\Delta_1/\Delta_2)}$, which is ${\rm reconstructed-true}$) and do not represent the statistical fluctuation.  The statistical fluctuation is large due to a small slope.}
 \label{to4}
 \end{table} 
  
The effects discussed above can also occur when $\Delta_1$ is very small.  To explore this case we generated events according to the spectrum in Table~\ref{tpg6}.  In addition, the width of particle $X$ is changed to about $4$ GeV to enhance the off-shell effects.  
 The scatter plot and the peak-point plot are shown in Fig.~\ref{gp6}.  The result of mass reconstruction is shown in Table~\ref{tg6}.  Although the value of $\Delta_1$ is very small, the result seems to be acceptable.  This is due to the following reasons: 1) although in the rest frame of particle $X$ the available energy for particle 1 is small, in the lab frame particle $X$ is more boosted on average than particle $Y$ so that particle 1 is less likely to be very soft; 2) the $E_T$ of particle 1 is less correlated with the relative direction of particle $Y$ because it does not directly come from the $Y$ decay; 3) the slope is large and the fluctuation is small compared with the previous case.

\begin{table}
\centering
\begin{tabular}{|c|c|c|c|}
\hline
     & $Y$ & $X$ & $N$ \\ \hline
   particle & heavy neutralino & slepton & lightest neutralino  \\ \hline
   mass[GeV] & 468 & 171 & 140.5 \\ \hline
 \end{tabular}
 \caption{A summary of the decay chain for the small $\Delta_1$ case in Appendix~\ref{o4}.}
 \label{tpg6}
 \end{table} 
\begin{figure}
\centering
\includegraphics[width=7cm]{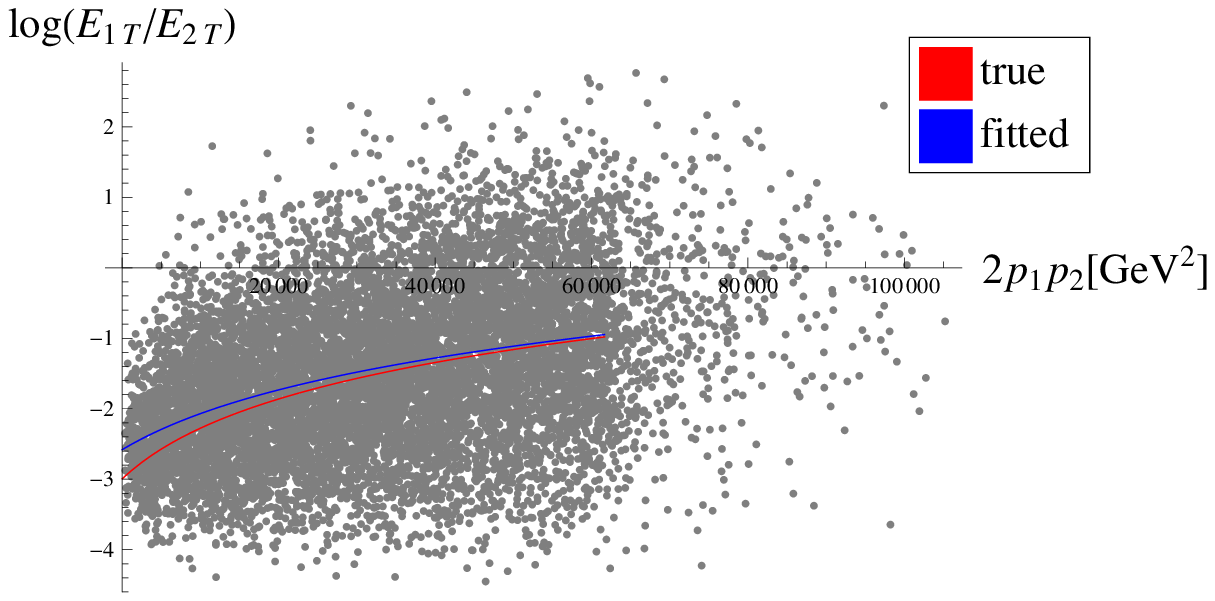}
\includegraphics[width=7cm]{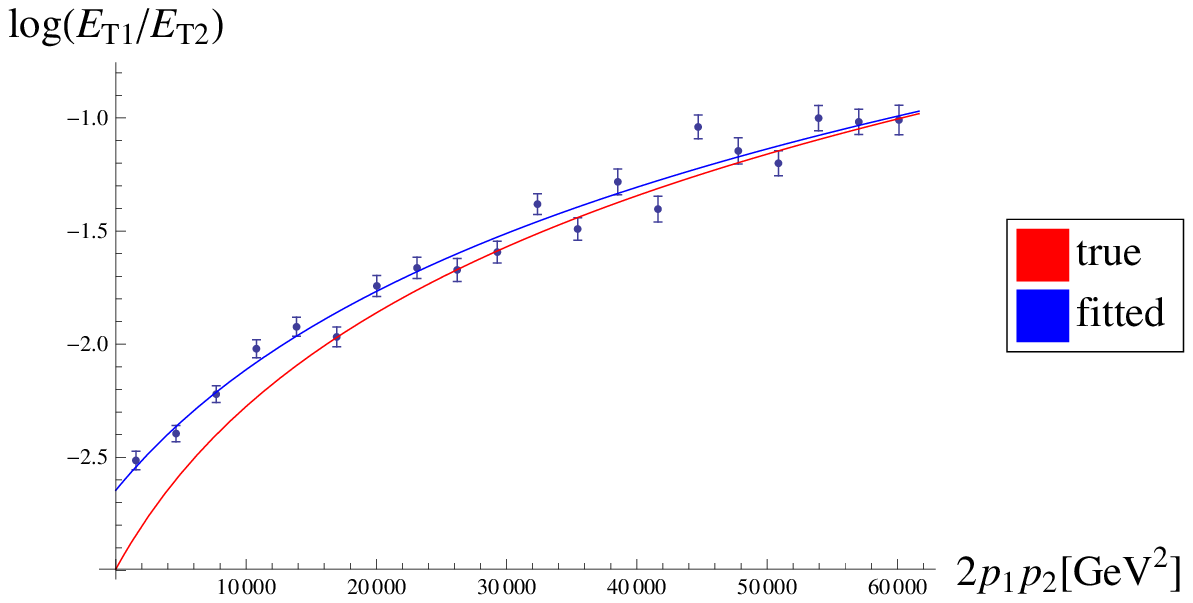}
\caption{The scatter plot (left) and the peak-point plot (right) of $\log{(E_{1T}/E_{2T})}$ vs. $2p_1p_2$ for the small $\Delta_1$ case in Appendix~\ref{o4}.  In both case the red curve is $\log{(E_{1T}/E_{2T})}=\log{(\frac{\Delta_1+2p_1p_2}{\Delta_2})}$ with true $\Delta_1$ and $\Delta_2$ and the blue curve is the fitted one.}
\label{gp6}
\end{figure}
\begin{table}
\begin{tabular}{|c|c|c|c|c|c|c|}
\hline
   & $\Delta_1[\mbox{GeV}^2]$ & $\Delta_2[\mbox{GeV}^2]$ & $\log{(\Delta_1/\Delta_2)}$ & $m_Y$[GeV] & $m_X$[GeV] & $m_N$[GeV] \\ \hline \hline
   
 true &  $9.501\times 10^3$  & $1.898\times 10^5$  & $-2.99$ & $468$ & $171$ & $140.5$ \\ \hline
 reconstructed &  $1.420\times 10^4$  & $2.001\times 10^5$  & $-2.65$ & $496$ & $215$ & $179$ \\ \hline
error &  $+49\%$  & $+5.4\%$  & $+35\%$ & $+6.0\%$ & $+26\%$ & $+27\%$ \\ \hline 

 \end{tabular}
 \caption{The result of mass reconstruction from fitting the peak points for the small $\Delta_1$ case in Appendix~\ref{o4}.  The errors are calculated using $\frac{\rm reconstructed-true}{\rm true}$ (except for $\log{(\Delta_1/\Delta_2)}$, which is ${\rm reconstructed-true}$) and do not represent the statistical fluctuation.}
 \label{tg6}
 \end{table}


\end{document}